\documentclass[12pt]{article}
\usepackage{amssymb}
\usepackage{amsfonts}
\usepackage{amsmath}
\usepackage{amsmath,amsthm}
\usepackage{subfigure}
\usepackage{graphicx,epsfig, color }
\usepackage{cite}
\oddsidemargin 10mm \numberwithin{equation}{section}
\begin{document}
\begin{center}
{{\bf {Bianchi I metric solutions with nonminimally coupled
Einstein-Maxwell gravity theory}}
 \vskip 1 cm {Hossein
Ghaffarnejad
 \footnote{E-mail address: hghafarnejad@semnan.ac.ir} and Hoda Gholipour
\footnote{E-mail address: gholipour.hoda@semnan.ac.ir}}
   \vskip 0.1 cm
   {\textit{Faculty of Physics, Semnan
            University, Semnan, 35131-19111, IRAN}}}
    \end{center}
\begin{abstract}
By using Bianchi I type of homogenous and anisotropic background
metric having cylindrical symmetry in $x$ direction of a local
cartesian coordinates system, we solve metric field equations for
a non-minimally coupled Einstein-Maxwell gravity. To do so we
choose long wavelength EM waves where spatial dependence of the
waves are negligible at the expansion duration. Motivation of this
work is study directional effects of the EM vector potential on
anisotropy trajectory of the space time expansion scale factor.
This problem is checked just for de Sitter inflationary epoch
because of its importance in the expansion of the universe. By
applying the dynamical system approach we investigate stability
nature of the cosmic system in two different cases: (a) direction
of the EM vector potential is parallel to $x$ direction of the
spacetime and (b) it is perpendicular to $x$ direction. We
obtained at stable critical points for some of obtained
inflationary solutions the EM waves behave as dark energy in cases
$a$ and $b$ where the anisotropy is negligible by expanding of the
universe while there are some stable critical points with
inflationary stable solutions in case $b$ for which the EM waves
behave as baryonic visible matter with non-vanishing anisotropy.
\end{abstract}
\section{Introduction}
Isotropy property of our universe on the large scales is known as
one of fundamental assumption in the standard cosmological model.
The well known Bianchi \cite{BI} cosmological solutions ,\cite{PO}
of the Einstein metric equations are used usually to break this
assumption. These models are investigated also via observational
data from Plank probe. For instance, Saadeh et al \cite{Da}
considered all degrees of freedom in the Bianchi solutions for the
first time, to conduct a general test of isotropy using cosmic
microwave background temperature and polarization data from Planck
probe. By considering the vector mode associated with vorticity,
they
 obtained a limit on the anisotropic expansion of the universe which is an order of magnitude tighter
 than previous Planck results that used cosmic microwave background temperature
 only. They also placed upper limits on other modes of anisotropic expansion with the weakest limit arising from the regular tensor mode.
  By including all degrees of freedom simultaneously for the first time they
  obtained statement where anisotropic expansion of the Universe
  may be strongly disfavored. But from point of view of the theoretical physics the anisotropy and other assumptions in the
  standard cosmology can be still considered as open problem.
  Applying the dynamical system approach the Aluri et al \cite{Pa} studied a
  Bianchi I universe in presence of the anisotropic sources and
  obtained some stable critical points in the extended phase space.
  They also checked the obtained solutions with the observational data,
   correspondence between analytical solutions  with numerical solutions and the de Sitter phase.
  They obtained also that the CMB anisotropy maps due to shear are also generated in this scenario,
  assuming that the universe contains anisotropic matter along with the usual matter and vacuum energy and their dark sector since decoupling.
Their solutions have also contributions dominantly to the CMB
quadrupole and possible any cosmic preferred axis present in the
data. We know now that the observational data from the Plank probe
predicts an anisotropy axis close to the mirror symmetry axis seen
in the cosmic microwave background (the axis of Evil). Sharif and
Waheed \cite{Sh} are considered the Brans Dicke scalar tensor
gravity with self-interacting potential by using magnetized
anisotropic perfect fluids model to study a Bianchi I type
cosmology. They assumed that the expansion scalar is proportional
with the shear scalar and also take a power law ansatz for the
scalar field and concluded that contrary to the universe model the
anisotropic fluid approaches isotropy at later times in all cases
which is consistent with observational data. Shamir \cite{Fa} used
Gauss-Bonnet topological invariant together with the trace of the
energy-momentum tensor as an alternative $f(F,T)$ gravity to study
anisotropic universe and concluded that presence of term $T$ in
the bivariate function f(G,T)  may gives many cosmologically
important solutions of the field equations.
 In general one can
seek in the literature to obtain numerous gravity models where the
anisotropy property of the universe is investigated  to reach to
great achievements (see for instance some of recent works as
\cite{Je}, \cite{Vi}, \cite{MM}, \cite{Ma}, \cite{Mr}, \cite{Am}.
\cite{Ta}, \cite{Ot}, \cite{Al},\cite{Ri},\cite{Vi2}, \cite{Mi} ,
\cite{Ro}, \cite{Me}, \cite{GH} and references therein). Today, we
understood this inevitable fact where the magnetic fields are
present throughout the Universe and play an important role in a
multitude of astrophysical situations. For instance the solar
winds are effect on shape of magnetosphere of the Earth and other
planets  in the solar system of our galaxy. Many other spiral
galaxies are endowed with coherent magnetic fields. They are also
affect on dynamics of the pulsars, white dwarfs and even black
holes. Theoretically the Einstein-Maxwell gravity is well known to
study cosmological systems where the both of electromagnetic and
gravity have high intensity. In usual way the latter model is
obtained by inducing a 5 dimensional Kaluza-Klein gravity into a 4
dimension (for a good review one can see \cite{we}). According to
the work \cite{TW} we apply an alternative non-minimal coupling of
the Einstein-Maxwell gravity
 to study inflationary phase of a Bianchi I type of
anisotropic cosmology via dynamical system approach. Non minimal
coupling lagrangian terms in this model are made by contraction of
the electromagnetic four vector potential and Recci tensor and
Recci scalar. We should point that at the minimal coupling regime
the Einstein-Maxwell gravity is gauge invariance while at the
non-minimal coupling model under consideration this property is
broken reaching to violate the charge conservation. Because these
additional terms break conformal invariance property of the
electromagnetic fields which is more important for amplification
of the weak cosmic magnetic fields. Authors of the work \cite{TW}
showed importance of these non-minimal coupling terms where an
inflationary expansion of the isotropic and homogenous FRW
cosmology can be produced just by including high intensity cosmic
magnetic fields. We should point that the cosmic magnetic fields
produced are small always unless the conformal invariance property
of the EM fields is broken. However, due to the physical
importance of this model both in particle physics and in
large-scale physics regimes, we like to analyze it in this paper
for Bianchi I cosmology.
Layout of this work is as follows.\\
In section 2, we define the gravity model under consideration and
calculate general form of the field equations. In section 3, we
define general form of the background metric for the Bianchi I
cosmology and generate metric field equations for this background.
In section 4, we apply to obtain metric solutions and whose
stabilizations for the Bianchi I space time via the dynamical
system approach. Section 5 denotes to concluding remark.
 \section{The Model}
Let us we start with the  non-minimally coupled Einstein-Maxwell
gravity action functional \cite{TW}
 \begin{equation}\label{action}
 I=-\int dx^4 \sqrt{g}\{\frac{1}{4}F_{\mu\nu}F^{\mu\nu}+\frac{\alpha}{2}RA^2+\frac{\beta}{2}R_{\mu\nu}A^{\mu}A^{\nu}\}
 \end{equation}
  where
 $F_{\mu\nu}=\nabla_{\mu}A_{\nu}-\nabla_{\nu}A_{\mu}$ is the antisymmetric Maxwell tensor
 field, $A_{\mu}$ is four vector potential with norm
 $A^2=g_{\mu\nu}A^{\mu}A^{\nu}.$ $R$ and $R_{\mu\nu}$ are the Recci scalar
 and the Recci tensor respectively. $g$ is absolute value of determinant of the background metric
 $g_{\mu\nu}.$
$\alpha$ and $\beta$ are non-minimally coupling constants of the
metric field $g_{\mu\nu}$ and the vector gauge field $A_{\mu}.$
$\frac{1}{4}F_{\mu\nu}F^{\mu\nu}$ is vacuum sector of the Maxwell
field Lagrangian density while other two terms are non-minimally
interacting parts. They come usually from QED in curved space
times. They break explicitly the conformal invariance of $U(1)$
group through the gravitational coupling. For $\alpha=\beta=0$ the
EM stress tensor given by (\ref{stressEM}) is traceless which
means that $T^{EM}_{\mu\nu}$ satisfies mass-less photon dynamics
while for $(\alpha,\beta)\neq0$ trace of total stress tensor given
in the right hand side of the metric field equation (\ref{Ein})
become non-zero which describes dynamics of a massive photon. This
mass is created from interaction of the photon with the gravity
background in the QED approach. In the other word the Recci scalar
in the above action behaves as mass of the photon while in the
second interacting term $R_{\mu\nu}A^{\nu}$ behaves as an
alternative EM current $J_{\mu}=R_{\mu\nu}A^{\nu}$ coming from
interaction between the photon and the geometry. However by
varying (\ref{action}) with respect to the four vector potential
$A^{\mu}$ we obtain equation of motion for the Maxwell tensor
field as follows.
\begin{equation}\label{vec}\nabla^{\mu}F_{\mu\nu}-\alpha RA_{\nu}-\beta
A_{\mu}R^{\mu}_{\nu}=0
\end{equation}
in which
\begin{equation}\label{div}\nabla^{\mu}F_{\mu\nu}=\frac{\partial^{\mu}(\sqrt{g}F_{\mu\nu})}{\sqrt{g}}\end{equation} and
for antisymmetric Maxwell tensor field $F_{\mu\nu}$ we can use the identities
\begin{equation}\label{sym}\partial_{\mu}F_{\lambda \kappa}+\partial_{\lambda}F_{\kappa\mu}+\partial_{\kappa}F_{\mu\lambda}=0.\end{equation}
 Varying the action functional (\ref{action}) with respect to the metric field
 $g^{\mu\nu}$ and removing divergence less terms after integrating by parts we obtain the metric field equation such that
\begin{equation}\label{Ein}\alpha A^2 G_{\mu\nu}
=-\alpha RA_\mu
A_\nu-\frac{1}{2}\bigg\{T^{(EM)}_{\mu\nu}+\nabla_\xi P^\xi
g_{\mu\nu}+W_{\mu\nu}\bigg\}\end{equation} where we defined
\begin{equation}\label{stressEM}
T^{(EM)}_{\mu\nu}=-\frac{1}{4}\bigg\{F_{\nu\lambda}F_{\mu}^{~~\lambda}+F_{\lambda\nu}F^{\lambda}_{~~\mu}-\frac{1}{2}g_{\mu\nu}\big(
F_{\epsilon\eta}F^{\epsilon\eta}\big)\bigg\}\end{equation}
\begin{equation}P^{\mu}=\frac{\nabla_{\nu}(\sqrt{g}\theta^{\mu\nu})}{\sqrt{g}},~~~
Q_{\lambda}^{~~\mu\nu}=\frac{\nabla_{\lambda}(\sqrt{g}\theta^{\mu\nu})
}{\sqrt{g}},~~~\theta^{\mu\nu}=\alpha A^2 g^{\mu\nu}+\beta A^\mu
A^\nu\label{stressalpha}\end{equation} and
\begin{equation}W_{\mu\nu}=\frac{\nabla_{\lambda}(\sqrt{g}K^{\lambda}_{~\mu\nu})}{\sqrt{g}},~~~K^\lambda_{~~\mu\nu}=
Q_{\mu\nu}^{~~\lambda}+Q_{\mu}~^{\lambda}~_{\nu}-Q^{\lambda}_{~~\mu\nu}\label{stressbeta}\end{equation}
To obtain the metric equation (\ref{Ein}) we used the following
identities.
\begin{equation} \delta R_{\mu\nu}=\nabla_\nu(\delta\Gamma^{\lambda}_{\mu\lambda})-\nabla_{\lambda}(\delta \Gamma^\lambda_{\mu\nu}),~~
~\Gamma_{\mu\lambda}^\lambda=\nabla_{\mu}\ln\sqrt{g},~~~\frac{\delta\sqrt{g}}{\sqrt{g}}=-\frac{1}{2}g_{\mu\nu}\delta
g^{\mu\nu}\label{identity}\end{equation}
$$\delta\Gamma^\lambda_{\mu\nu}=\frac{1}{2}g^{\lambda\rho}[\nabla_\nu\delta g_{\rho\mu}+\nabla_\mu \delta g_{\rho\nu}-
\nabla_\rho\delta g_{\mu\nu}],~~~\delta
g_{\eta\sigma}=-g_{\mu\eta}g_{\nu\sigma}\delta g^{\mu\nu }.$$ It
is easy to check $T^{\mu{(EM)}}_{\mu}=0$ because of massless
property of the photon in case $\alpha=\beta=0$ for which the
model reduces to the linear Einstein-Maxwell gravity itself.
Substituting the definition
$F^{\mu\nu}=\nabla^{\mu}A^{\nu}-\nabla^{\nu}A^{\mu}$ and identity
$(\nabla_{a}\nabla_{b}-\nabla_{b}\nabla_{a})A^c=R^c_{abd}A^d$ for
which $\nabla_{\mu}\nabla^\nu A^\mu=\nabla^\nu\nabla_\mu
A^\mu-R^\nu_{\lambda}A^\lambda$ the equation (\ref{vec}) leads to
the wave equation
\begin{equation}\label{wave}\Box A^{\nu}=\alpha RA^\nu+(\beta-1)A^\mu R_{\mu}^\nu+ \nabla^\nu (\nabla_\mu A^{\mu}).\end{equation}
Looking at the above wave equation one can infer that the first
two terms in right side are source to create the EM vector
potential $A^{\mu}$ even if we remove the Lorentz gauge invariance
condition $\nabla_\mu A^\mu.$ However the action functional
(\ref{action}) is the gauge non-invariant and so we should not
remove the last term in right side of the wave equation
(\ref{wave}). One should note that mass of the photon is related
to the curvature scalar $R$ \cite{TW} which in the cosmological
models reduces to prediction of time dependent photon mass. In the
next section we want to study this model for particular
anisotropic homogenous cosmological space time via the Bianchi I
line element.
\section{Bianchi I cosmology}
To redefine the covariant form of the Maxwell equations
(\ref{vec}) and (\ref{sym}) versus the electric and magnetic
vector fields $\overrightarrow{E}$ and $\overrightarrow{B}$
 we should fix the background metric where we will use time-dependent homogenous and anisotropic Bianchi I line element as follows.
 \begin{equation}
ds^2=-dt^2+e^{2a(t)}\{e^{-4b(t)}dx^2+e^{2b(t)}(dy^2+dz^2)\}\label{Bianchi}\end{equation}
which has cylindrical symmetry in $x$-direction.  Applying
(\ref{Bianchi}) one can infer the electromagnetic antisymmetric
tensor field can be written as follows (see appendix I).
\begin{equation}\label{Fmunu}
F_{\mu\nu}=\left(%
\begin{array}{cccc}
  0 & -e^{a-2b}E_x & -e^{a+b}E_y & -e^{a+b}E_z \\
  e^{a-2b}E_x & 0 & e^{2a-b}B_z & -e^{2a-b}B_y \\
 e^{a+b}E_y & -e^{2a-b}B_z & 0 & e^{2a+2b}B_x \\
 e^{a+b}E_z & e^{2a-b}B_y & -e^{2a+2b}B_x & 0 \\
\end{array}%
\right)
\end{equation}
 where $E_{x,y,z}$ and $B_{x,y,z}$ are
Cartesian components of the electric and magnetic vector fields
respectively. Substituting (\ref{Bianchi}) into the equation
(\ref{vec}) and (\ref{sym}) we obtain the Maxwell equations as
follows.
\begin{equation}\vec{\nabla}\cdot\vec{E^*}=-e^a\phi_b-3A_tM^2e^{2(a+b)},
\label{at}\end{equation}
\begin{equation}e^{-3a}\partial_t[e^{3a}\vec{E^*}]-e^{-2(a+b)}\vec{\nabla}\times\vec{B^*}+N^2\vec{A}=\psi_b\hat{k}
\label{EEs}\end{equation} and
\begin{equation}\partial_t
\vec{B^*}+\vec{\nabla}\times\vec{E^*}=0\label{Bs}\end{equation}
where we defined
\begin{equation}E^*_x=e^{a-2b}E_x,~~~E^*_{y,z}=e^{a+b}E_{y,z},~~~B^*_x=e^{2(a+b)}B_x,~~~B^*_{y,z}=e^{2a-b}B_{y,z}\end{equation}
\begin{equation}\phi_b=2e^b\sinh3b\partial_xE_x,~~~\psi_b=2e^{2b}\sinh b\partial_xB_y,~~~\vec{A}=A_x\hat{i}+A_y\hat{j}+A_z\hat{k}\end{equation}
\begin{equation}N^2=3(\beta+4\alpha)\dot{a}^2+6\alpha\dot{b}^2+(\beta+6\alpha)\ddot{a}
+3\beta\dot{a}\dot{b}+\beta\ddot{b}\label{N}\end{equation} and
\begin{equation}M^2=(\beta+4\alpha)\dot{a}^2+2(\alpha+\beta)\dot{b}^2+(\beta+2\alpha)\ddot{a}.\label{M}\end{equation}
The equation (\ref{at}) is the Gauss law.  The equation
(\ref{EEs}) is  the Ampere`s law and the equation (\ref{Bs}) is
the Faraday`s law. The forth equation of the EM Maxwell equation
is
\begin{equation}\vec{\nabla}\cdot\vec{B}^*=0.\end{equation}  Taking the curl of the equation (\ref{EEs}) and using the
equation (\ref{Bs}) to eliminate $\vec{E^*}$ we find wave equation
for the cosmic magnetic field as
\begin{equation}e^{-3a}\partial_t[e^{3a}\partial_t\vec{B^*}]-e^{-2(a+b)}\nabla^2\vec{B^*}-N^2\vec{B}=\hat{k}\times\vec{\nabla}\psi_b\label{bs}\end{equation}
where we set
\begin{equation}\label{curl}\vec{B}=\vec{\nabla}\times\vec{A},~~~~\vec{\nabla}\cdot\vec{B^*}=0.\end{equation}
Taking the curl of the equation (\ref{Bs}) and using the equations
(\ref{at}) and (\ref{EEs}) to eliminate $\vec{B^*}$ and
$\vec{\nabla}\cdot\vec{E^*}$ we find wave equation for the cosmic
electric field as
\begin{equation}\partial_t\{e^{2b-a}\partial_t(e^{3a}\vec{E^*})\}-\nabla^2\vec{E^*}=2e^a\vec{\nabla}\phi_b+3
e^{2(a+b)}M^2\vec{\nabla}A_t+\partial_t(\psi_b\hat{k}-N^2\vec{A})\label{es}\end{equation}
Now we are in position to solve the metric field equation
(\ref{Ein}) for which we can choose two ways namely short and long
wave-length electromagnetic waves. In fact the cosmic inflation
provides very long wavelength EM waves at very early times through
microphysical processes operating on scales less than the Hubble
radius. In this regime we can ignore spatial dependence of the
electromagnetic fields and consider just time dependence for the
EM fields. While when we want to study expansion of the universe
before than the inflation where the wavelength of the EM waves are
comparable with the Hubble radius and so we should consider
spatial dependence of the EM waves in the cosmic dynamics together
with their time dependence. In the latter approach we can consider
the time dependent EM fields made from fourier transform of
wavelets which are normalized to the Hubble radius at the
presentence. This approach is studied in ref. \cite{TW} for the
FRW metric by applying the alternative Einstein-Maxwell gravity
(\ref{action}). According to this short description about
importance of spatial and time dependence of the EM waves which
behave before the beginning of the inflationary epoch of the
expansion of the universe we will study short wave-length EM waves
effects on the expansion of the universe, namely radiation and
dust dominant in our next work. Here we seek just time dependent
long wavelength EM waves to produce inflationary metric solutions
of the Bianchi I cosmology in presence of the anisotropy property.
\section{Metric solutions by long wavelength EM waves} According to the description of the previous part for long
wavelength EM fields we choose
\begin{equation}A^{\mu}=A^{\mu}(t)\label{time}.\end{equation} for which the EM tensor field $F_{\mu\nu}(t)$ reads
\begin{equation}\label{bt}\vec{B}(t)=0,~~~~\vec{E}^*(t)=-\vec{\dot{A}}(t)\end{equation} where over dot denotes to time derivative.
   Applying (\ref{time}) and (\ref{bt}) the Gauss`s law (\ref{at}) reads to the following condition
\begin{equation}A_t(t)=0\label{At0}\end{equation} and the equation (\ref{EEs}) reads
\begin{equation}\ddot{v}-(\beta+6\alpha)\ddot{a}-\beta\ddot{b}=-3\dot{v}\dot{a}-\dot{v}^2+3(\beta+4\alpha)\dot{a}^2+6\alpha\dot{b}^2
+3\beta\dot{a}\dot{b} \label{Es1}\end{equation} and
\begin{equation}\ddot{u}-(\beta+6\alpha)\ddot{a}-\beta\ddot{b}=-3\dot{u}\dot{a}-\dot{u}^2+3(\beta+4\alpha)\dot{a}^2+6\alpha\dot{b}^2
+3\beta\dot{a}\dot{b}. \label{Es2}\end{equation} where we defined
\begin{equation}A_x=A_{\|}=e^{v(t)},~~~A_y=A_z=A_{\perp}=e^{u(t)}.\end{equation}
 We assumed $A_y=A_z$ because of cylindrical symmetry of the
space time in $x$ direction which reads $G_{yy}=G_{zz}$ for the
Einstein tensor transverse components. By regarding this symmetry
in the right side stress tensors of the metric field equation
(\ref{Ein}) we must solve the metric field equation (\ref{Ein})
for two different situations of the EM waves as
$(A_{||}\neq0,A_{\perp}=0)$ or vice versa. In this case
non-diagonal terms of the stress tensors in right side of the
metric equation (\ref{Ein}) are eliminated and also $yy$ component
will has similar form with the $zz$ component in the right side
stress tensor satisfying  $G_{yy}=G_{zz}.$ In fact we have three
field equations which should be determined as $\{v(t),a(t),b(t)\}$
for case $A_{\|}\neq0,A_{\perp}=0$ and $\{u(t),a(t),b(t)\}$ for
case $A_{\|}=0,A_{\perp}\neq0$ respectively. To obtain
corresponding dynamical field equations it will be useful we
assume that the matter stress tensor in right side of the Einstein
metric equation (\ref{Ein}) behaves as anisotropic perfect fluid
meaning that they have no viscosity or heat flow. This assumption
is usual way in the cosmological studies. In this case we can use
the following correspondence.
\begin{equation}G^\mu_{\nu}=-\{T^{\mu(EM)}_{\nu}+\alpha T^{\mu(\alpha)}_{\nu}+\frac{\beta}{2}T^{\mu(\beta)}_\nu\}/\alpha
A^2=8\pi diag\{\rho,-p_x,-p_y,-p_z\} \label{perfect}\end{equation}
where $\rho$ is the mass density and $p_{x}=p_{||}$ and
$p_y=p_z=p_{\perp}$ are the directional pressures. In this case we
define directional barotropic indexes as follows.
\begin{equation}\label{baro}-\gamma_{\|}=\frac{G^x_x}{G^t_t}=\frac{p_{\|}}{\rho},~~~-\gamma_{\perp}=\frac{G^y_y}{G^t_t}=\frac{p_{\perp}}{\rho}\end{equation}
which by substituting the metric line element (\ref{Bianchi}) we
can obtain their differential equations form given in the appendix
II and they reduce to the following formes respectively in limits
$e^{a(t)}>>1$ (the inflation epoch).
\begin{equation}2\alpha\ddot{v}-\alpha\ddot{a}-4\alpha\ddot{b}=4\alpha\dot{a}^2+16\alpha\dot{b}^2+2\alpha\dot{a}\dot{b}+2(6\alpha+\beta)\dot{a}\dot{v}+2(2\beta-6\alpha)\dot{v}\dot{b}\label{ev1}\end{equation} and
\begin{equation}-2\ddot{v}+\ddot{a}+4\ddot{b}=4\dot{v}^2-4\dot{a}^2-16\dot{b}^2-2\dot{a}\dot{b}\label{ev2}\end{equation}
in case $A_{\|}\neq0,A_{\perp}=0$ and
\begin{equation}2\alpha\ddot{u}+[2\alpha-\beta+\gamma_{\|}(\alpha-\beta)]\ddot{a}-[\beta+4\alpha+\gamma_{\|}(\beta+2\alpha)]\ddot{b}\label{eu1}\end{equation}
$$=-[4\alpha+1/4-\gamma_{\|}/8]\dot{u}^2+[2\beta-14\alpha-2(4\alpha+\beta)\gamma_{\|}]\dot{a}\dot{u}$$$$
+[28\alpha+2\beta+\gamma_{\|} (14\alpha+\beta)]\dot{b}\dot{u}
-[\beta+48\alpha-\gamma_{\|}(\beta+6\alpha)]\dot{a}\dot{b}$$$$-[\beta+18\alpha-\gamma_{\|}(\beta+4\alpha)]\dot{a}^2+[24\alpha-2\beta-
2(2\alpha-\beta)\gamma_{\|}]\dot{b}^2$$ and
\begin{equation}
2\alpha\ddot{u}+[7\alpha-5\beta/2+\gamma_{\perp}(\alpha-\beta)]\ddot{a}-(\beta+2\alpha)(1+\gamma_{\|})\ddot{b}\label{eu2}\end{equation}
$$
=-[3\beta-\gamma_{\perp}(6\alpha+\beta)]\dot{a}\dot{b}-[17\alpha-17\beta/2+\gamma_{\perp}(2\beta+8\alpha)]\dot{u}\dot{a}$$$$+
[16\alpha+6\beta+\gamma_{\perp}(14\alpha+\beta)]\dot{u}\dot{b}
-(4\alpha-\gamma_{\perp}/8)\dot{u}^2$$$$-[12\alpha-6\beta+\gamma_{\perp}(\beta+4\alpha)]\dot{a}^2
-[6\alpha+2\gamma_{\perp}(2\alpha-\beta)]\dot{b}^2$$
 in case $A_{\|}=0,A_{\perp}\neq0$ where over dot means the differentiation with respect to cosmic
time $`t`.$ Now we investigate the inflationary epoch solutions of
the above field equations in two different situations for the time
dependent EM vector potential $\vec{A}(t)$ as follows.
\subsection{Metric solutions for $A_{\|}\neq0,A_{\perp}=0$}
In this case we should obtain $\{v(t), a(t), b(t)\}$ by solving
the equations (\ref{Es1}), (\ref{ev1}) and (\ref{ev2}). These
equations are nonlinear two order differential equations for which
we can obtain analytic solutions for the fields near the critical
points in phase space. This can be done via dynamical system
approach. To do so we first define
\begin{equation} V(t)=\dot{v},~~~~X(t)=\dot{b},~~~~H(t)=\dot{a}\label{timdim}\end{equation}
 and substitute (\ref{Es1}) into (\ref{ev2}) to eliminate second order differentiation of the fields such that
\begin{equation}2(2\beta-6\alpha)X+4\alpha V+(6\alpha+\beta)H=0.\label{hubb}\end{equation}
 here $H$ is the
Hubble parameter. Substituting the identity (\ref{hubb}) and the
definitions (\ref{timdim}) into the equations (\ref{Es1}) and
(\ref{ev2})
 they reduce to the following forms respectively.
\begin{equation}\dot{V}=C_{VV}V^2+C_{XV}XV+C_{XX}X^2
\label{V1}\end{equation}
\begin{equation}\dot{X}=D_{VV}V^2+D_{XV}XV+D_{XX}X^2\label{V2}\end{equation}
where explicit forms of the coefficients $C_{ij}(\alpha,\beta)$
and $D_{ij}(\alpha,\beta)$ with $\{i,j\}=\{V,X\}$ are given versus
the  $\alpha$ and the $\beta$ parameters of the theory in the
appendix III.
 Now we should investigate solutions of the equations (\ref{V1}) and (\ref{V2}) near some
 possible critical points where
 these equations should behave as
linear differential equations. What is theoretically important for
our solutions is to find the stability conditions of the solutions
near their critical points $\{V_c,X_c\}$. To do so we should first
obtain critical points of the above equations for which we assume
two  dimensional phase space defined by a vector field
$\vec{\xi}(t)=\{V(t),X(t)\}$ which is a constant vector field at
the critical points. Thus one can infer that the critical points
in the phase space are determined by solving the equations
$\vec{\dot{\xi}}=0.$ Near the critical points one can obtain time
evolutions of the vector field $\vec{\xi}(t)$ by the linear
equations $\dot{\xi}_i=\sum_{j=1}^{2}J_{ij}\xi_j$ for $2$
dimensional phase space where
$J_{ij}=\frac{\partial\dot{\xi}_i}{\partial\xi_j}_{\mid{\vec{\xi}_c}}$
is Jacobian matrix of the dynamical field equations at the
critical points $\vec{\xi}_c$ in which the above equations behave
as linear one order differential equations. By solving the
Jacobian secular equation one can obtain eigenvalues where
negative (positive) real eigenvalues show stable (unstable) nature
of the obtained solutions. In general, if the obtained eigenvalues
have imaginary part then the nature of the solutions near the
critical points will be spiral stable (unstable) if their real
part become negative (positive). Usually in the dynamical system
approach stable (unstable) state called as sink (source) for
absolutely real eigenvalues and spiral stable (unstable) for
eigenvalues with complex valued (see \cite{GH3} and references
therein). After some descriptions about stability nature of the
solutions via dynamical system approach, we investigate to obtain
critical points by solving $\dot{V}=\dot{X}=0$ for de Sitter phase
$\dot{H}=0$ in the equations (\ref{V1}) and (\ref{V2}).
 This is down to
obtain the critical points given by
\begin{equation}\label{XCVC}\frac{X_c}{V_c}=\chi(\alpha,\beta)=\frac{C_{XV}D_{VV}-C_{VV}D_{XV}}{C_{VV}D_{XX}-C_{XX}D_{VV}}
\end{equation}
 with
\begin{equation}\label{cond}f(\alpha,\beta)=C_{VV}\chi^2+C_{XV}\chi+C_{XX}=0\end{equation}
where all real possible values for $(\alpha,\beta)$ are given in
figure 1 by plotting the equations (\ref{cond}).
 Near the critical point we obtain
metric components of the line element (\ref{Bianchi}) as follows.
\begin{equation}e^{a(t)}=e^{Ht},~~~~e^{b(t)}=e^{X_Ct},~~~A_{\|}(t)=e^{V_C t}\label{solpar}\end{equation}
where
\begin{equation}\frac{X_C}{H}=\frac{(6\alpha+\beta)\chi}{4\alpha(3\chi-1)-4\beta\chi},~~~\frac{V_C}{H}=\frac{6\alpha+\beta}{4\alpha(3\chi-1)-4\beta\chi}
\end{equation}
 and observational value for the Hubble constant is
$\frac{1}{H}=4.55\times10^{17}sec.$  For the critical points given
in the equations (\ref{XCVC}) and (\ref{cond}) one can solve
secular equation $det\{J_{ij}-\lambda \delta_{ij}\}=0$ of the
Jacobi matrix of the dynamical field equations (\ref{V1}) and
(\ref{V2}) given by
\begin{equation}\label{Jacob}J_{ij}=\Delta\left(%
\begin{array}{cc}
  2C_{VV}+\chi C_{XV} & C_{XV}+2\chi C_{XX} \\
  2D_{VV}+\chi D_{XV} & D_{XV}+2\chi D_{XX} \\
\end{array}%
\right)\end{equation} as
\begin{equation}\bigg(\frac{\lambda}{\Delta}\bigg)^2-\Omega\bigg(\frac{\lambda}{\Delta}\bigg)+\Pi=0\label{secular}\end{equation}
with eigenvalues
\begin{equation}\label{lambda}\frac{\lambda_{\pm}}{\Delta}=\frac{\Omega\pm\sqrt{\Omega^2-4\Pi}}{2}\end{equation}
where we defined
\begin{equation}\label{delta}\Delta=\frac{(6\alpha+\beta)H}{4\alpha(3\chi-1)-4\beta\chi},\end{equation}
\begin{equation}\label{Omega}\Omega=2C_{VV}+D_{XV}+\chi(2D_{XX}+C_{XV})\end{equation}
and
\begin{equation}\label{Pi}\Pi=(C_{VV}+\chi C_{XV})(D_{XV}+2\chi D_{XX})-(C_{XV}+2\chi C_{XX})(2D_{VV}+\chi
D_{XV})\end{equation} Eigenvalues (\ref{lambda}) are parametric
solutions which their negativity/positivity sign are dependent to
numerical values of the $\alpha$ and the $\beta$ parameters of the
model. We use numerical method and set the ansatz
$\epsilon=-10....10,$ to collect numerical solutions for different
values of these theory parameters by defining
\begin{equation}\epsilon=\frac{\beta}{\alpha}\end{equation} for which the equation (\ref{cond})
reduces to multiplication of two different algebraic equations as
$g(\alpha,\beta)h(\alpha,\beta)=0$ for which we defined
\begin{equation}\label{g}g(\alpha,\beta)=-\alpha+\frac{-9(\epsilon^2-48\epsilon-36)}{5\epsilon^4+78\epsilon^3+244\epsilon^2-504\epsilon-3168}=0
\end{equation} and
\begin{equation}\label{h}h(\alpha,\beta)=\eta_3(\epsilon)\alpha^3+\eta_2(\epsilon)\alpha^2+\eta_1(\epsilon)\alpha+\eta_0(\epsilon)=0\end{equation}
where $\eta_{0,1,2,3}(\epsilon)$ are given in the appendix III. By
fixing numerical values for $\alpha$ and $\beta$ obtained from
(\ref{g}) and (\ref{h}) we can determine which of the eigenvalues
(\ref{lambda}) have negative numerical value showing stable state
of our metric solutions. This is done by numerical calculations
via MATHEMATICA software where we collected results of the
calculations in the table 1 for (\ref{g}) and in the table 2 for
(\ref{h}) respectively. In these tables we used the following
equations for the directional barotropic indexes
\begin{equation}\label{barc}\gamma_{\|}=-\bigg(\frac{1+2X_c/H+X_c^2/H^2}{1-X_c^2/H^2}\bigg),\gamma_{\perp}=
-\bigg(\frac{1-X_c/H+X_c^2/H^2}{1-X_c^2/H^2}\bigg).\end{equation}
which are obtained by substituting (\ref{Gtt}), (\ref{Gxx}) and
(\ref{Gyy}) into (\ref{baro}) at the critical points
$\{X_c,V_c\}.$ Their numerical values for the stable points given
in the table 1 are collected in the table 3 where
$\bar{\gamma}=(\gamma_{\|}+2\gamma_{\perp})/3$ at the last column
is average barotropic index. Looking at the numerical values of
the barotropic indexes given in the table 3 one can infer that in
the case $A_{\|}\neq0,A_{\perp}=0$ the EM fields behave as dark
sector of the cosmic source in the inflation.
 In figure 1 we plot arrow diagrams of the stable
points given in the table 1.
  In general, stable nature of the system is just for case where all of
eigenvalues have negative numeric values. According to this
definition we collected directional barotropic indexes just for
stable critical points in phase space $\{X,V\}.$ By looking at the
table 1 we infer $\beta\neq0$ corresponded to  stable inflationary
solutions shows essential behavior of the last term in the action
functional (\ref{action}) to produces possible anisotropic
trajectories if and only if $A_{\|}\neq0,A_{\perp}=0.$ Looking at
the obtained critical points $X_{c}$ in the table 1 we see that
the anisotropy of the space time is negligible at large scales
because from the obtained solutions (\ref{solpar}) one can see
\begin{equation}\lim_{t\to\infty}\frac{e^{b(t)}}{e^{a(t)}}=\lim_{t\to\infty}e^{(\delta-1)Ht}\to0\end{equation}
where we defined $\delta=\frac{X_c}{H}$ which their numerical
values as $0<\delta<1$ are addressed from the table 1 for the
stable solutions. In this case the EM waves behave as dark sector
of the cosmic source because of negativity sign of the numeric
values of the directional barotropic indexes given in the table 3.
In the next subsection we check other possible stable inflationary
solutions of the background metric for the case
$A_{\|}=0,A_{\perp}\neq0$.
\subsection{Metric solutions for $A_{\|}=0,~A_{\perp}\neq0$}
In this case for large scale regime of the space time
$\exp{a(t)}>>1$ we substitute $U=\dot{u}$ and $X=\dot{b}$ and
$H=\dot{a}$ into the equations (\ref{Es2}), (\ref{eu1}) and
(\ref{eu2}) to obtain first order nonlinear differential equations
of the cosmic system as follows.
\begin{equation}\dot{U}=\frac{E_1U^2+E_2H^2+E_3X^2+E_4HU+E_5HX+E_6UX}{J(\alpha,\beta, \gamma_{\|},\gamma_{\perp})}\label{eu11},\end{equation}
\begin{equation}\dot{H}=\frac{M_1U^2+M_2H^2+M_3X^2+M_4HU+M_5HX+M_6UX}{J(\alpha,\beta, \gamma_{\|},\gamma_{\perp})}\label{eu22}\end{equation}
\begin{equation}\dot{X}=\frac{N_1U^2+N_2H^2+N_3X^2+N_4HU+N_5HX+N_6UX}{J(\alpha,\beta, \gamma_{\|},\gamma_{\perp})}\label{eu33}\end{equation}
where explicit form of the functions $E_i,M_i,N_i$ and $J_i$ with
$i=1,2...,6$ are given in the appendix IV. To obtain critical
points $\{U_C,X_C\}$ it is useful we define
\begin{equation}W=\frac{U_C}{H}\label{Uc}\end{equation}
\begin{equation}S=\frac{X_C}{H}\label{Xc}\end{equation}
for which the equations of the critical points given by $E=0=M=N$
reduce to the following equations.
\begin{equation}W(S,\alpha,\epsilon)=\frac{P_1(S,\alpha,\epsilon)}{P_2(S,\alpha,\epsilon)}\end{equation}
\begin{equation}Q_1(S,\alpha,\epsilon)K(S,\alpha,\epsilon)=0\end{equation} and \begin{equation}Q_2(S,\alpha,\epsilon)K(S,\alpha,\epsilon)=0
\end{equation}
where explicit forms of the functions $Q_{1,2}$ and $K$ are given
in the appendix IV. According to assumption
$-10\leq\epsilon\leq10$ in the previous subsection they are used
again in the present subsection to calculate numerical solutions
of the critical point equations $Q_1=0=Q_2.$ We collected all
numeric values of the critical points in the table 4 and
corresponding eigenvalues and directional barotropic indexes are
collected in the tables 5 and 6 respectively. Looking at these
tables we infer that the obtained solutions predict some stable
metric solutions which for some of them the anisotropy vanishes by
expansion of the space time but not for some others. In fact for
our obtained critical points in the table 4 where $S<1$ the
anisotropy vanishes by raising the scale factor which can be
checked easily as follows.
\begin{equation}\label{zero}\lim_{t\to\infty}\frac{e^{b(t)}}{e^{a(t)}}=e^{(S-1)Ht}\to0\end{equation}
where $S<1$ should be substituted from the table 6.  However the
table 4 shows that there is still some stable nature critical
points for which $S>1$ and so the anisotropy property of the space
time rises by increasing the scale factor. The latter solutions
may not be physical metric solutions because they do not satisfy
the condition (\ref{zero}) corresponding to the observational data
about the anisotropy at the present epoch of the universe. But
they have positive numeric values for the corresponding barotropic
indexes which behave as visible baryonic ordinary matter EM waves
in case $A_{\|}=0,A_{\perp}\neq0.$ This let us to claim that the
massive time dependent photons described by the therm $RA^2$ in
the action functional (\ref{action}) behaves as an baryonic
visible matter and it can still support inflation of the universe
instead of the unknown dark matter/energy if and only
 if $A_{\|}=0,A_{\perp}\neq0.$ This is a promising result for our solutions where
a non-minimal interacting Einstein-Maxwell gravity can produce an
anisotropic exponentially inflation for the universe without to
use the unknown additional cosmological constant or dark sector of
the cosmic matter.
\section{Concluding Remarks}
We showed non-minimally coupled Einstein-Maxwell gravity produces
some stable inflationary expansion of metric solutions for the
Bianchi I cosmology where the anisotropy is negligible at large
scales of the spacetime if  direction of the EM vector potential
be parallel to the cylindrical symmetry of the spacetime and its
stress tensor behaves as dark sector of the cosmic perfect fluid
(see tables 1 and 3 and figures 1). When direction of the EM
vector potential is perpendicular to the cylindrical symmetry axis
of the spacetime we obtained two different class of stable
inflationary metric solutions which for some of them the
anisotropy is negligible again and their stress tensor behave as
dark energy while some other solutions behave as baryonic visible
matter for which the anisotropy factor of the space time increases
at large scale of the space time. We used dynamical system
approach to study stability nature of these obtained solutions and
applied numerical methods via Mathematica software to produce
numeric values of the physical and geometrical quantities.\\
Outlook of this work can be pointed as follows: Interacting
massive photons with the geometry described by the action
functional under consideration can support the inflation of the
universe with negligible anisotropy, instead of the unknown dark
energy. As a future work we like to investigate this problem for
short wavelength EM waves by considering spatial dependence of the
EM fields containing viscose terms which are applicable to study
the Big Bang singularity regime before than the inflation. One of
the authors was checked previously removing the naked cosmic
singularity by alternative gravity models with and without
anisotropy property of the space time via canonical quantum
cosmology approach. In fact quantum uncertainty on the cosmic
dynamical quantities may to be resolve the cosmic naked
singularity (see \cite{GH} and \cite{GH1}). This encourage us to
investigate quantum cosmological behavior of the model under
consideration to study relationship between anisotropy of the
spacetime and the naked singularity which is predicted from the
standard FRW classical cosmology. In fact the anisotropy in
space-time seems to explain the naked singularity predicted by FRW
cosmology.

 \vskip .5cm
\noindent
\begin{figure}[ht] \centering {\label{1}}
\includegraphics[width=0.32\textwidth]{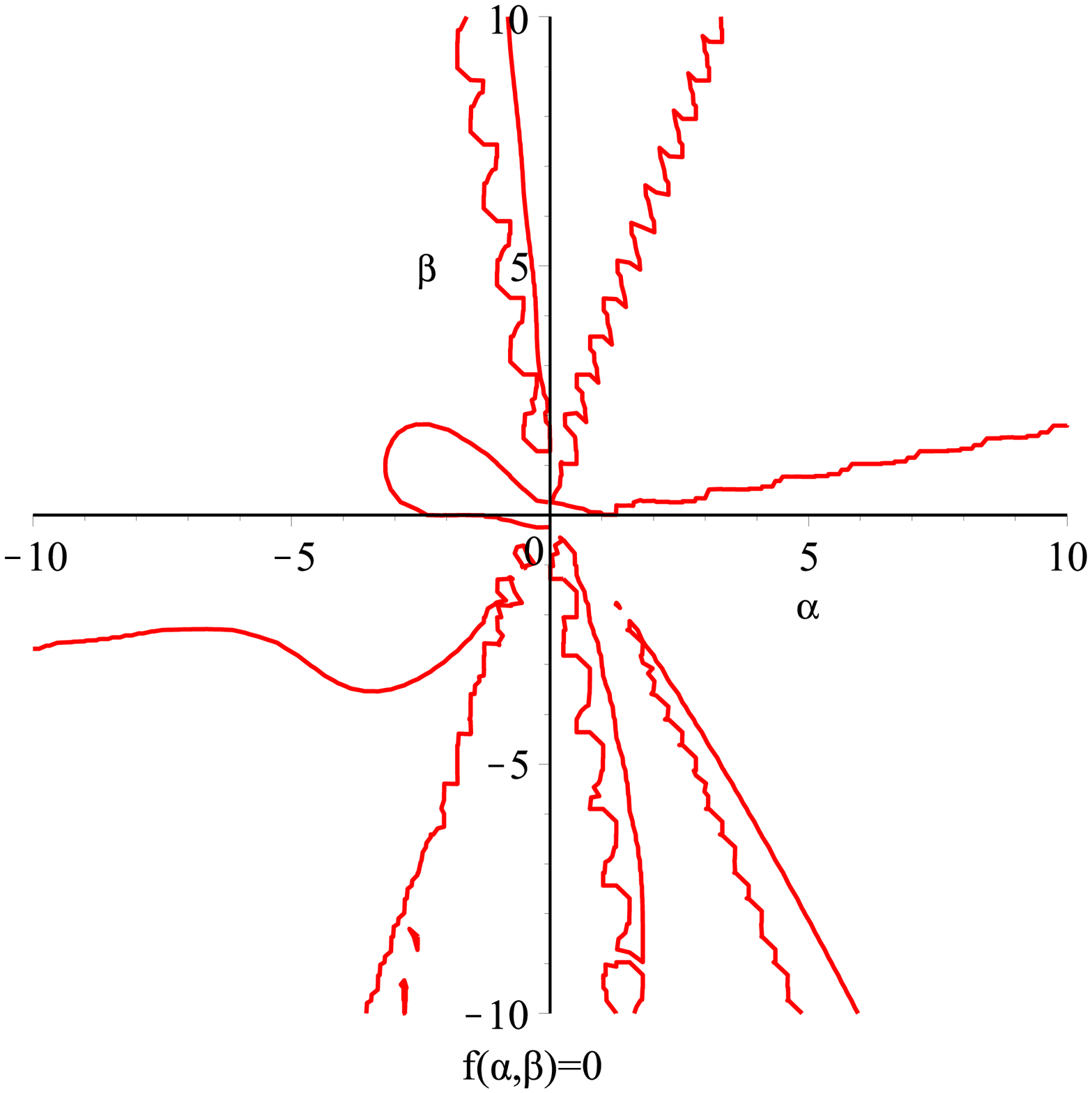}
\includegraphics[width=0.32\textwidth]{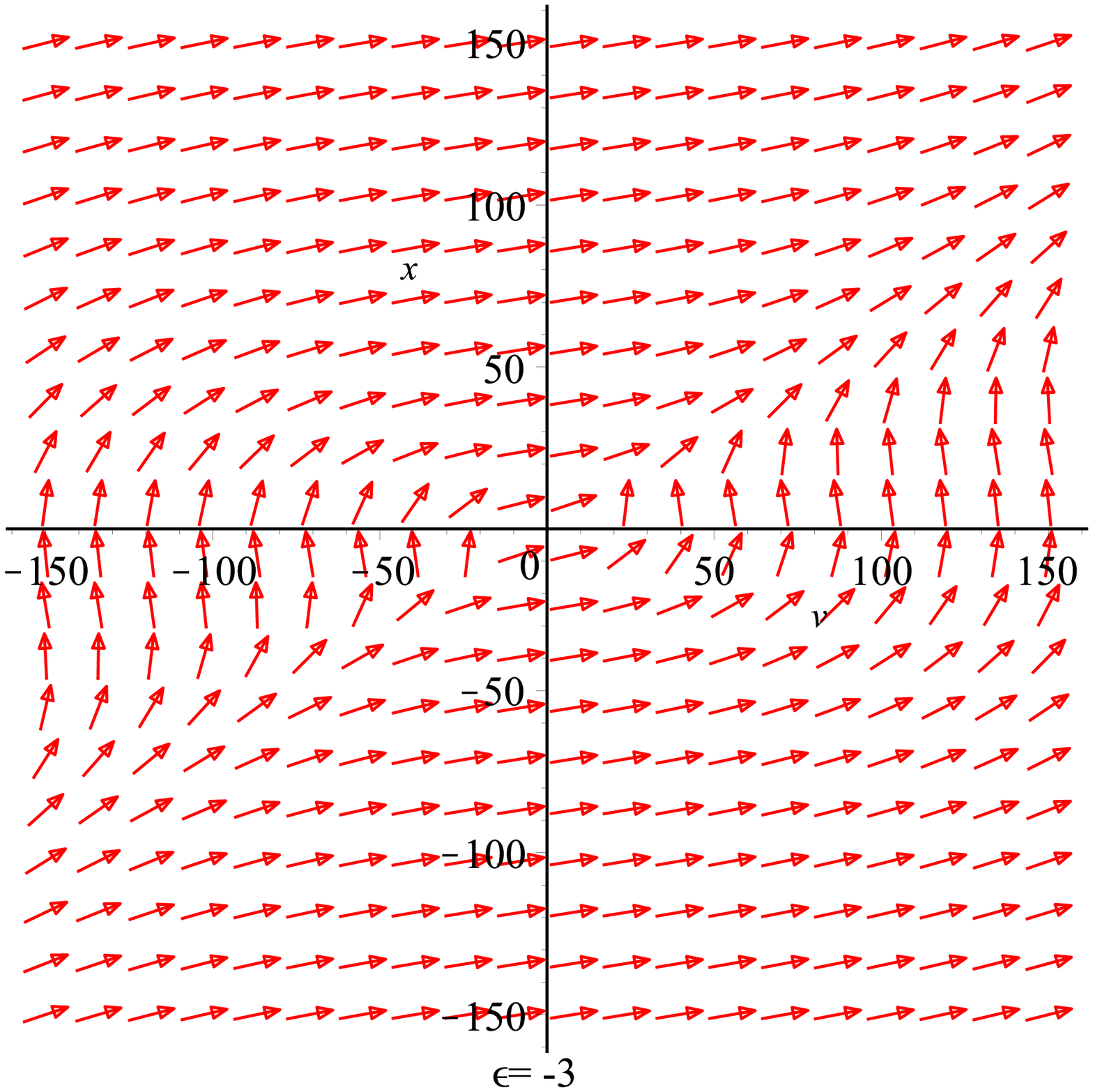}
\includegraphics[width=0.32\textwidth]{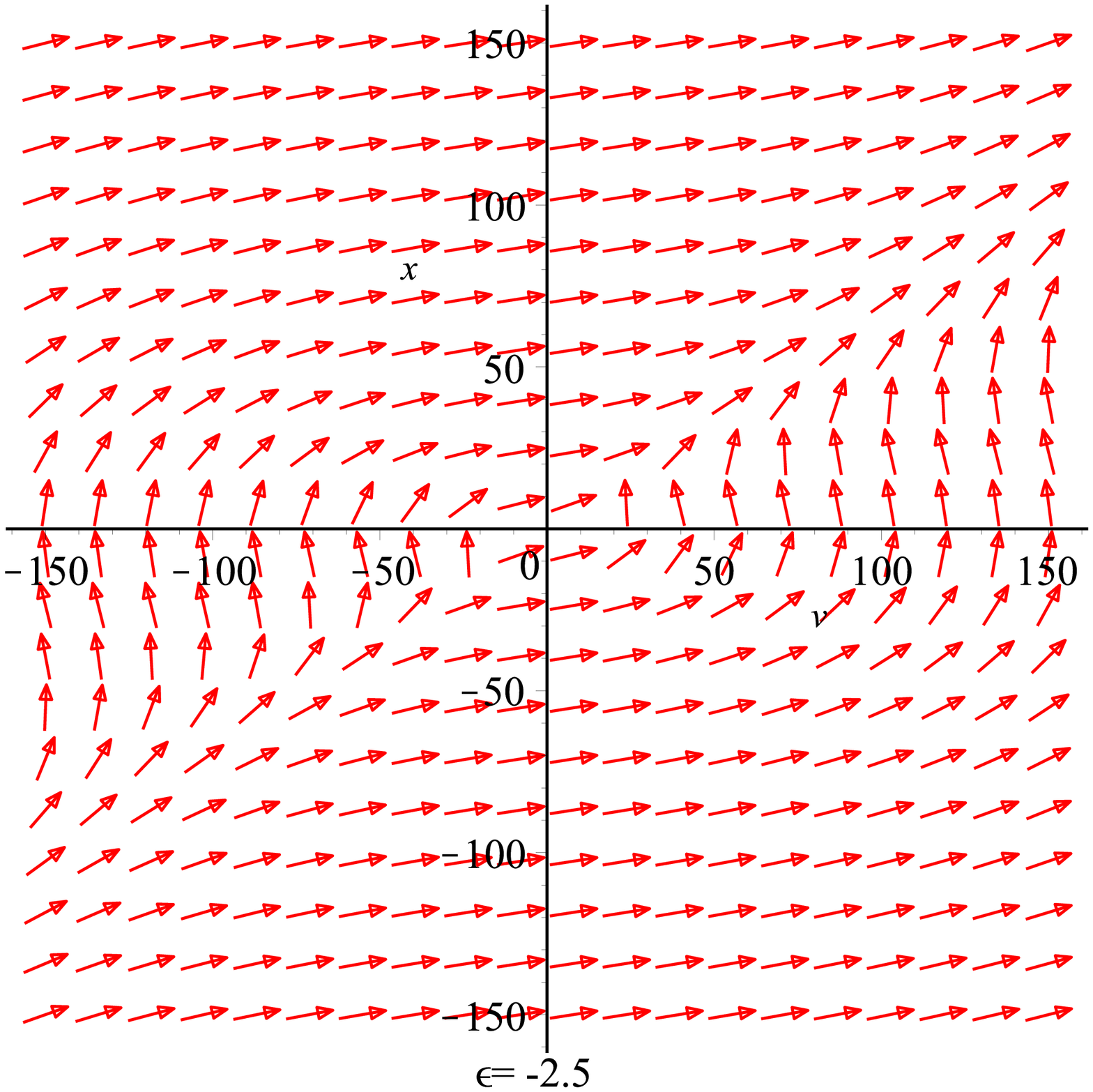}
\includegraphics[width=0.32\textwidth]{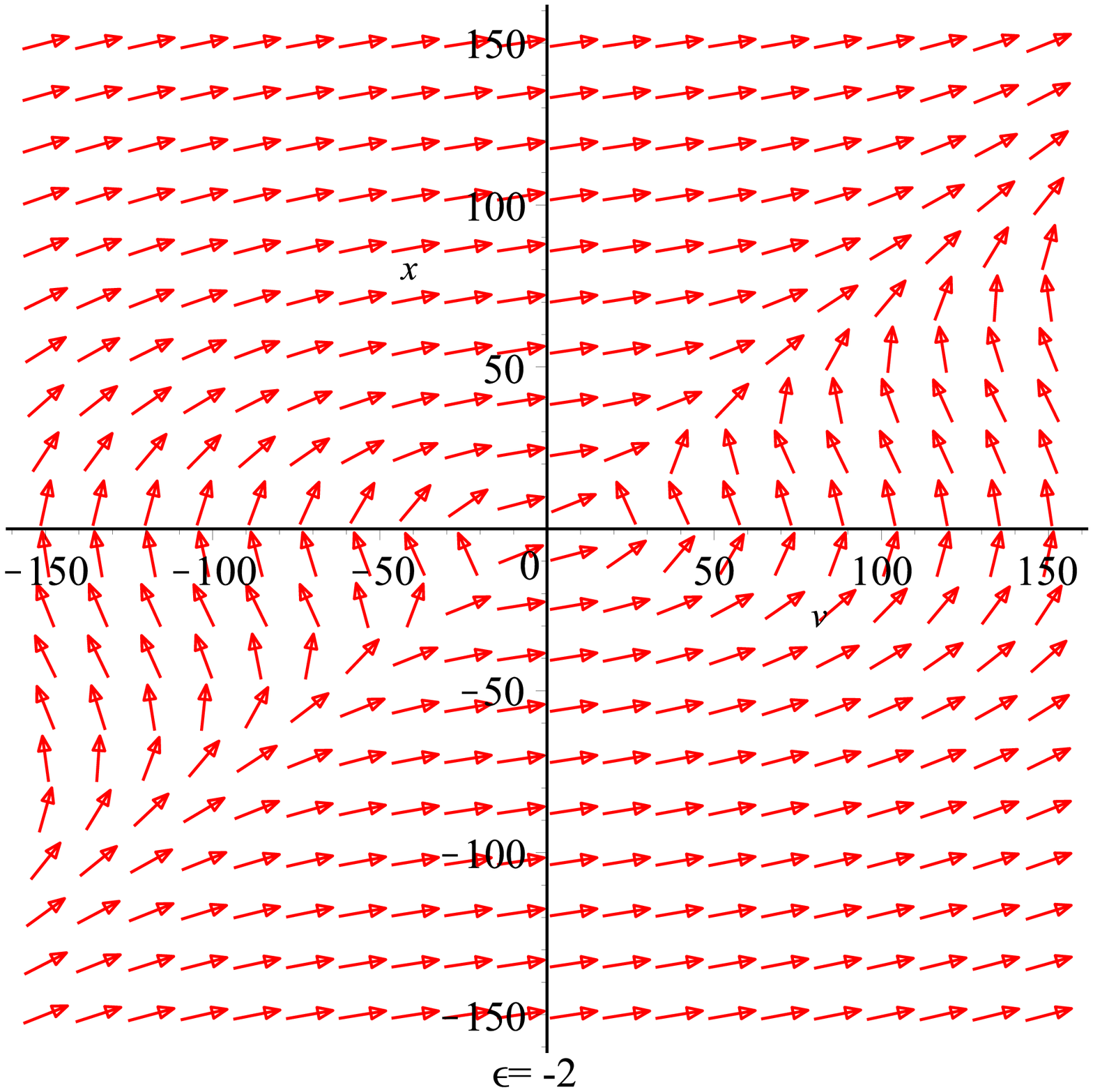}
\includegraphics[width=0.32\textwidth]{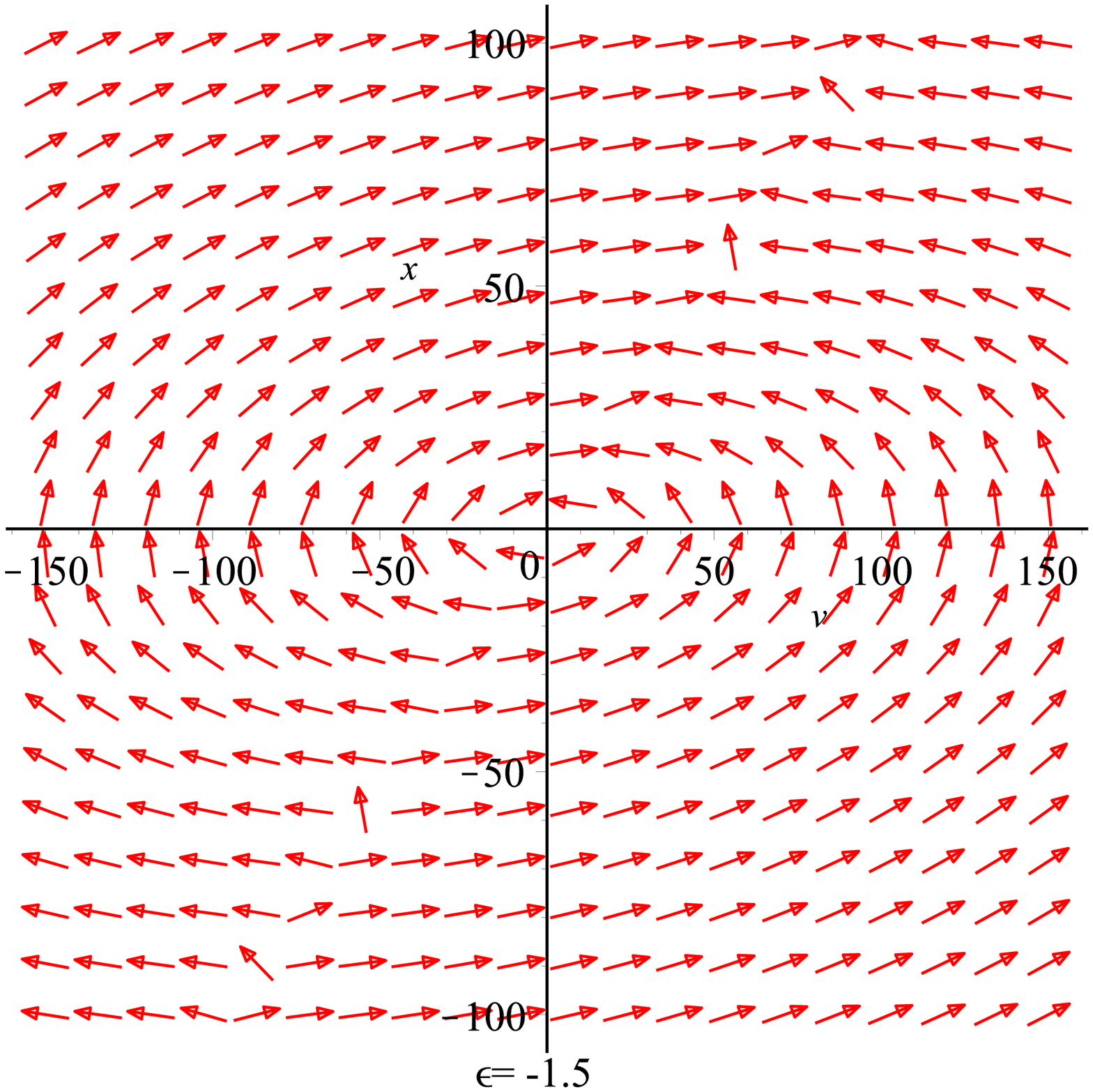}
\includegraphics[width=0.32\textwidth]{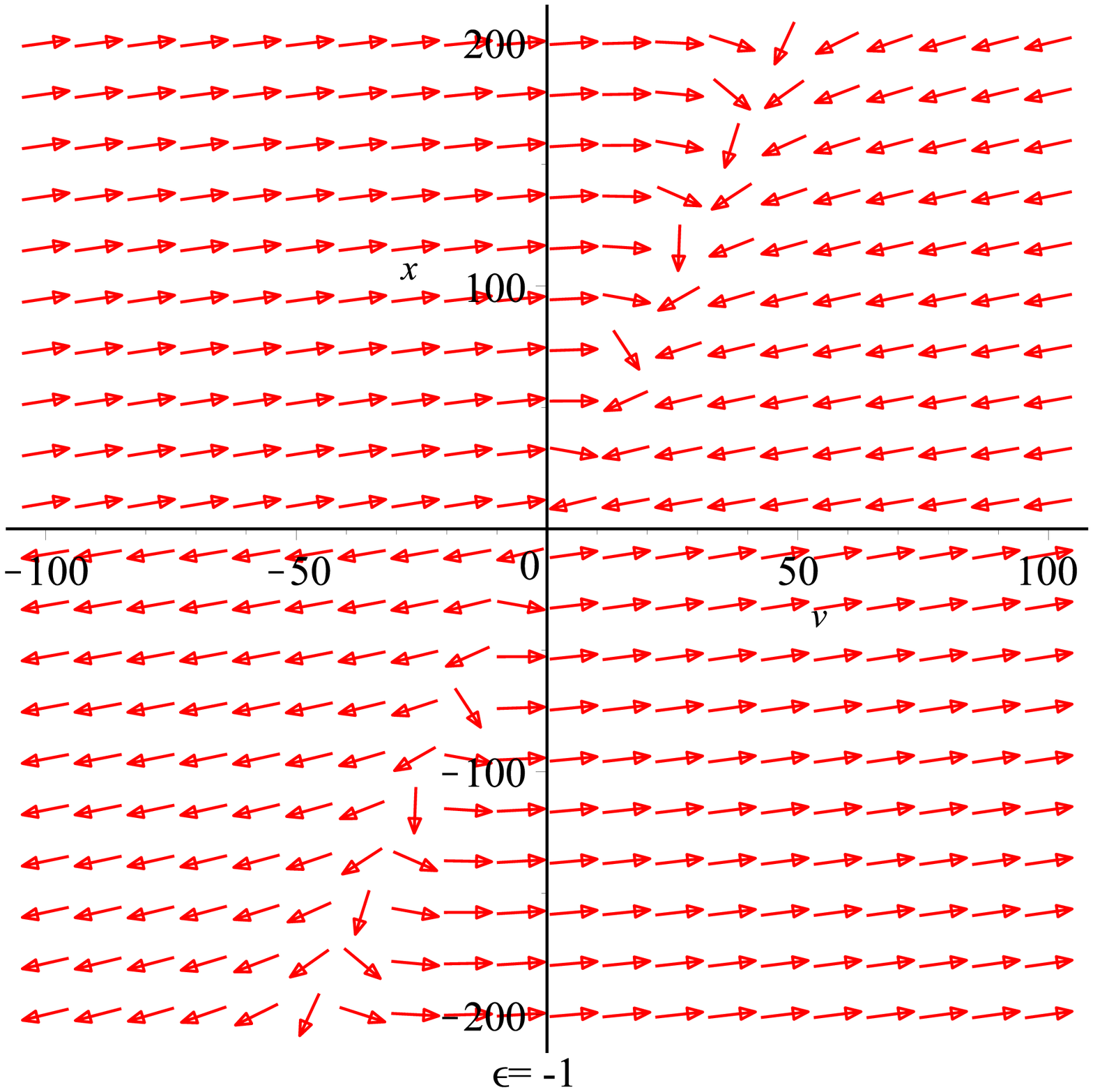}
\includegraphics[width=0.32\textwidth]{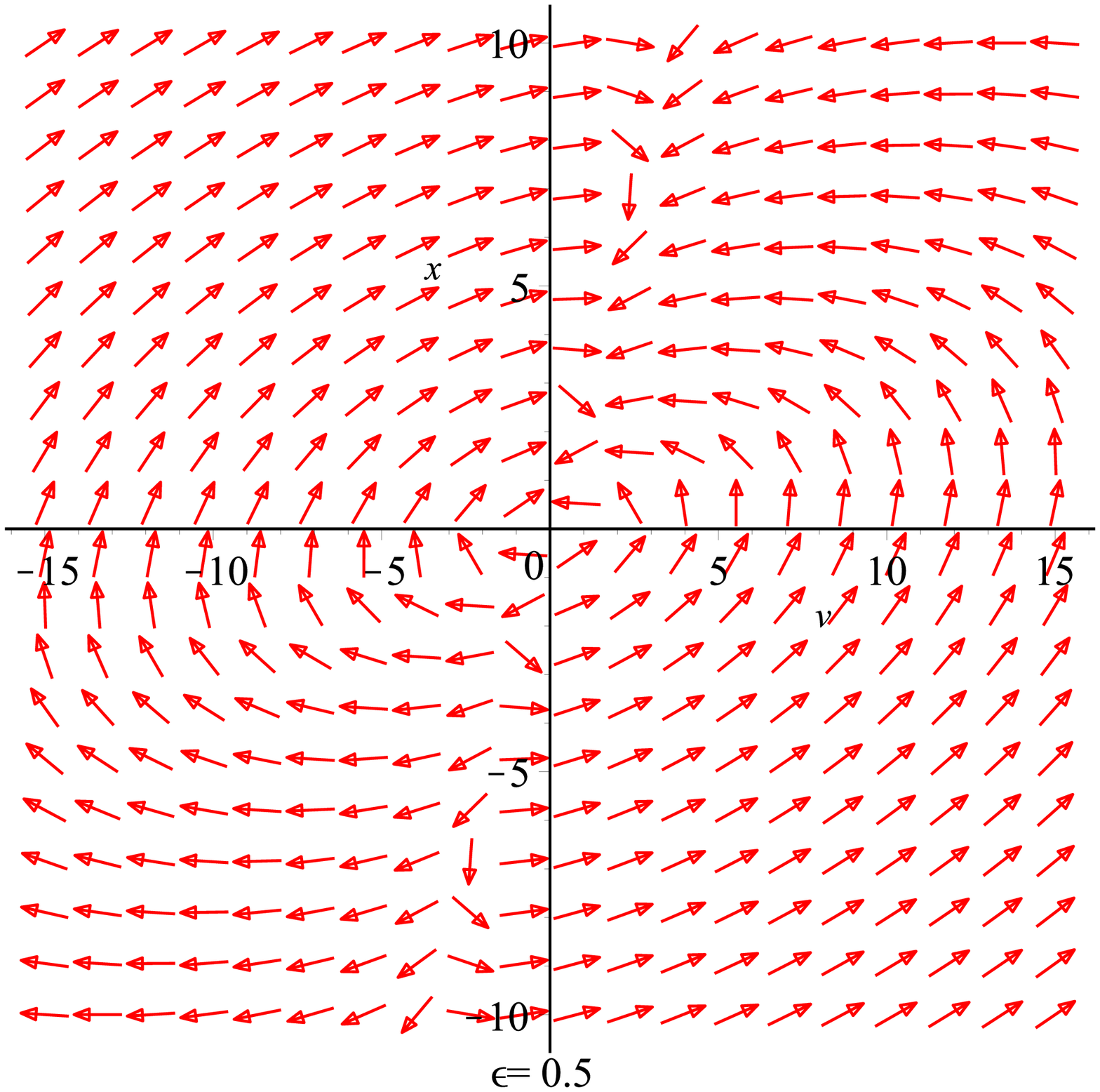}
\includegraphics[width=0.32\textwidth]{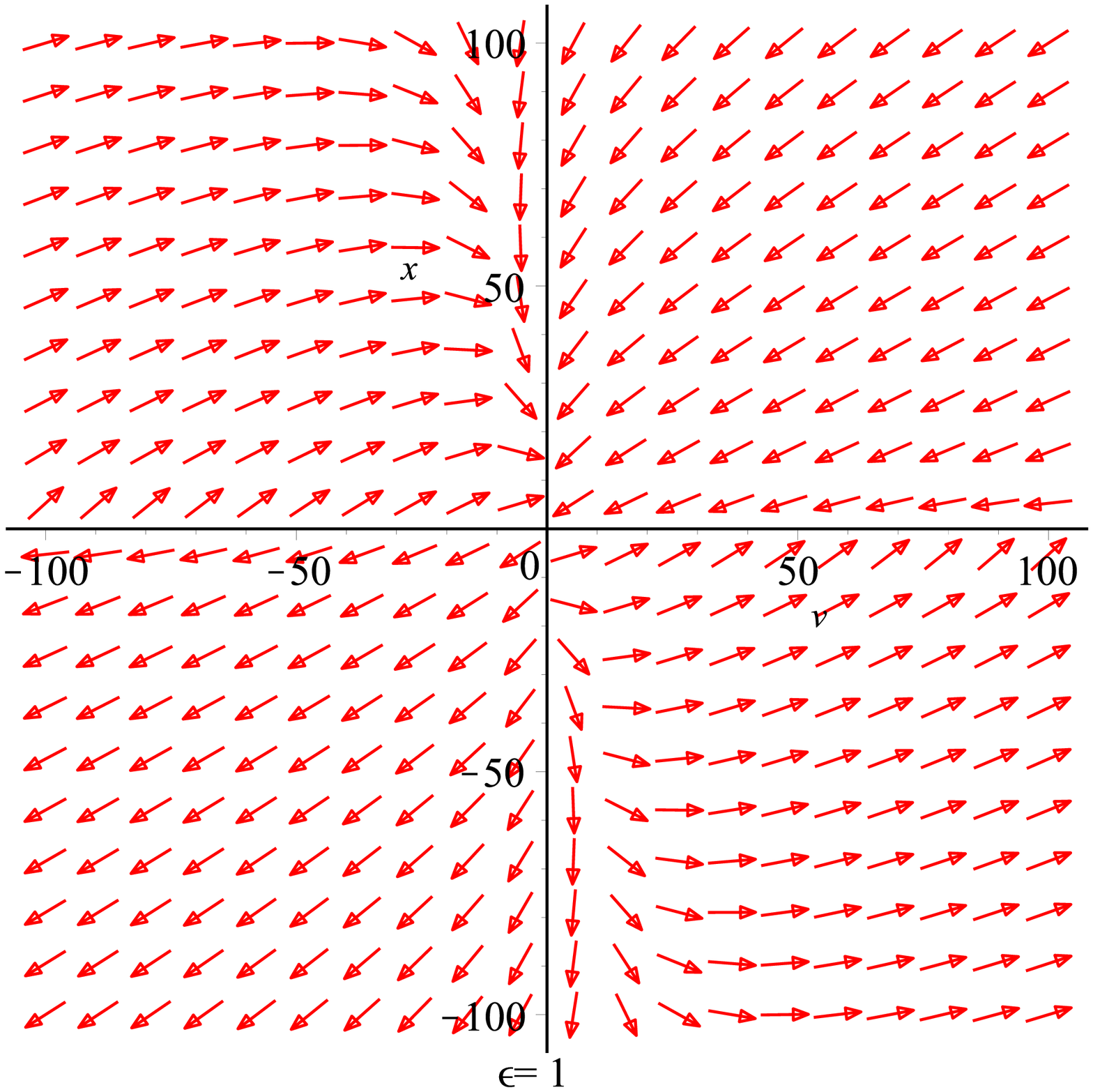}
\includegraphics[width=0.32\textwidth]{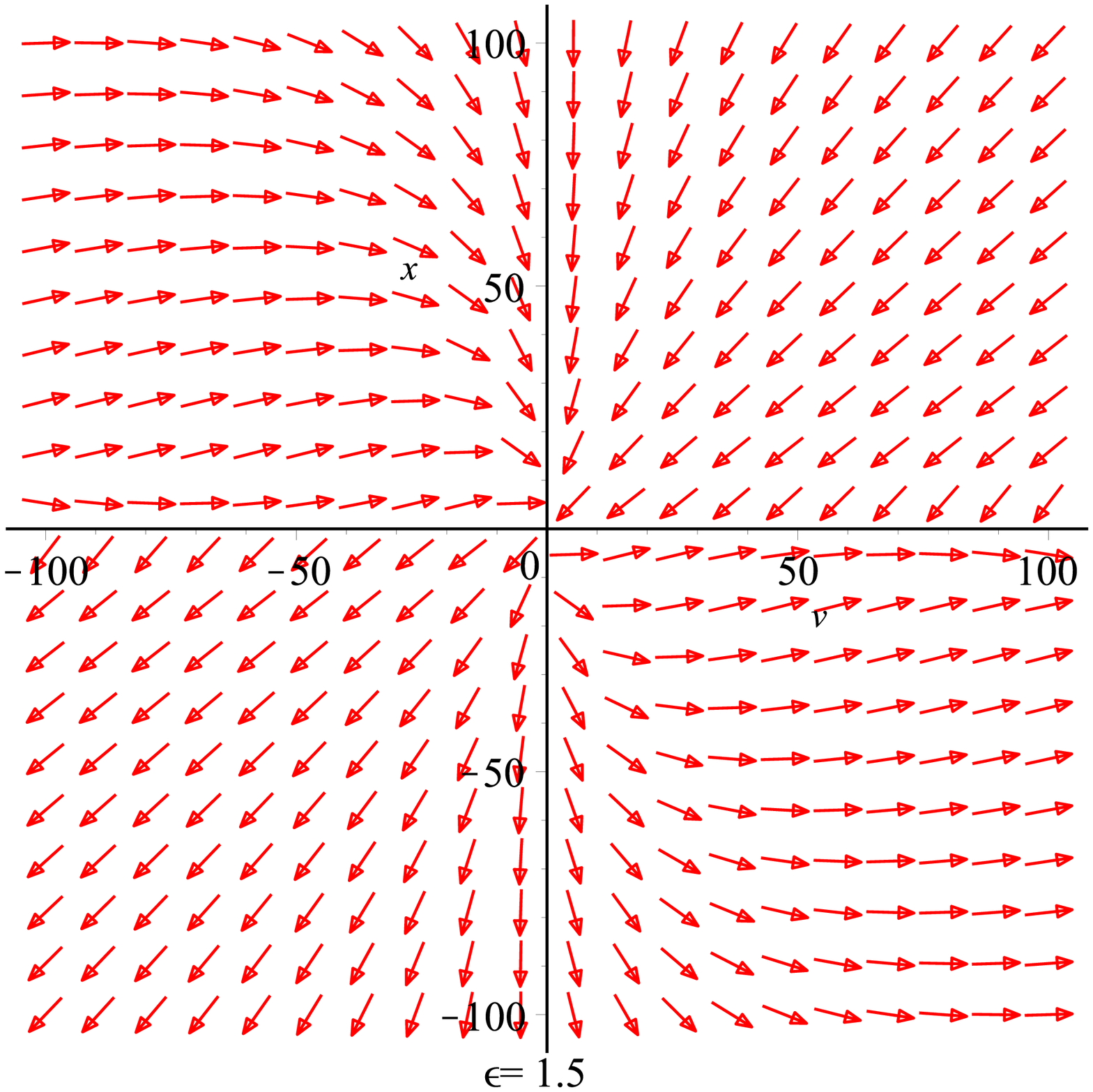}
\includegraphics[width=0.32\textwidth]{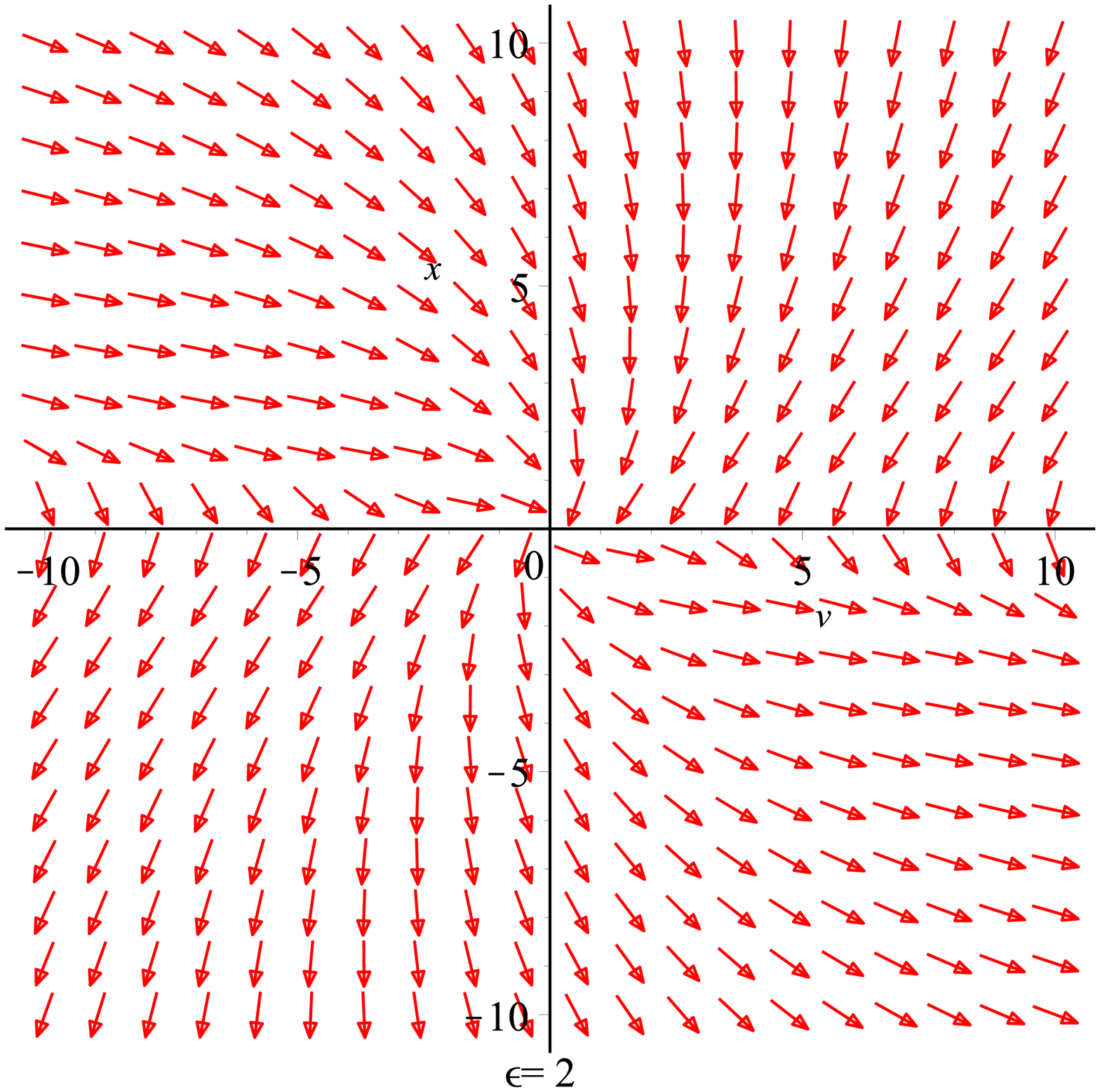}
\includegraphics[width=0.32\textwidth]{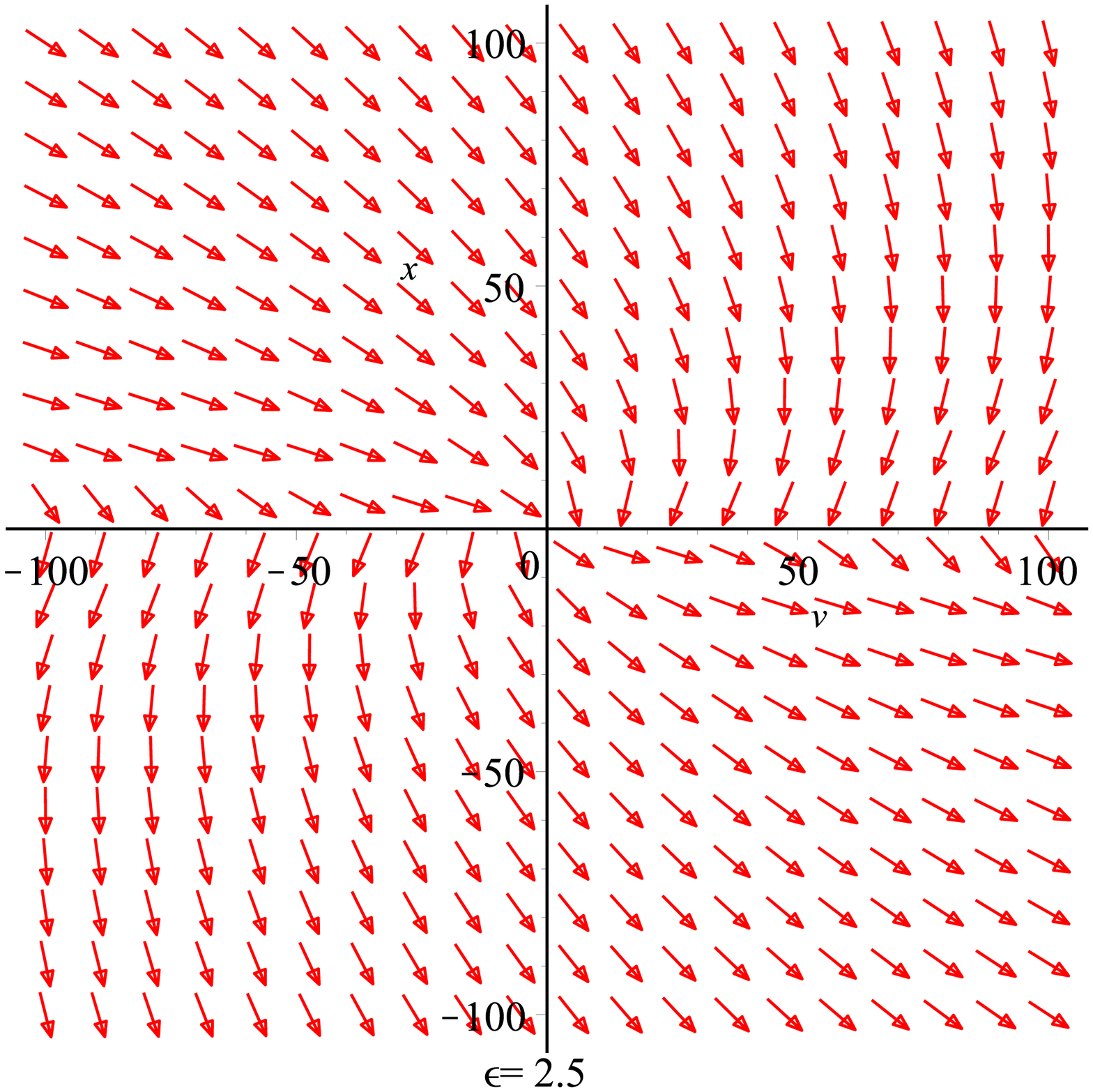}
\includegraphics[width=0.32\textwidth]{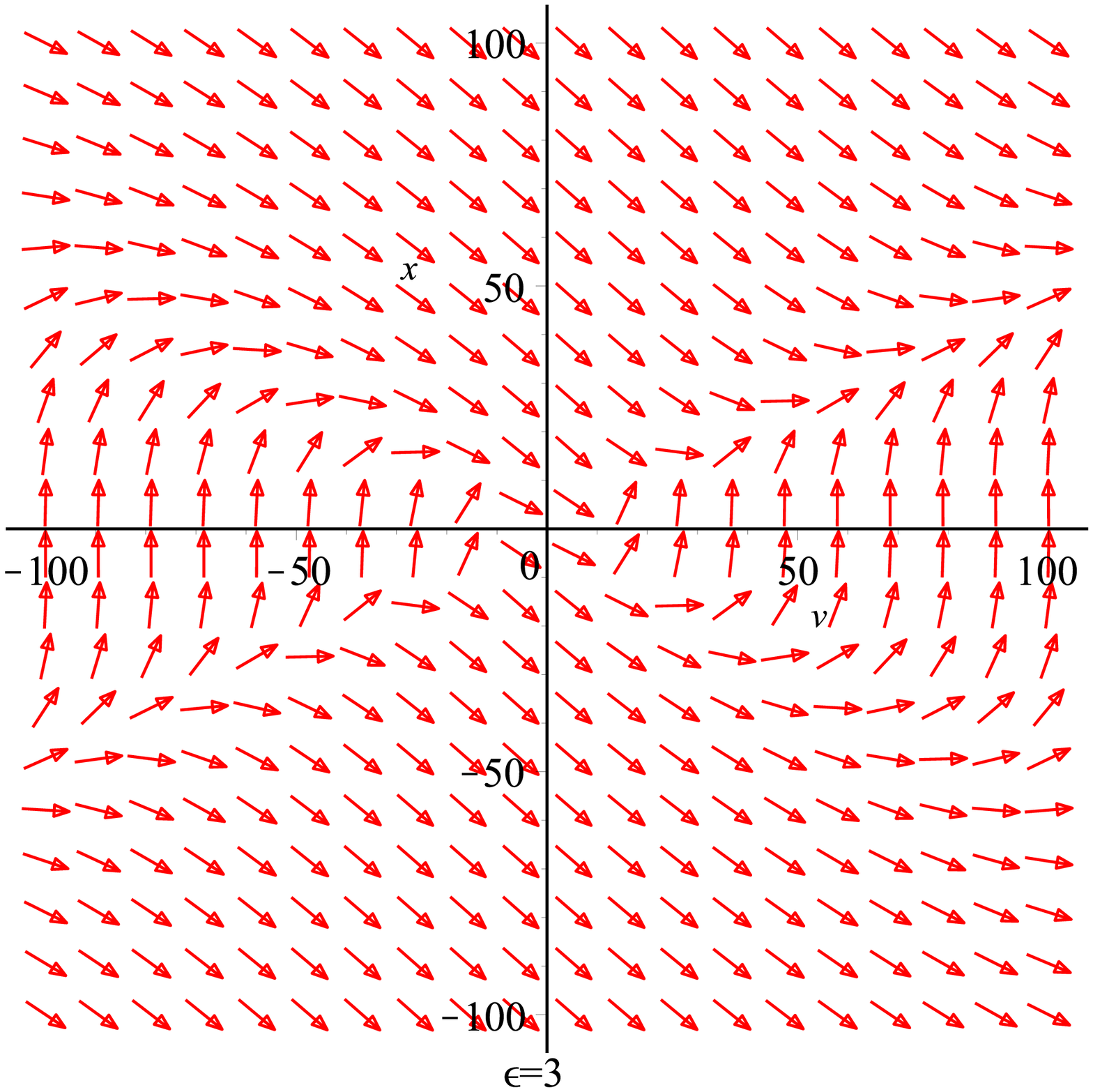}
 \caption{Arrow diagrams for de Sitter inflationary phase for $A_{\|}\neq0,A_{\perp}=0$. The arrows in the diagram show stable
attraction points.}
\end{figure}
\begin{figure}[ht] \centering {\label{1}}
\includegraphics[width=0.32\textwidth]{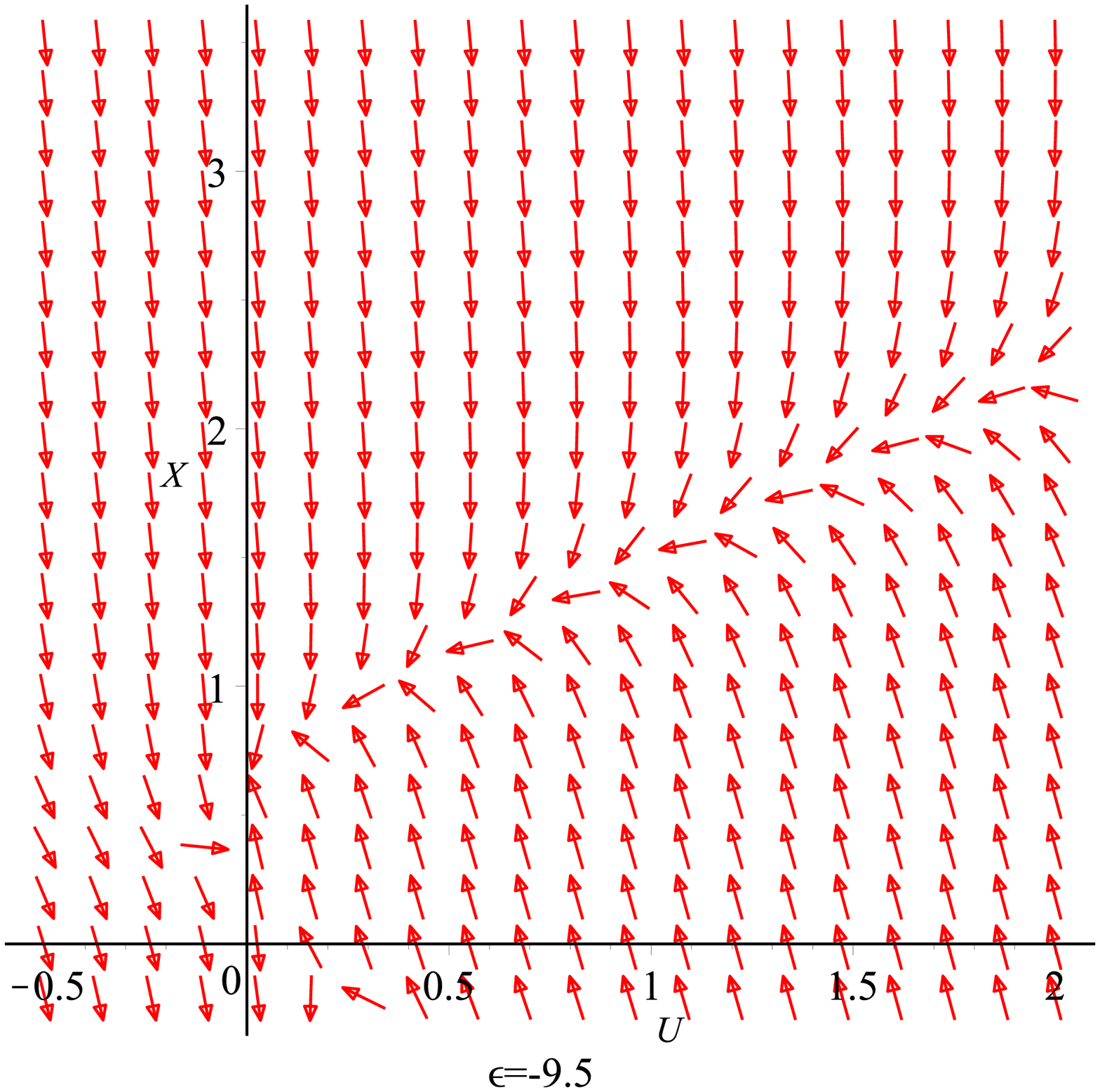}
\includegraphics[width=0.32\textwidth]{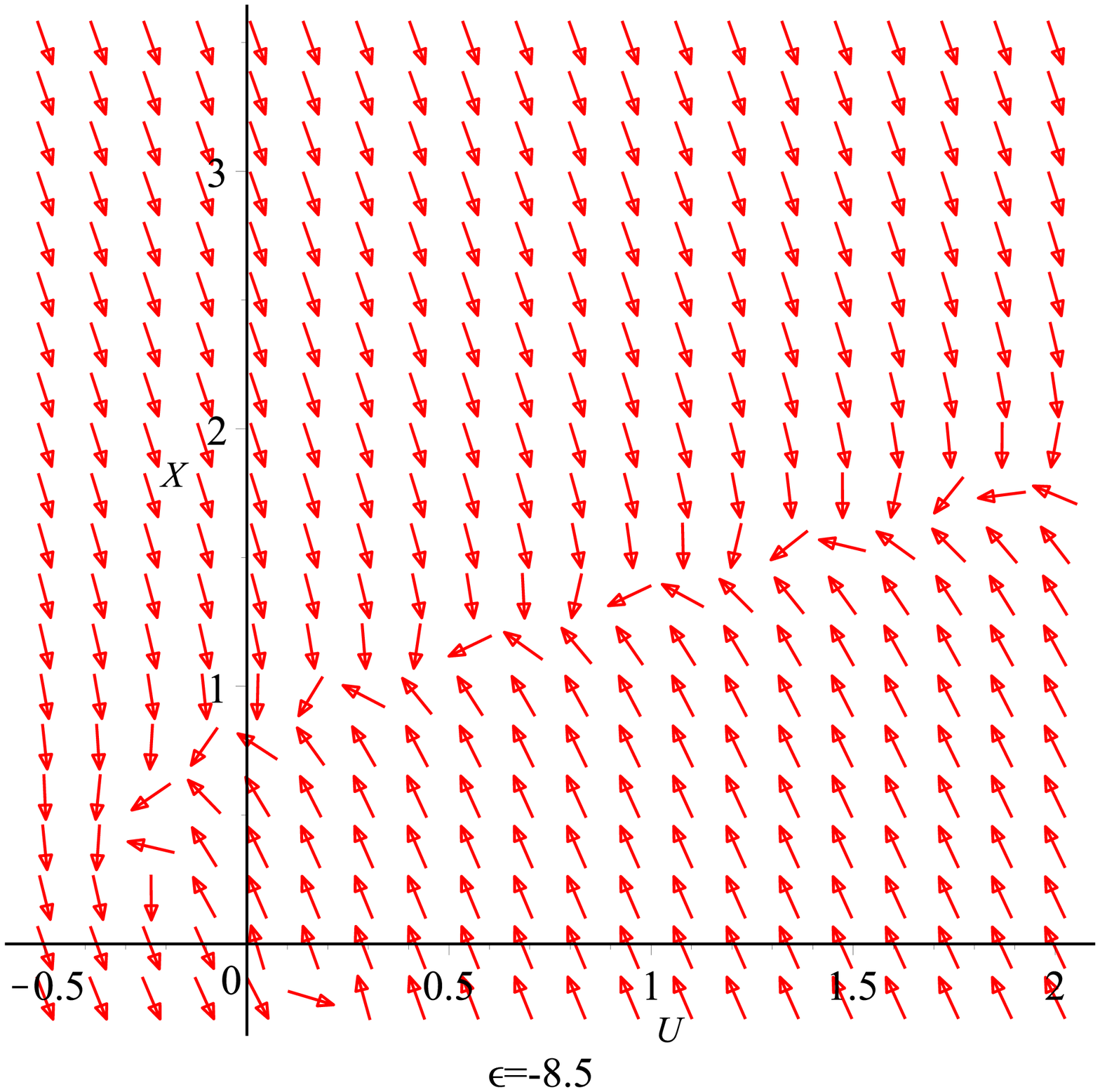}
\includegraphics[width=0.32\textwidth]{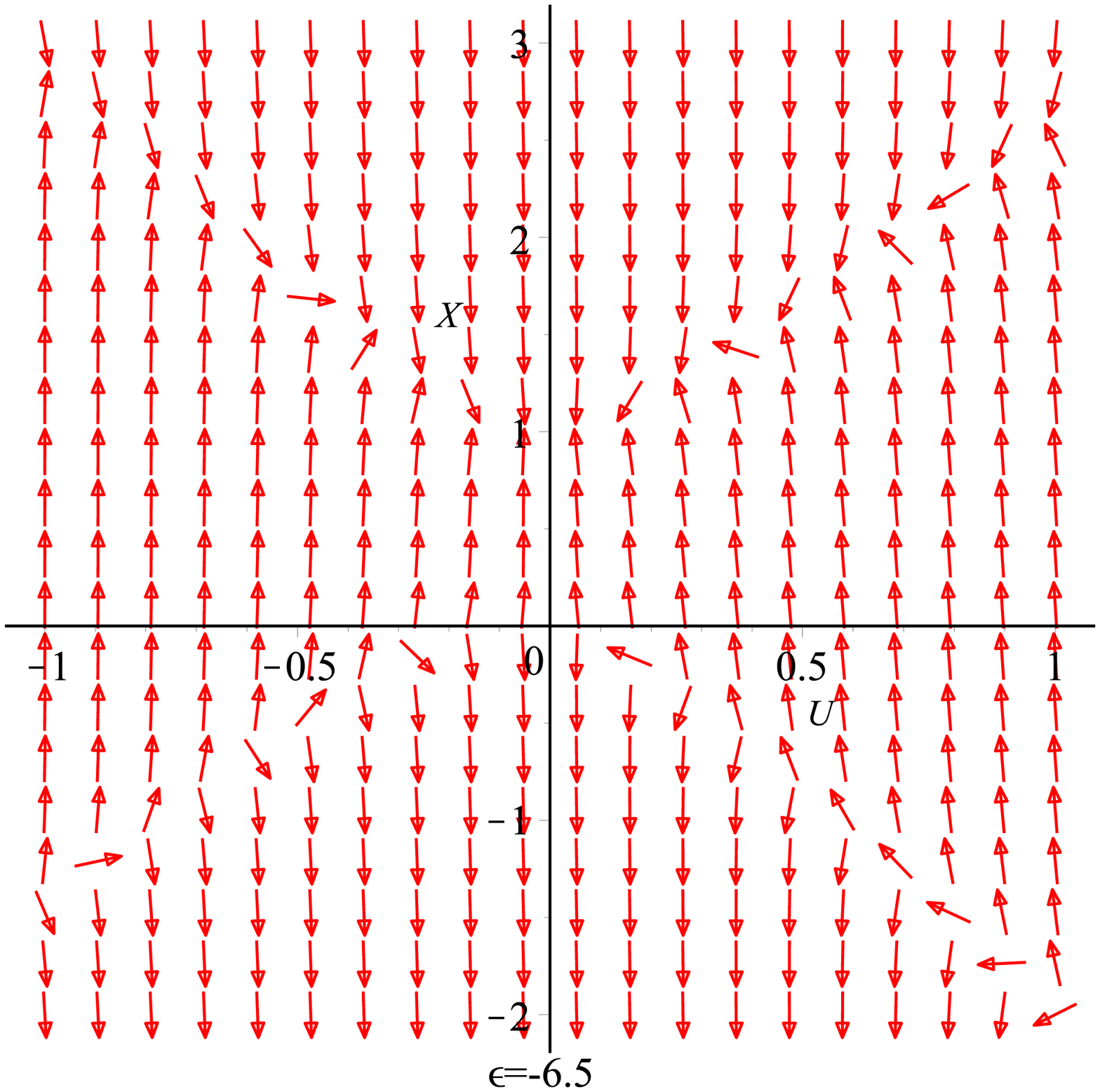}
\includegraphics[width=0.32\textwidth]{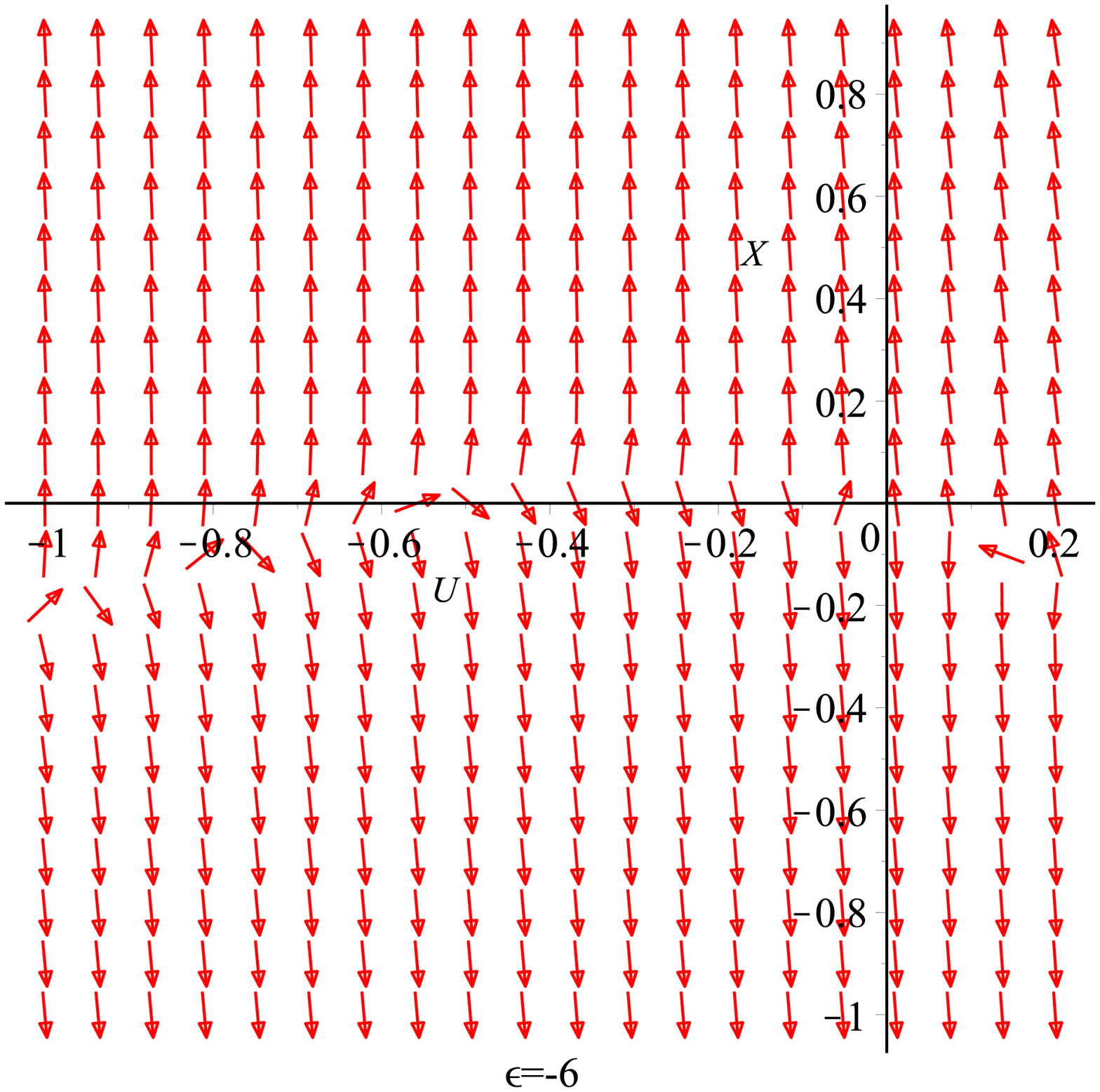}
\includegraphics[width=0.32\textwidth]{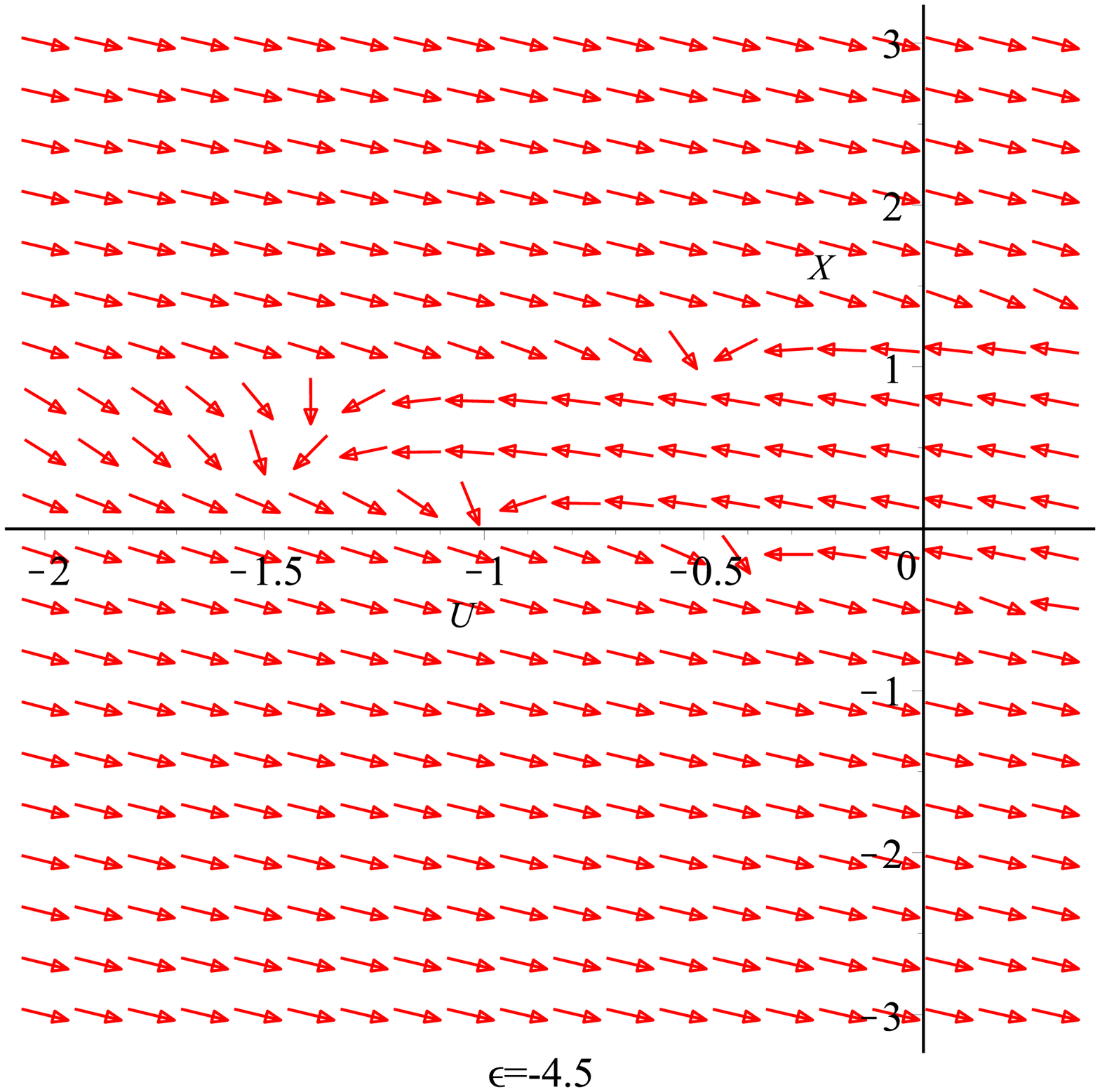}
\includegraphics[width=0.32\textwidth]{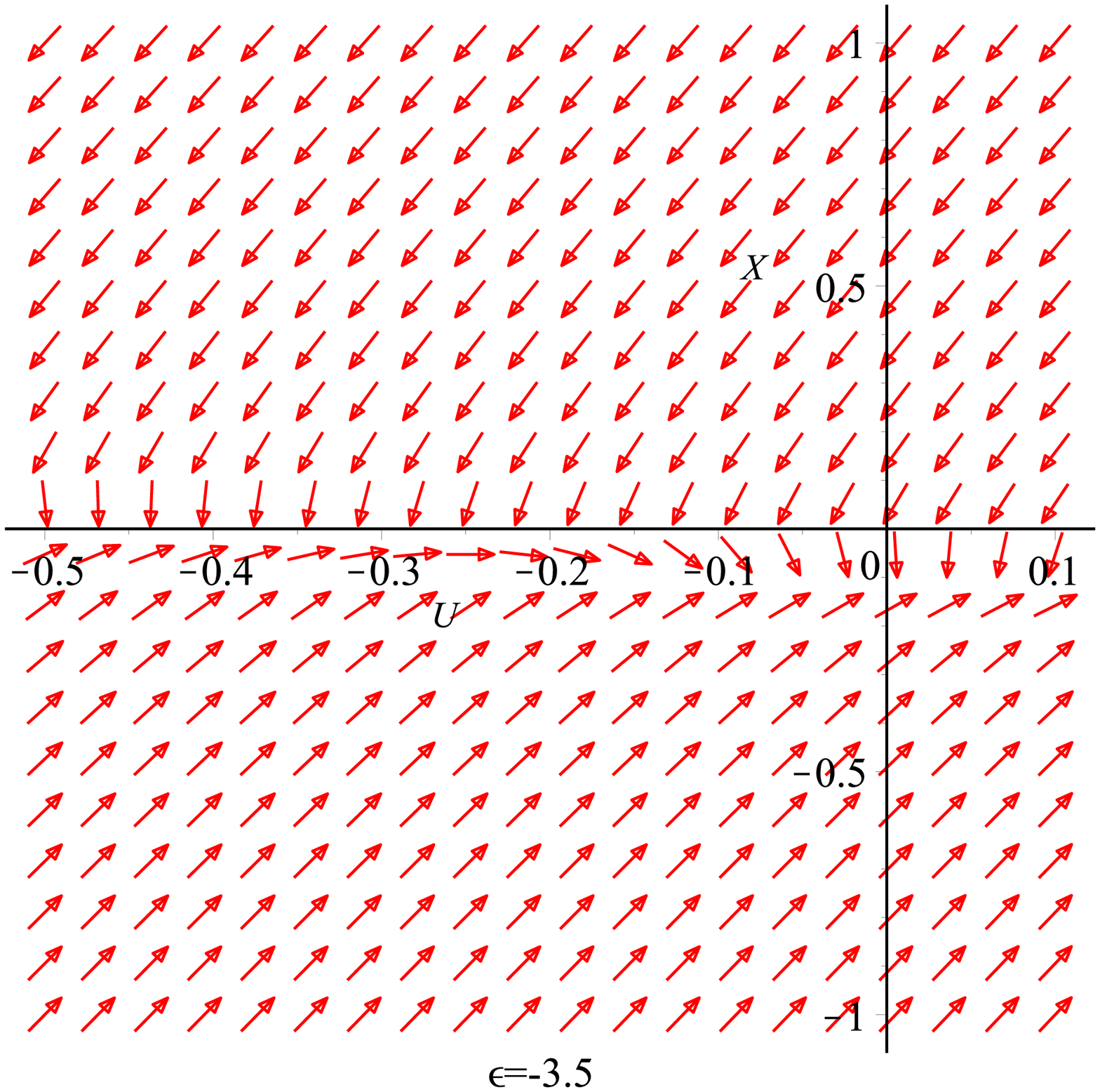}
\includegraphics[width=0.32\textwidth]{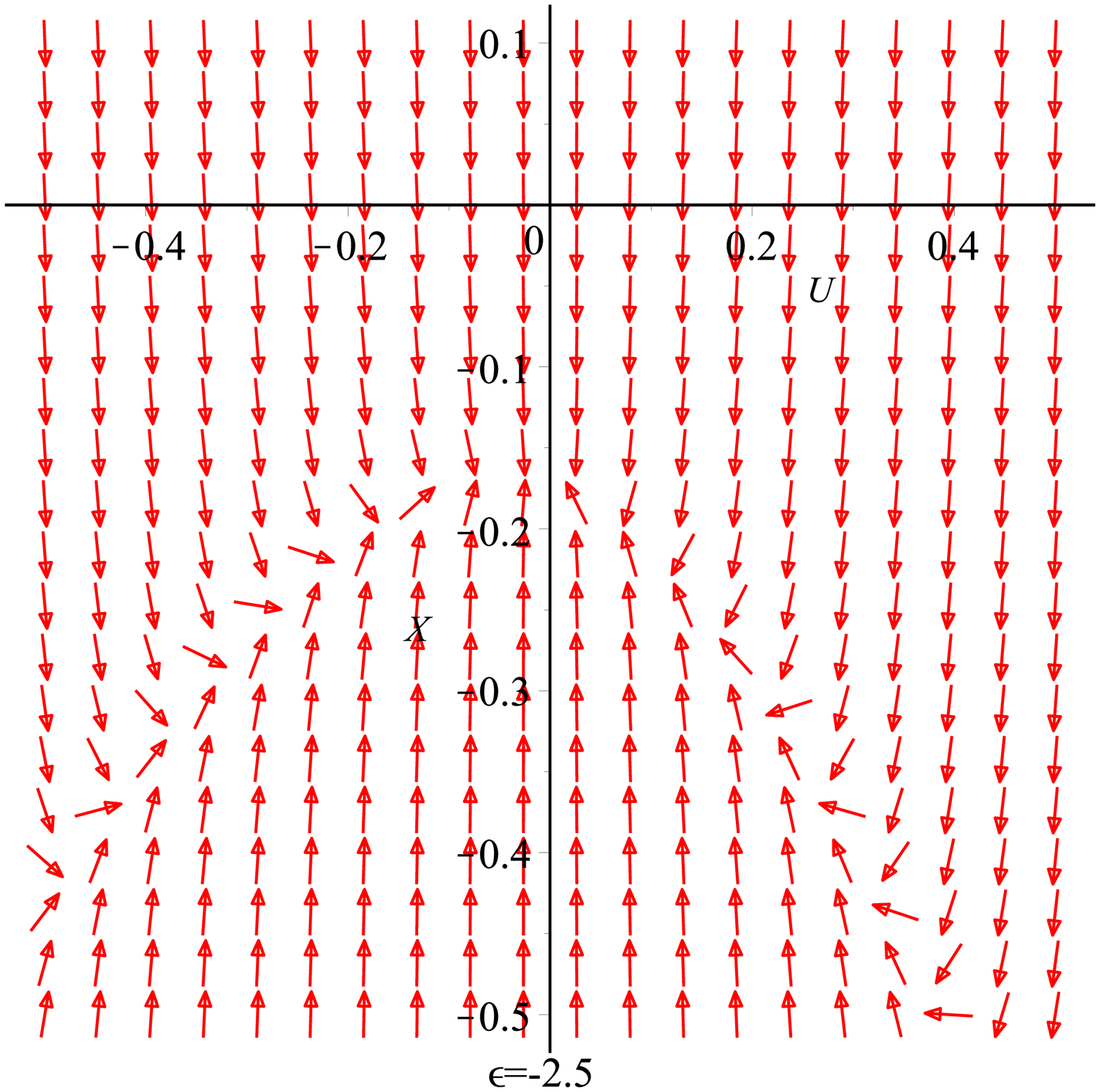}
\includegraphics[width=0.32\textwidth]{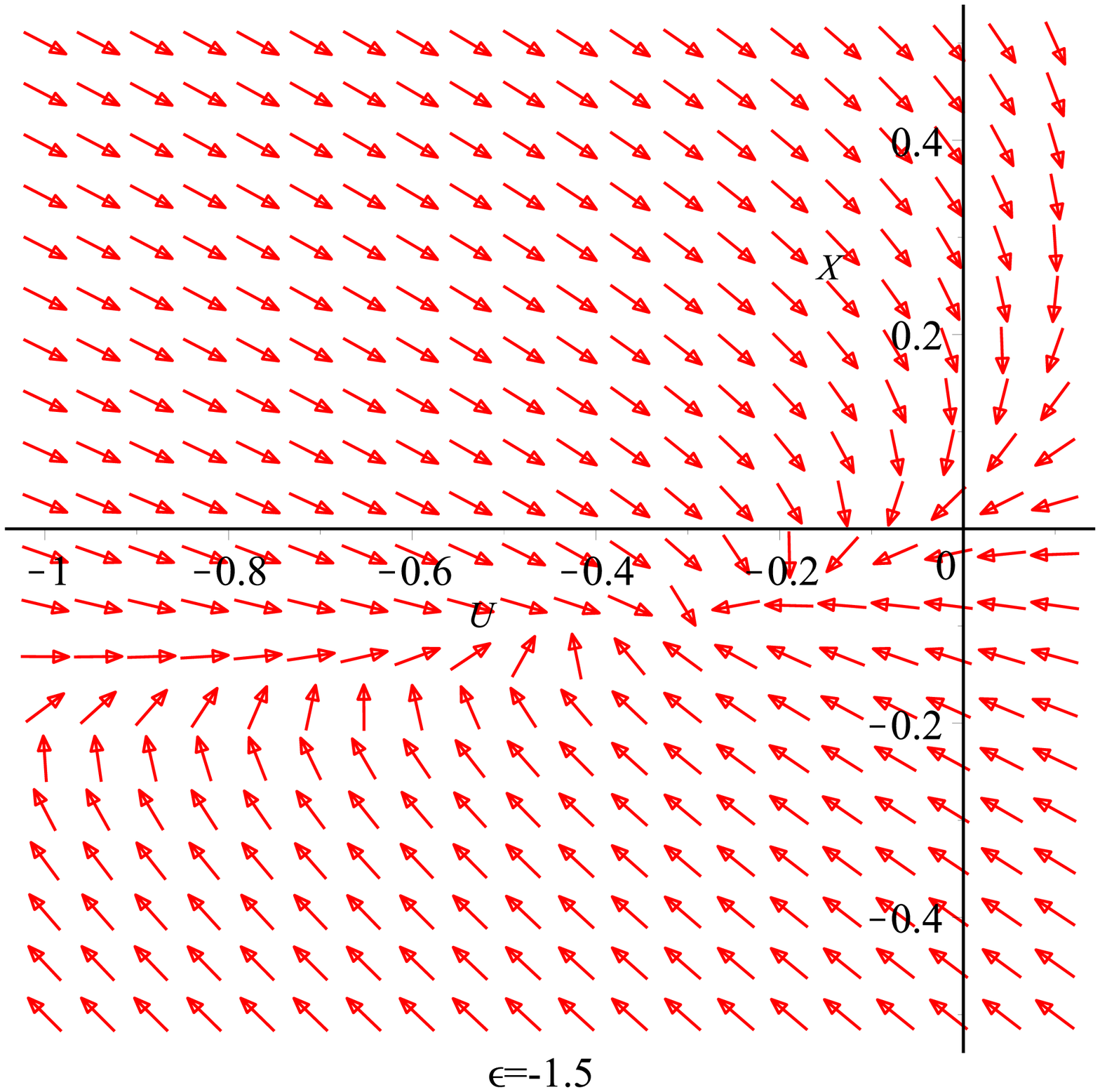}
\includegraphics[width=0.32\textwidth]{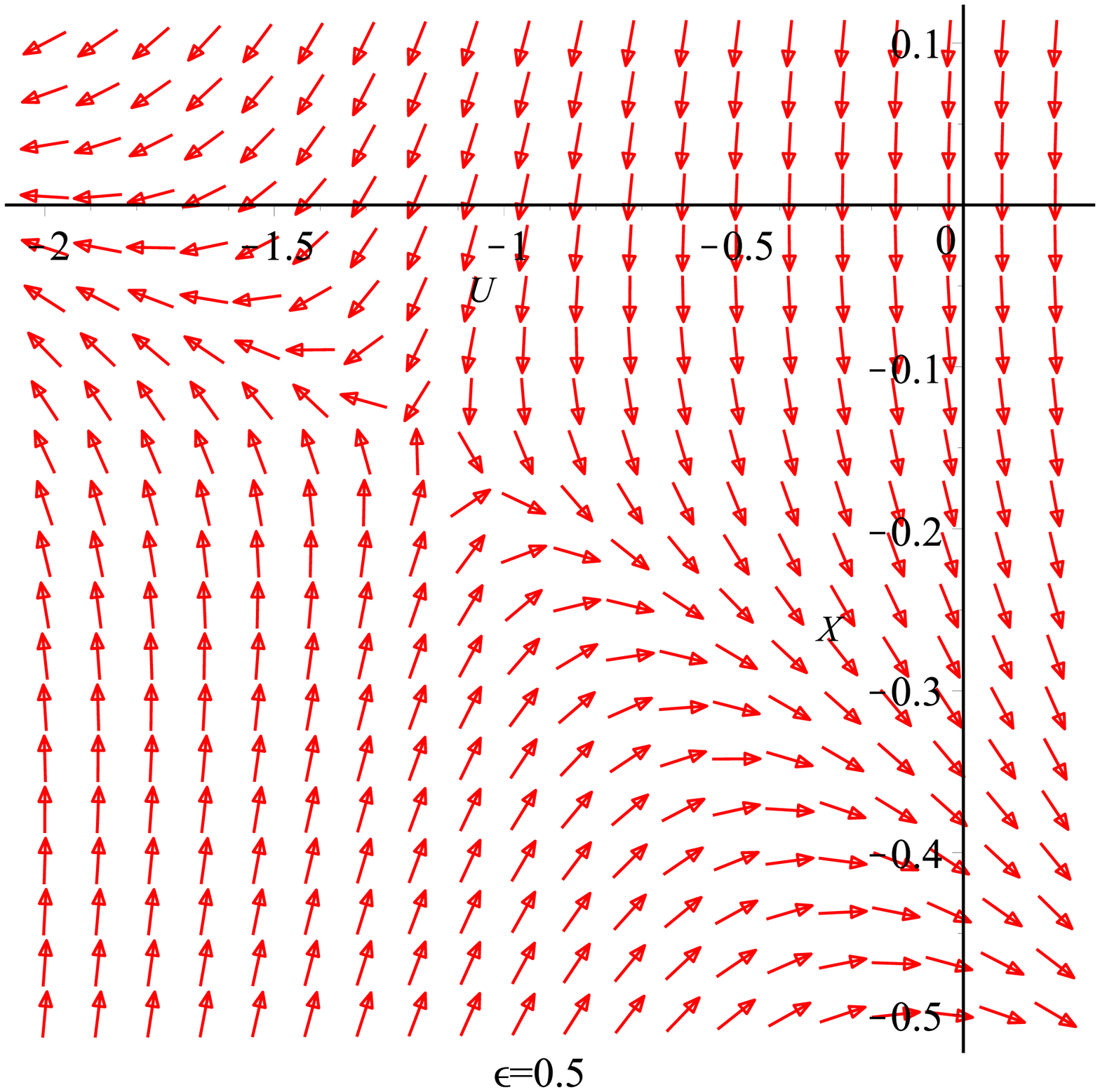}
\includegraphics[width=0.32\textwidth]{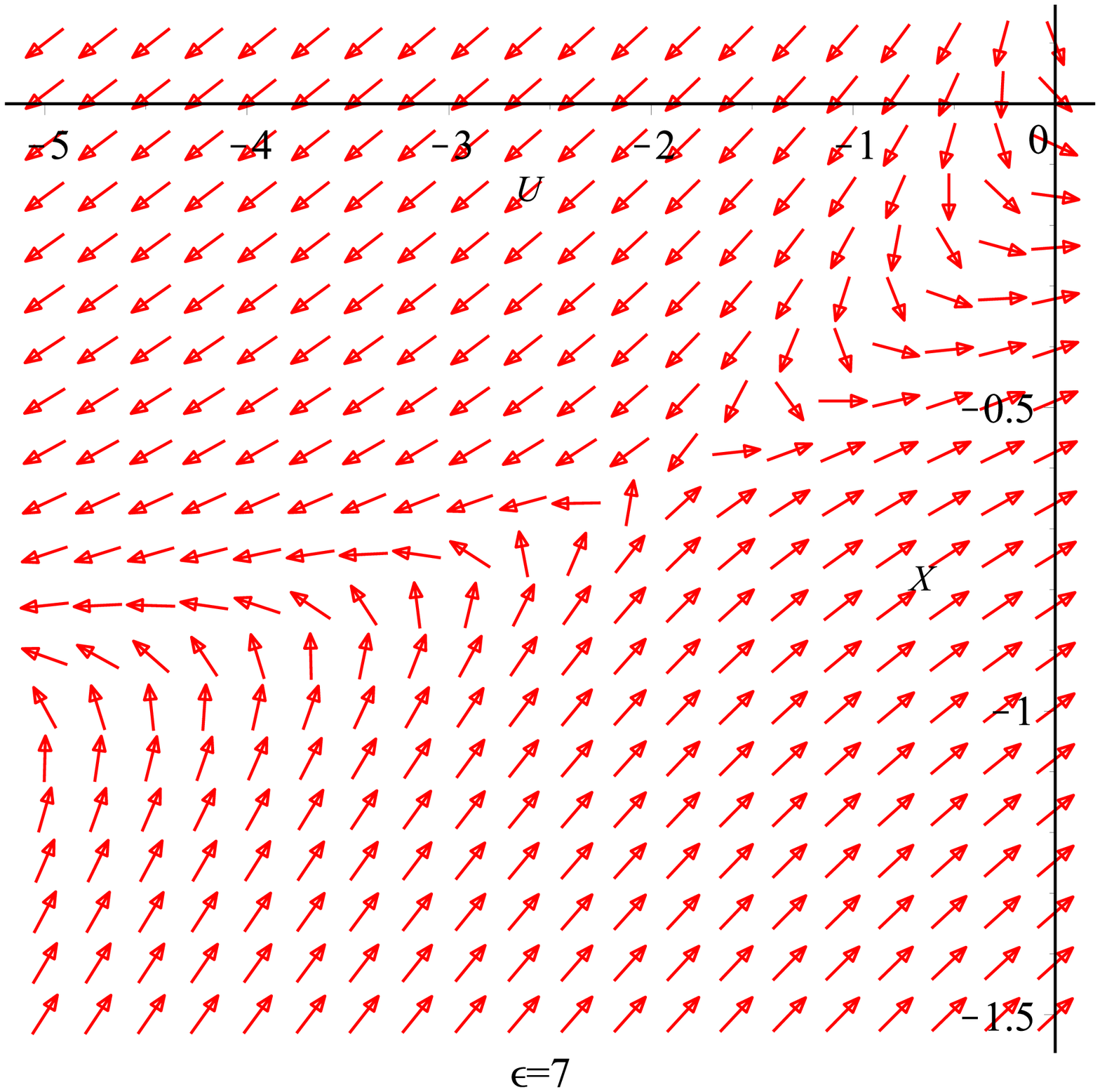}
\includegraphics[width=0.32\textwidth]{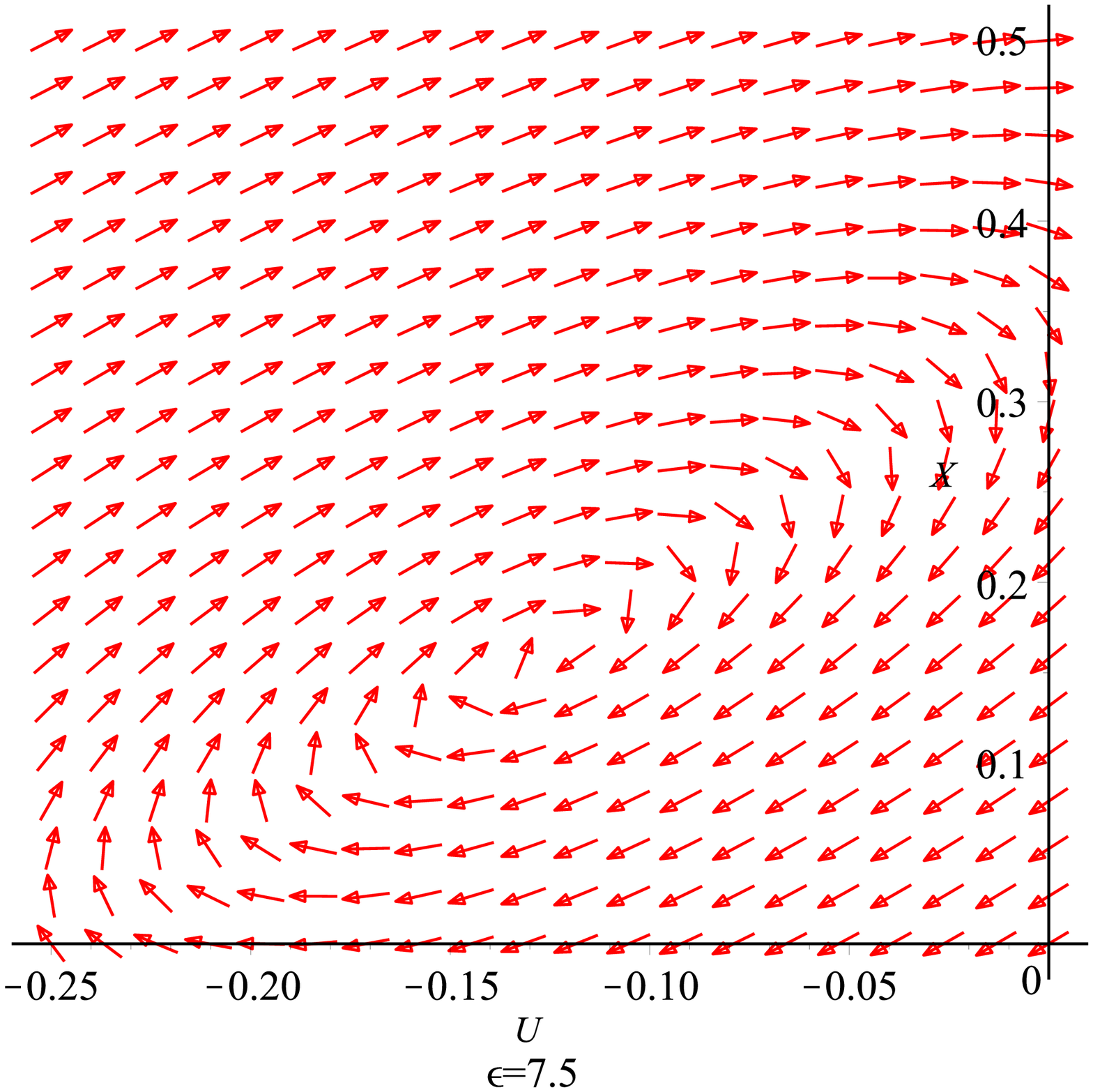}
\includegraphics[width=0.32\textwidth]{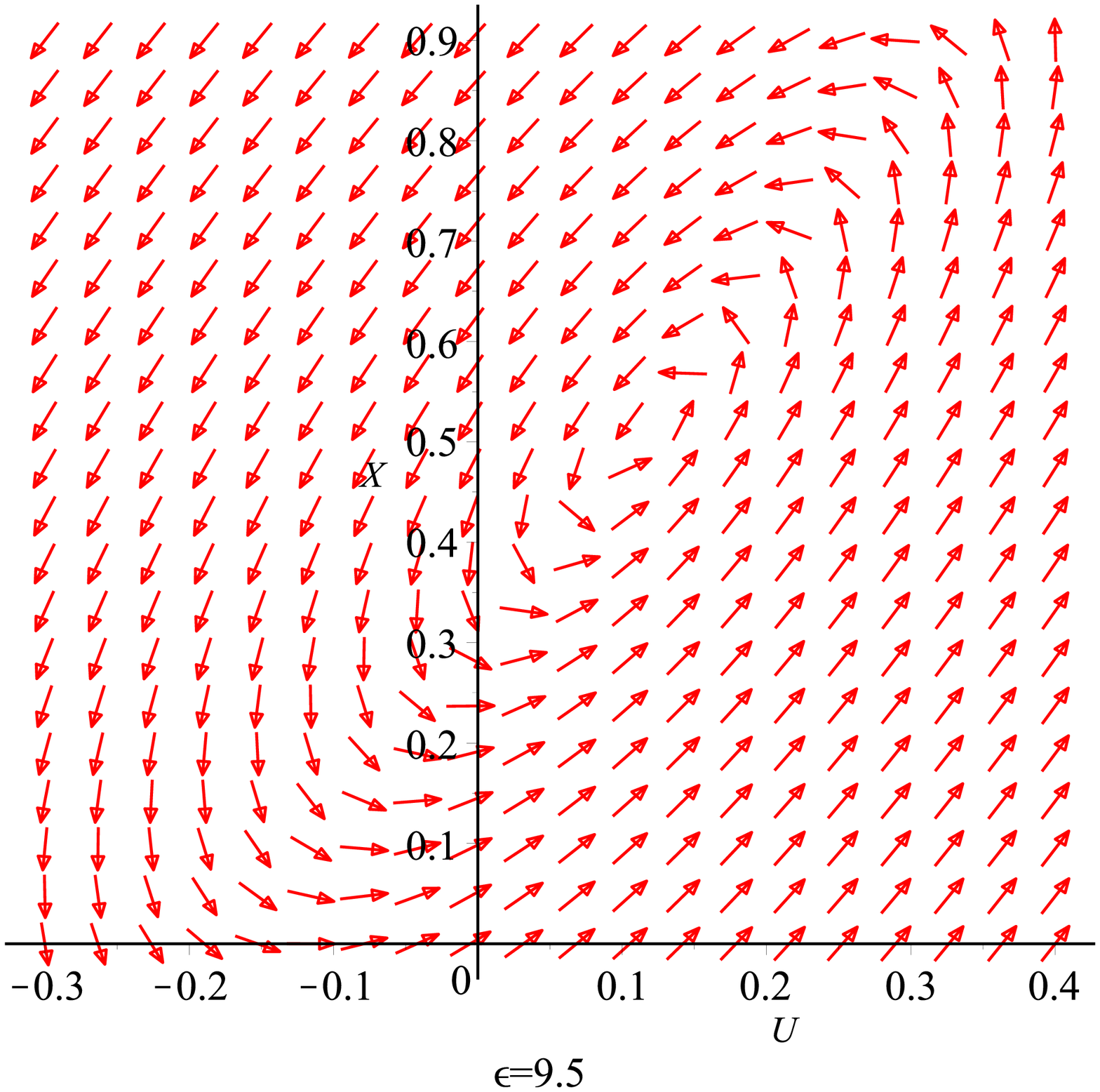}
 \caption{Arrow diagrams for de Sitter inflationary phase for $A_{\|}=0,A_{\perp}\neq0$. The arrows in the diagram show stable
attraction points.}
\end{figure}
\textbf{Appendix I}\\

 We obtain the EM tensor field
(\ref{Fmunu}) as follows. At the first step we choose a local
coordinate transformation
\begin{equation}\label{trans} d\tilde{t}=dt,~~~d\tilde{x}=e^{a-2b}dx,~~~d\tilde{y}=e^{a+b}dy,~~~d\tilde{z}=e^{a+b}dz
\end{equation} in which the coordinates  $(\tilde{t},\tilde{x},\tilde{y},\tilde{z})$
are characterized by a flat Minkonski space time
$ds^2=-d\tilde{t}^2+d\tilde{x}^2+d\tilde{y}^2+d\tilde{z}^2$ for
which
\begin{equation}\label{Fmunuprime}
F_{\tilde{\mu}\tilde{\nu}}=\left(%
\begin{array}{cccc}
  0 & -E_{\tilde{x}} & -E_{\tilde{y}} & -E_{\tilde{z}} \\
  E_{\tilde{x}} & 0 & B_{\tilde{z}} & -B_{\tilde{y}} \\
 E_{\tilde{y}} & -B_{\tilde{z}} & 0 & B_{\tilde{x}} \\
 E_{\tilde{z}} & B_{\tilde{y}} & -B_{\tilde{x}} & 0 \\
\end{array}%
\right)
\end{equation} is well known but the  coordinates $(t,x,y,z)$ correspond to the background metric (\ref{Bianchi}). At the second step
we apply (\ref{trans}) to transform (\ref{Fmunuprime}) as
$F_{\alpha\beta}=\frac{\partial x^{\tilde{\mu}}}{\partial
x^{\alpha}} \frac{\partial x^{\tilde{\nu}}}{\partial
x^{\beta}}F_{\tilde{\mu}\tilde{\nu}}$ which reduces to the
equation (\ref{Fmunu}) and for simplicity we drop over tilde
$`\tilde{~}`$ in the indexes in (\ref{Fmunu}).
\vskip .5cm
\noindent
 \textbf{Appendix II}\\

 Applying the metric
equation (\ref{Bianchi}) one can calculate simply the
non-vanishing Einstein tensor components $G_{\mu}^{\nu}$ as
\begin{equation}\label{Gtt}
G^t_t=3(\dot{a}^2-\dot{b}^2),\end{equation}\begin{equation}\label{Gxx}G^x_x=3\dot{a}^2+6\dot{a}\dot{b}+3\dot{b}^2+2\ddot{a}+2\ddot{b}
,\end{equation}
\begin{equation}\label{Gyy}G^y_y=G^z_z=3\dot{a}^2-3\dot{a}\dot{b}+3\dot{b}^2+2\ddot{a}-\ddot{b}\end{equation}
for which the Ricci scalar is
\begin{equation}R=-12\dot{a}^2-6\dot{b}^2-6\ddot{a}\label{Ricci}\end{equation} and for the stress
tensors (\ref{stressEM}), (\ref{stressalpha}) and
(\ref{stressbeta}) we obtain respectively
\begin{equation}\label{TEMtt}T^{t~(EM)}_t=\frac{1}{4}\{E^2+B^2\},\end{equation}
\begin{equation}\label{TEMxx}T^{x~(EM)}_{x}=\frac{1}{2}\{E_x^2+B_x^2\}-\frac{1}{4}\{E^2+B^2\},
\end{equation}
\begin{equation}\label{TEMyy}T^{y~(EM)}_{y}=\frac{1}{2}\{E_y^2+B_y^2\}-\frac{1}{4}\{E^2+B^2\},
\end{equation}
\begin{equation}\label{TEMzz}T^{z~(EM)}_{z}=\frac{1}{2}\{E_z^2+B_z^2\}-\frac{1}{4}\{E^2+B^2\},
\end{equation}
 \begin{equation}\label{Talphatt}T^{(EM)}_{tx}=e^{a-2b}\frac{S_x}{2},~~~T^{(EM)}_{ty}=e^{a+b}\frac{S_y}{2},~~~T^{(EM)}_{tz}=e^{a+b}\frac{S_z}{2}
 \end{equation} where
\begin{equation}\label{Talphaxx}~~~\vec{S}=\vec{E}\times\vec{B}
\end{equation}
is the poynting vector and for both components of the stress
tensor $W_{\mu\nu}$ referred by  $\|,\perp$, we obtain their
relationship with the $\theta_{\mu\nu}$ components as follows.
\begin{equation}W_{tt}=-e^{-6a}\frac{d}{dt}[e^{3a}\frac{d}{dt}(e^{3a}\theta_{tt})]-2(\dot{a}-2\dot{b})e^{-2a+4b}[\dot{\theta}_{xx}+(\dot{a}+4\dot{b})\theta_{xx}]
\label{wtt}\end{equation}
$$-2(\dot{a}+\dot{b})e^{-2(a+b)}[(\dot{a}-2\dot{b})(\theta_{yy}+\theta_{zz})+\dot{\theta}_{yy}+\dot{\theta}_{zz}]$$
\begin{equation}W_{xx}=2e^{-6a}\frac{d}{dt}[(\dot{a}-2\dot{b})e^{8a-4b}\theta_{tt}+(\dot{a}-2\dot{b})e^{6a}\theta_{xx}]
-4(\dot{a}-2\dot{b})\label{wxx}\end{equation}
$$
\times[(\dot{a}-2\dot{b})e^{2a-4b}\theta_{tt}+\dot{\theta}_{xx}+2(\dot{a}+\dot{b})\theta_{xx}
]
$$
\begin{equation}W_{yy}=e^{-6a}\frac{d}{dt}[e^{6a}\dot{\theta}_{yy}+3\dot{a} e^{6a}\theta_{yy}+2(\dot{a}+\dot{b})e^{8a+2b}\theta_{tt}]
\label{wyy}\end{equation}
$$
-4(\dot{a}+\dot{b})[\dot{\theta}_{yy}+(2\dot{a}-\dot{b})\theta_{yy}+(\dot{a}+\dot{b})e^{2(a+b)}\theta_{tt}]$$
and
\begin{equation}W_{zz}=e^{-6a}\frac{d}{dt}[e^{6a}\dot{\theta}_{zz}+3\dot{a} e^{6a}\theta_{zz}+2(\dot{a}+\dot{b})e^{8a+2b}\theta_{tt}]
\label{wzz}\end{equation}
$$
-4(\dot{a}+\dot{b})[\dot{\theta}_{zz}+(2\dot{a}-\dot{b})\theta_{zz}+(\dot{a}+\dot{b})e^{2(a+b)}\theta_{tt}]$$
By looking at the above equations
 we obtain for EM field stress tensor
\begin{equation}T_t^{t(EM_{||})}=\frac{1}{4}({E_{||}^2+B^2_{||}}),~~~~T_t^{t(EM_\perp)}=\frac{1}{2}(E^2_{\perp}+B^2_{\perp})\label{EMtpar}\end{equation}
\begin{equation}T_x^{x(EM_{||})}=\frac{1}{4}({E_{||}^2+B^2_{||}}),~~~~T_x^{x(EM_\perp)}=-\frac{1}{2}(E^2_{\perp}+B^2_{\perp})\label{EMxpar}\end{equation}
\begin{equation}T_y^{y(EM_{||})}=-\frac{1}{4}({E_{||}^2+B^2_{||}})=T_z^{z(EM_{||})},~~~~T_y^{y(EM_\perp)}=0=T_z^{z(EM_{\perp})}\label{EMypar}\end{equation}
\begin{equation}T_{tx}^{(EM_{||,\perp})}=0=T_{ty}^{(EM_{||,\perp})}=T_{tz}^{(EM_{||,\perp})}\label{EMtxpar}\end{equation}
Also we obtain for the vector $P^{\mu}$ and the tensor
$\theta^{\mu\nu}$ fields :
\begin{equation}\theta_{||}^{tt}=-\alpha e^{-2a+4b}A_{||}^2,~~~\theta_{\perp}^{tt}=-2\alpha e^{-2a-2b}A^2_{\perp}\end{equation}
\begin{equation}\theta_{||}^{xx}=(\alpha+\beta) e^{-4a+8b}A_{||}^2,~~~\theta_{\perp}^{xx}=2\alpha e^{-4a+2b}A^2_{\perp}\end{equation}
\begin{equation}\theta_{||}^{yy}=\alpha e^{-4a-4b}A_{||}^2=\theta^{zz}_{||},~~~\theta_{\perp}^{yy}=(2\alpha+\beta)e^{-4a-4b}A^2_{\perp}=
\theta^{zz}_{\perp}\end{equation}
\begin{equation}P^t_{||}=-2\alpha A_{||}\dot{A}_{||}e^{-2a+4b}+\{2\alpha(\dot{a}+\dot{b})e^{-2(a+b)}-4\alpha(\dot{a}+\dot{b})e^{-2a+4b}\label{Ptpar}\end{equation}
$$
+(\alpha+\beta)(\dot{a}-2\dot{b})e^{-6a+4b} \}A_{||}^2$$
\begin{equation}P^t_{\perp}=-4\alpha A_{\perp}\dot{A}_{\perp}e^{-2(a+b)}+\{2[(\beta-2\alpha)\dot{a}+(\beta+4\alpha)\dot{b}]
e^{-2(a+b)}$$$$+2\alpha(\dot{a}-2\dot{b})e^{-2(3a+b)}\}A^2_{\perp}\label{ptperp}\end{equation}
and
\begin{equation}P^{x,y,z}_{{||,\perp}}=0\label{pxyz}\end{equation}
\begin{equation}\nabla_\xi P^\xi_{||}=-2\alpha e^{-2a+4b}\{\dot{A}_{||}^2+A_{||}\ddot{A}_{||}+(5\dot{a}+8\dot{b})A_{||}\dot{A_{||}}\}\label{nabpapxi}
\end{equation}$$+2\alpha
e^{-2a+2b}\{(\ddot{a}+\ddot{b}+\dot{a}^2+3\dot{a}\dot{b}+2\dot{b}^2)A^2_{||}+2(\dot{a}+\dot{b})A_{||}\dot{A}_{||}\}$$
$$+(\alpha+\beta)e^{-6a+4b}\{(\ddot{a}-2\ddot{b}-3\dot{a}^2+10\dot{a}\dot{b}-8\dot{b}^2)A^2_{||}+2(\dot{a}-2\dot{b})A_{||}\dot{A}_{||}\}$$
 and
 \begin{equation}\nabla_{\xi}P_{\perp}^{\xi}=e^{-2(a+b)}\{-4\alpha(\dot{A}^2_{\perp}+A_{\perp}\ddot{A}_{\perp})
\label{nablapxiperp}\end{equation}
$$
+4A_{\perp}\dot{A}_{\perp}[(\beta-3\alpha)\dot{a}+(\beta+6\alpha)\dot{b}]
+A^2_{\perp}[2(\beta-2\alpha)\ddot{a}+2(\beta+4\alpha)\ddot{b}$$
$$+2(\beta-2\alpha)\dot{a}^2+2(8\alpha-\beta)\dot{a}\dot{b}-4(\beta+4\alpha)\dot{b}^2]\}
$$$$
+2\alpha
e^{-2(3a+b)}\{2(\dot{a}-2\dot{b})A_\perp\dot{A}_{\perp}+A^2_{\perp}(\ddot{a}-2\ddot{b}-3\dot{a}^2+4\dot{a}\dot{b}+4\dot{b}^2)\}.$$
For components of the stress tensors $W^{\|}_{\mu\nu}$ and
$W^{\perp}_{\mu\nu}$ we obtain
\begin{equation}W^{||}_{tt}=2\alpha
e^{10a+4b}(\dot{A}^2_{||}+A_{||}\ddot{A}_{||})\label{wttpar}\end{equation}
$$+\{\alpha
e^{10a+4b}(\ddot{a}+4\ddot{b}+4\dot{a}^2+20\dot{a}\dot{b}+16\dot{b}^2)
-2(\alpha+\beta)e^{-2a+4b}(\dot{a^2}+2\dot{a}\dot{b}-8\dot{b}^2)$$
$$-4\alpha
e^{-2(a+b)}(\dot{a}^2-\dot{a}\dot{b}-2\dot{b}^2)\}A_{||}^2+\{2\alpha
e^{10a+4b}(5\dot{a}+8\dot{b})
$$
$$-4(\alpha+\beta)(\dot{a}-2\dot{b})e^{-2a+4b}-8\alpha(\dot{a}+\dot{b})e^{-2(a+b)}\}A_{||}\dot{A}_{||}$$
\begin{equation}W^{\perp}_{tt}=e^{-2(a+b)}\{4\alpha(\dot{A}_{\perp}^2+A_{\perp}\ddot{A}_{\perp})
\label{wttperp}\end{equation}
$$-2[2(\alpha+2\beta)\dot{a}+(2\alpha+\beta)\dot{b}]A_{\perp}\dot{A}_{\perp}
$$
$$+[2\alpha\ddot{a}-4\alpha\ddot{b}-4(\alpha+\beta)\dot{a}^2+4(\beta-\alpha)\dot{a}\dot{b}+8(\alpha+\beta)\dot{b}^2]A_{\perp}^2\}$$
\begin{equation}W^{||}_{xx}=-4(2\alpha+\beta)(\dot{a}+2\dot{b})A_{||}\dot{A}_{||}+\{2\beta(\ddot{a}-2\ddot{b})\label{wxxpar}\end{equation}
$$
+4(\beta-\alpha)\dot{a}^2-8(\alpha+2\beta)\dot{a}\dot{b}+16(2\alpha+\beta)\dot{b}^2
\}A^2_{||}
$$
\begin{equation}W^{\perp}_{xx}=8\alpha e^{-6b}\{2(2\dot{b}-\dot{a})A_{\perp}\dot{A}_{\perp}+(5\dot{a}^2-14\dot{a}\dot{b}+8\dot{b}^2)A^2_{\perp}\}
\label{wxxperp}\end{equation}
\begin{equation}W^{||}_{yy}=W^{||}_{zz}=2\alpha[9\dot{a}-2(\dot{a}+\dot{b})e^{6a}]A_{||}\dot{A}_{||}+2\alpha \dot{A}^2_{||}+2\alpha A_{||}\ddot{A}_{||}
\label{wyypar}\end{equation}
$$+
\alpha\{3\ddot{a}+18\dot{a}^2-2e^{6a}[\ddot{a}+\ddot{b}+12\dot{a}\dot{b}+12\dot{a}^2]\}A_{||}^2$$
and
\begin{equation}W^{\perp}_{yy}=W^{\perp}_{zz}=2\{4(\alpha+\beta)\dot{b}+(22\alpha+13\beta)
\dot{a}\}A_{\perp}\dot{A}_{\perp}\label{wyyperp}\end{equation}
$$+\{(2\alpha+3\beta)\ddot{a}-4\alpha \ddot{b}
+2(2\alpha+5\beta)\dot{a}^2+4(\beta+4\alpha)\dot{b}(\dot{b}-\dot{a})\}A_{\perp}^2$$
By substituting the above relations into the equation (\ref{baro})
we obtain for the directional barotropic indexes $\gamma_{\|}$ and
$\gamma_{\perp}$ respectively
\begin{equation}\label{gammapar}\alpha[(1+\gamma_{\|})e^{-12a}+\gamma_{\|}]\ddot{v}$$$$
+[\alpha(2-\gamma_{\|})e^{-12a-2b}+(6\alpha+\beta)e^{-12a}
+(\alpha+\beta)(1-\gamma_{\|})e^{-16a}-\alpha\gamma_{\|}/2]\ddot{a}$$$$
+[\alpha(2-\gamma_{\|})e^{-12a-2b}-(\alpha+\beta)(2-\gamma_{\|})e^{-16a}
+2\beta
e^{-12a}-2\alpha\gamma_{\|}]\ddot{b}+([[(9\alpha+2\beta)$$$$+\gamma_{\|}(7\alpha+2\beta)]\dot{a}+4[(4\alpha+\beta)-\gamma_{\|}
(\beta-\alpha)]\dot{b}]e^{-12a}$$$$
+2\alpha(2-\gamma_{\|})(\dot{a}+\dot{b})e^{-12a-2b}+2(\alpha+\beta)(\dot{a}-2\dot{b})e^{-16a}-\gamma_{\|}(\alpha+\beta)(\dot{a}-2\dot{b})$$$$
-\gamma_{\|}\alpha(5\dot{a}+8\dot{b})+4\gamma_{\|}\alpha(\dot{a}+\dot{b})e^{-12a-6b})\dot{v}+e^{-12a}[2\alpha(1+2\gamma_{\|})-(\gamma_{\|}+1)/8
]\dot{v}^2$$$$+\{\alpha(2-\gamma_{\|})e^{-12a-2b}+[14\alpha-2\beta+\gamma_{\|}(\alpha+\beta)]e^{-12a}$$$$
-3(\alpha+\beta)(1-\gamma_{\|}/2)e^{-16a}
$$$$-2\alpha\gamma_{\|}+2\alpha\gamma_{\|}e^{-12a-6b}\}\dot{a}^2+\{2\alpha(2-\gamma_{\|})e^{-12a-2b}$$$$
-[10\alpha+8\beta+8\gamma_{\|}(\alpha+\beta)]e^{-12a}+4\gamma_{\|}(\alpha+\beta)e^{-16a}-8\alpha\gamma_{\|}$$$$
-4\alpha\gamma_{\|}e^{-12a-6b}\}\dot{b}^2+\{3\alpha(2-\gamma_{\|})e^{-12a-2b}+5(\alpha+\beta)(2-\gamma_{\|})e^{-16a}$$$$
+2(\alpha+\beta)(2+\gamma_{\|})e^{-12a}-\alpha\gamma_{\|}-2\alpha\gamma_{\|}e^{-12a-6b}\}\dot{a}\dot{b}=0
\end{equation}
and
\begin{equation}\alpha\{(1+\gamma_{\perp})e^{-12a}-e^{-12a-6b}+\gamma_{\perp}\}\ddot{v}$$$$
-\{\alpha(\gamma_{\perp}+1)e^{-12a-2b}+(\alpha+\beta)(1-\gamma_{\perp}/2)e^{-16a}-3\alpha
e^{-12a-6b}/2+\alpha e^{-6a-6b}$$$$
-\alpha\gamma_{\perp}/2\}\ddot{a}+\{\alpha
e^{-6a-6b}+(1+\gamma_{\perp})(\alpha+\beta)e^{-16a}-\alpha(1+\gamma_{\perp})e^{-12a-2b}$$$$
-2\alpha\gamma_{\perp}\}\ddot{b}+\{[(\alpha(5+\gamma_{\perp})+2\beta\gamma_{\perp})\dot{a}+(4\alpha(2+\gamma_{\perp})-
4\beta\gamma_{\perp})\dot{b}]e^{-12a} \}\dot{v}$$$$
+\{[(1-\gamma_{\perp})/8+2\alpha(1+\gamma_{\perp})]e^{-12a}-2\alpha
e^{-12a-6b}+2\alpha\gamma_{\perp}\}\dot{v}^2$$$$ +\{12\alpha
e^{-6a-6b}-\alpha(\gamma_{\perp}+1)e^{-12a-2b}+3(\alpha+\beta)(1+\gamma_{\perp})e^{-16a}/2-2\alpha\gamma_{\perp}$$$$
+\gamma_{\perp}(\alpha+\beta)e^{-12a}-\alpha(9-2\gamma_{\perp})e^{-12a-6b}\}\dot{a}^2+\{4(\alpha+\beta)(1+\gamma_{\perp})e^{-16a}$$$$
-2\alpha(\gamma_{\perp}+1)e^{-12a-2b}-8\alpha\gamma_{\perp}-8\gamma_{\perp}(\alpha+\beta)e^{-12a}-4\alpha\gamma_{\perp}e^{-12a-6b}\}\dot{b}^2
$$$$+\{12\alpha
e^{-6a-6b}-5(\gamma_{\perp}+1)(\alpha+\beta)e^{-16a}-3\alpha(\gamma_{\perp}+1)e^{-12a-2b}-\alpha\gamma_{\perp}
$$$$+2\gamma_{\perp}(\alpha+\beta)e^{-12a}-2\alpha\gamma_{\perp}e^{-12a-6b}\}\dot{a}\dot{b}=0\label{gammaperp}\end{equation}
respectively for  case $(A_{\|}\neq0,A_{\perp}=0)$ and
\begin{equation}2\alpha \ddot{u}+[2\alpha-\beta+\gamma_{\|}(\alpha-\beta)]\ddot{a}$$$$-[\beta+4\alpha+\gamma_{\|}(\beta+2\alpha)]\ddot{b}
+[4\alpha+1/4-\gamma_{\|}/8]\dot{u}^2$$$$
+[14\alpha-2\beta+2(4\alpha+\beta)\gamma_{\|}]\dot{a}\dot{u}-[28\alpha+2\beta+\gamma_{\|}(14\alpha+\beta)]\dot{u}\dot{b}
+[\beta+48\alpha$$$$
-\gamma_{\|}(\beta+6\alpha)]\dot{a}\dot{b}-[\beta+18\alpha-\gamma_{\|}(\beta+4\alpha)]\dot{a}^2+[2\beta-24\alpha+2(2\alpha-\beta)\gamma_{\|}]
\dot{b}^2$$$$
\alpha(2-\gamma_{\|})e^{-2(2a+b)}[2\dot{u}(\dot{a}-2\dot{b})+\ddot{a}-2\ddot{b}-3\dot{a}^2+4\dot{a}\dot{b}+4\dot{b}^2]=0\label{vert1}\end{equation}
and
\begin{equation}2\alpha\ddot{u}+[7\alpha-5\beta/2+\gamma_{\perp}(\alpha-\beta)]\ddot{a}-(\beta+2\alpha)(1+\gamma_{\perp})\ddot{b}$$$$+[3\beta-
\gamma_{\perp} (6\alpha+\beta)]\dot{a}\dot{b}
+[17\alpha-17\beta/2+\gamma_{\perp}(2\beta+8\alpha)]\dot{u}\dot{a}$$$$-[16\alpha+6\beta+\gamma_{\perp}(14\alpha+\beta)]\dot{u}\dot{b}+(4\alpha-
\gamma_{\perp} /8
)\dot{u}^2$$$$+[12\alpha-6\beta+\gamma_{\perp}(\beta+4\alpha)]\dot{a}^2+[6\alpha+2\gamma_{\perp}(2\alpha-\beta)]\dot{b}^2$$$$+\alpha(2-
\gamma_{\perp})e^{-4a} [2\dot{u}(\dot{a}-2\dot{b})
+\ddot{a}-2\ddot{b}-3\dot{a}^2+4\dot{a}\dot{b}+4\dot{b}^2]=0\label{vert2}\end{equation}
respectively for  case $A_{\|}=0,A_{\perp}\neq0.$ The above
equations in limits $e^{a(t)}>>1$ reduce to the simpler forms
given by the equations (\ref{ev1}) and (\ref{ev2}) respectively
for $A_{\|}\neq0,A_{\perp}=0$ and (\ref{eu1}) and (\ref{eu2})
respectively for $A_{\|}=0,A_{\perp}\neq0.$
 \vskip .5cm \noindent
\textbf{Appendix III}\\

 The coefficients of the dynamical
equations (\ref{V1}) and (\ref{V2}) are obtained respectively as
follows.
\begin{equation}C_{VV}=2(3168\alpha^4+504\alpha^3\beta-244\alpha^2\beta^2-78\alpha \beta^3-5\beta^4+324\alpha^3+432\alpha^2\beta$$$$-9\alpha \beta^2)
/(6\alpha+\beta)^2(-6\alpha^2+28\alpha\beta+5\beta^2)\end{equation}
\begin{equation}C_{XV}=-4(1728\alpha^4+1584\alpha^3\beta-426\alpha^2\beta^2-80\alpha\beta^3-1404\alpha^3+198\alpha^2\beta$$$$+102\alpha\beta
^2+5\beta^3)/(6\alpha+\beta)^2(-6\alpha^2+28\alpha\beta+5\beta^2)\end{equation}
and\begin{equation}C_{XX}=-\frac{36(324\alpha^4-324\alpha^3\beta+99\alpha^2\beta^2-21\alpha\beta^3-5\beta^4)}{(6\alpha+\beta)^2(-6\alpha^2+28\alpha\beta+5\beta^2)}
\end{equation}
\begin{equation}D_{VV}=(2736\alpha^4+1112\alpha^3\beta+180\alpha^2\beta^2+10\alpha\beta^3+288\alpha^3$$$$+420\beta^2\beta+
40\alpha\beta^2-\beta^3)/(6\alpha+\beta)^2(-6\alpha^2+28\alpha\beta+5\beta^2)\end{equation}\begin{equation}
D_{XV}=-4(864\alpha^4+336\alpha^3\beta-112\alpha^2\beta^2-21\alpha\beta^3-612\alpha^3-42\alpha^2\beta$$$$+28\alpha\beta^2+
3\beta^3)/(6\alpha+\beta)^2(-6\alpha^2+28\alpha\beta+5\beta^2)
\end{equation}and \begin{equation}D_{XX}=-\frac{18(216\alpha^4+84\alpha^3\beta+110\alpha^2\beta^2+\alpha\beta^3-2\beta^4)}{(6\alpha+\beta)^2
(-6\alpha^2+28\alpha\beta+5\beta^2)}.\end{equation} The
coefficients in the equation (\ref{h}) have the following forms.
\begin{equation}
\eta_0(\epsilon)=-83800\epsilon^{13}-2587010
\epsilon^{12}-22948268\epsilon^{11}-60128728\epsilon^{10}$$$$-212135040\epsilon^9-2359178568\epsilon^8+3993113808\epsilon^7+79174050048\epsilon^6$$$$+
16352788032\epsilon^5-591326822016\epsilon^4+145023084288\epsilon^3$$$$+548045263872\epsilon^2-121315304448\epsilon+1451188224
\end{equation}
\begin{equation}
\eta_1(\epsilon)=39500025\epsilon^{14}+341537030\epsilon^{13}-803772011\epsilon^{12}-1532681912\epsilon^{11}$$$$
+35418699464\epsilon^{10}
-104741851680\epsilon^9+212404729332\epsilon^8-307964233008\epsilon^7$$$$-271585335024\epsilon^6+1562211831744\epsilon^5+1336466034432\epsilon^4$$$$-
1089013932288\epsilon^3-1111662247680\epsilon^2+249154237440\epsilon+6832677888
\end{equation}
\begin{equation}
\eta_2(\epsilon)=377100\epsilon^{15}+9214170\epsilon^{14}+58826066\epsilon^{13}-134067668\epsilon^{12}$$$$-2199922976\epsilon^{11}
-1551871744\epsilon^{10}
+35035595592\epsilon^9+122344000512\epsilon^8$$$$+101660228064\epsilon^7
-240137479872\epsilon^6-245324163456\epsilon^5-2037395172096\epsilon^4$$$$+
5424039562752\epsilon^3+6649445219328\epsilon^2-2625414488064\epsilon$$$$+214533992448
\end{equation}
\begin{equation}
\eta_3(\epsilon)=900\epsilon^{16}+36180\epsilon^{15}+517209\epsilon^{14}+2130900\epsilon^{13}-22734456\epsilon^{12}$$$$-
321272496\epsilon^{11}
-1473018732\epsilon^{10}-847693968\epsilon^9+20182695696\epsilon^8$$$$+89619111936\epsilon^7+139261389696\epsilon^6-18328467456
\epsilon^5-71728187904\epsilon^4$$$$+910217502720\epsilon^3+1721868420096\epsilon^2-825967964160\epsilon+89248075776
\end{equation}
\vskip .5cm \noindent
\textbf{Appendix IV}\\
  \begin{equation}P_1=64S^5\alpha\epsilon-1136S^5\alpha+4S^5\epsilon-344S^4\alpha\epsilon+28S^5+3128S^4\alpha
-12S^4\epsilon$$$$+292S^3\alpha\epsilon-116S^4-1592S^3\alpha+32S^3\epsilon+80S^2\alpha\epsilon+176S^3-1920S^2\alpha-18S^2\epsilon
$$$$-356S\alpha\epsilon-44S^2+1168S\alpha-4S\epsilon+264\alpha\epsilon-8S+352\alpha+46\epsilon-4
\end{equation}\begin{equation}P_2=-128S^4\alpha\epsilon+384S^4\alpha-20S^4\epsilon+16S^3\alpha\epsilon-144S^4+
1056S^3\alpha+39S^3\epsilon+48S^2\alpha\epsilon$$$$
+390S^3-2592S^2\alpha+39S^2\epsilon+272S\alpha\epsilon-354S^2+1056S\alpha-3S\epsilon
$$$$-208\alpha\epsilon+90S+96\alpha-55\epsilon+18\end{equation}
\begin{equation}Q_1=24576S^{10}\alpha^3\epsilon^2+3072S^{10}\alpha^2\epsilon^3+12288S^9\alpha^3\epsilon^3-147456S^{10}\alpha^3\epsilon
$$$$-62720S^{10}\alpha^2\epsilon^2+480S^{10}\alpha\epsilon^3-79872S^9\alpha^3\epsilon^2-19968S^9\alpha^2\epsilon^3+9216S^8\alpha^3\epsilon^3
$$$$+221184S^{10}\alpha^3-428800S^{10}\alpha^2\epsilon-5536S^{10}\alpha\epsilon^2-276480S^9\alpha^3\epsilon+75488S^9\alpha^2\epsilon^2
$$$$-4296S^9\alpha\epsilon^3-236160S^8\alpha^3\epsilon^2-11792S^8\alpha^2\epsilon^3-12096S^7\alpha^3\epsilon^3+1990720S^{10}\alpha^2$$$$
-158240S^{10}\alpha\epsilon
+4S^{10}\epsilon^2+1216512S^9\alpha^3+2511936S^9\alpha^2\epsilon-10600S^9\alpha\epsilon^2$$$$
+1525248S^8\alpha^3\epsilon+607064S^8\alpha^2\epsilon^2
+3856S^8\alpha\epsilon^3+242688S^7\alpha^3\epsilon^2+1108S^7\alpha^2\epsilon^3$$$$
-60096S^6\alpha^3\epsilon^3-672544S^{10}\alpha+56S^{10}\epsilon
-2697152S^9\alpha^2+703712S^9\alpha\epsilon-24S^9\epsilon^2$$$$-870912S^8\alpha^3-5531872S^8\alpha^2\epsilon+119672S^8\alpha\epsilon^2
-887040S^7\alpha^3\epsilon-478080S^7\alpha^2\epsilon^2$$$$+11016S^7\alpha\epsilon^3+578688S^6\alpha^3\epsilon^2+41572S^6\alpha^2\epsilon^3
-2880S^5\alpha^3\epsilon^3+196S^{10}$$$$+4037712S^9\alpha-400S^9\epsilon-15858960S^8\alpha^2-738904S^8\alpha\epsilon+100S^8\epsilon^2$$$$
-4561920S^7\alpha^3 +1839136S^7\alpha^2\epsilon
-162996S^7\alpha\epsilon^2-1688832S^6\alpha^3\epsilon-876292S^6\alpha^2\epsilon^2$$$$+1336S^6\alpha\epsilon^3-76032S^5\alpha^3
\epsilon^2
+10492S^5\alpha^2\epsilon^3+62784S^4\alpha^3\epsilon^3-1624S^9$$$$-10577216S^8\alpha+1496S^8\epsilon+57504736S^7\alpha^2
-635032S^7\alpha\epsilon
-228S^7\epsilon^2$$$$+10907136S^6\alpha^3+5613344S^6\alpha^2\epsilon-193924S^6\alpha\epsilon^2
+523008S^5\alpha^3\epsilon+128800S^5\alpha^2\epsilon^2
$$$$-8228S^5\alpha\epsilon^3-903552S^4\alpha^3\epsilon^2+14156S^4\alpha^2\epsilon^3
+55104S^3\alpha^3\epsilon^3+5828S^8$$$$+14927136S^7\alpha-3252S^7\epsilon
-75168336S^6\alpha^2+1035376S^6\alpha\epsilon$$$$
+276S^6\epsilon^2-21897216S^5\alpha^3+3502272S^5\alpha^2\epsilon+361200S^5\alpha\epsilon^2+5831424S^4\alpha^3
\epsilon$$$$
+664112S^4\alpha^2\epsilon^2-932S^4\alpha\epsilon^3-173568S^3\alpha^3\epsilon^2+10732\S^3\alpha^2\epsilon^3-44352S^2\alpha^3\epsilon^3
$$$$
-10824S^7-11085888S^6\alpha+3492S^6\epsilon+36085056S^5\alpha^2+1569752S^5\alpha\epsilon$$$$+68S^5\epsilon^2
+27993600S^4\alpha^3
-17729664S^4\alpha^2\epsilon+180476S^4\alpha\epsilon^2-9305856S^3\alpha^3\epsilon$$$$+824000S^3\alpha^2\epsilon^2
-7344S^3\alpha\epsilon^3+1183104S^2\alpha^3
\epsilon^2-54884S^2\alpha^2\epsilon^3-52416S\alpha^3\epsilon^3$$$$+10184S^6+2759264S^5\alpha
+948S^5\epsilon+7933920S^4\alpha^2-3349368S^4\alpha\epsilon
-899S^4\epsilon^2$$$$-15510528S^3\alpha^3+10608720S^3\alpha^2\epsilon$$$$
-252676S^3\alpha\epsilon^2+5541120S^2\alpha^3\epsilon-1020020S^2\alpha^2\epsilon^2
-10336S^2\alpha\epsilon^3-659712S\alpha^3\epsilon^2$$$$
-43836S\alpha^2\epsilon^3+32448\alpha^3\epsilon^3-3464S^5+1523680S^4\alpha-6248S^4\epsilon
-11764864S^3\alpha^2$$$$
+1583464S^3\alpha\epsilon+1132S^3\epsilon^2+1866240S^2\alpha^3+1220976S^2\alpha^2\epsilon-188892S^2\alpha\epsilon^2$$$$
-1002240S\alpha^3
\epsilon-49088S\alpha^2\epsilon^2+3668S\alpha\epsilon^3+99840\alpha^3\epsilon^2+49348\alpha^2\epsilon^3+12S^4$$$$
-993152S^3\alpha+
5024S^3\epsilon+1157152S^2\alpha^2+215184S^2\alpha\epsilon+70S^2\epsilon^2+608256S\alpha^3$$$$-1497712S\alpha^2\epsilon+108272S\alpha\epsilon^2
-112896\alpha^3\epsilon+186736\alpha^2\epsilon^2+10780\alpha\epsilon^3-176S^3$$$$
+21520S^2\alpha-800S^2\epsilon+756224S\alpha^2-205096S\alpha\epsilon
-1012S\epsilon^2+27648\alpha^3$$$$
-108336\alpha^2\epsilon+45004\alpha\epsilon^2+104S^2+55440S\alpha-96S\epsilon+61504\alpha^2-20848\alpha\epsilon+
$$$$529\epsilon^2+16S+4048\alpha-92\epsilon+4\end{equation}
\begin{equation}Q_2=6144S^{10}\alpha^3\epsilon^5-253952S^{10}\alpha^3\epsilon^4+8160S^{10}\alpha^2\epsilon^5-105216S^9\alpha^3\epsilon^5
$$$$-1902592S^{10}\alpha^3\epsilon^3-234400S^{10}\alpha^2\epsilon^4-430336S^9\alpha^3\epsilon^4-133896S^9\alpha^2\epsilon^5$$$$
-119296S^8\alpha^3\epsilon^5
+8212480S^{10}\alpha^3\epsilon^2-6215936S^{10}\alpha^2\epsilon^3+1156S^{10}\alpha\epsilon^4$$$$
+10249728S^9\alpha^3\epsilon^3-2447320S^9\alpha^2\epsilon^4
+2077696S^8\alpha^3\epsilon^4-261808S^8\alpha^2\epsilon^5$$$$
+57728S^7\alpha^3\epsilon^5+13541376S^{10}\alpha^3\epsilon-40224640S^{10}\alpha^2\epsilon^2+
22712S^{10}\alpha\epsilon^3$$$$
+7140352S^9\alpha^3\epsilon^2-8378320S^9\alpha^2\epsilon^3-16728S^9\alpha\epsilon^4-9789184S^8\alpha^3\epsilon^3
$$$$
-1721008S^8\alpha^2\epsilon^4-132864S^7\alpha^3\epsilon^4+27588S^7\alpha^2\epsilon^5+403712S^6\alpha^3\epsilon^5$$$$
+3686400S^{10}\alpha^3-81907200S^{10}
\alpha^2\epsilon+131956S^{10}\alpha\epsilon^2+3373056\S^9\alpha^3\epsilon$$$$
+36944736S^9\alpha^2\epsilon^2-379240S^9\alpha\epsilon^3-40166400S^8\alpha^3
\epsilon^2-4043304S^8\alpha^2\epsilon^3$$$$
+4652S^8\alpha\epsilon^4-528896S^7\alpha^3\epsilon^3+1689368S^7\alpha^2\epsilon^4-2399232S^6\alpha^3\epsilon^4
$$$$+517248S^6\alpha^2\epsilon^5$$$$
+188672S^5\alpha^3\epsilon^5-47655936S^{10}\alpha^2+200400S^{10}\alpha\epsilon-1474560S^9\alpha^3$$$$+124048320S^9\alpha^2\epsilon
$$$$
-2579920S^9\alpha\epsilon^2-24013824S^8\alpha^3\epsilon-90626336S^8\alpha^2\epsilon^2+448128S^8\alpha\epsilon^3$$$$
+135737344S^7\alpha^3\epsilon^2
+7465024S^7\alpha^2\epsilon^3-71364S^7\alpha\epsilon^4+5481728S^6\alpha^3\epsilon^3$$$$
+2902864S^6\alpha^2\epsilon^4-256S^5\alpha^3\epsilon^4
+429496S^5\alpha^2\epsilon^5-253568S^4\alpha^3\epsilon^5$$$$+90000S^{10}\alpha$$$$
+60042240S^9\alpha^2-5098560S^9\alpha\epsilon+7520256S^8\alpha^3
-239536416S^8\alpha^2\epsilon$$$$
+4928264S^8\alpha\epsilon^2+69556224S^7\alpha^3\epsilon+61286304S^7\alpha^2\epsilon^2-413444S^7\alpha\epsilon^3
$$$$
-232470528S^6\alpha^3\epsilon^2+6333656S^6\alpha^2\epsilon^3-45008S^6\alpha\epsilon^4+12833280S^5\alpha^3\epsilon^3$$$$
+2968136S^5\alpha^2\epsilon^4
+1385728S^4\alpha^3\epsilon^4-51716S^4\alpha^2\epsilon^5-285568S^3\alpha^3\epsilon^5$$$$-2881152S^9\alpha
-97712640S^8\alpha^2+12139536S^8\alpha\epsilon
+576S^8\epsilon^2$$$$-2949120S^7\alpha^3+184487616S^7\alpha^2\epsilon
+3377376S^7\alpha\epsilon^2$$$$+576S^7\epsilon^3-137892864S^6\alpha^3\epsilon
+41340800S^6\alpha^2\epsilon^2-1515888S^6\alpha\epsilon^3$$$$
+146562048S^5\alpha^3\epsilon^2+21912144S^5\alpha^2\epsilon^3+176488S^5\alpha\epsilon^4
-43125504S^4\alpha^3\epsilon^3$$$$
+1574072S^4\alpha^2\epsilon^4+1281792S^3\alpha^3\epsilon^4-301452S^3\alpha^2\epsilon^5-35584S^2\alpha^3\epsilon^5
$$$$+8660160S^8\alpha
+5760S^8\epsilon+98878464S^7\alpha^2$$$$+19415424S^7\alpha\epsilon+5760S^7\epsilon^2-10764288S^6\alpha^3$$$$
+46298528S^6\alpha^2\epsilon
-12410180S^6\alpha\epsilon^2+2304S^6\epsilon^3+115292160S^5\alpha^3\epsilon$$$$
+44002656S^5\alpha^2\epsilon^2+353992S^5\alpha\epsilon^3-13218816S^4\alpha^3
\epsilon^2-8373416S^4\alpha^2\epsilon^3$$$$
+205437S^4\alpha\epsilon^4+29578752S^3\alpha^3\epsilon^3-1545016S^3\alpha^2\epsilon^4-2917376S^2\alpha^3\epsilon^4
$$$$
-217224S^2\alpha^2\epsilon^5+61440S\alpha^3\epsilon^5+5184S^8+22019616S^7\alpha-8064S^7\epsilon$$$$
-14608896S^6\alpha^2-36238000S^6\alpha\epsilon
+19008S^6\epsilon^2+4423680S^5\alpha^3$$$$-5504064S^5\alpha^2\epsilon
-3138432S^5\alpha\epsilon^2$$$$+3456S^5\epsilon^3-51391488S^4\alpha^3\epsilon
-75845920S^4\alpha^2\epsilon^2+1633188S^4\alpha\epsilon^3$$$$
-25483264S^3\alpha^3\epsilon^2-1767488S^3\alpha^2\epsilon^3+4168S^3\alpha\epsilon^4
-3055872S^2\alpha^3\epsilon^3$$$$
-2991008S^2\alpha^2\epsilon^4+1217024S\alpha^3\epsilon^4-68136S\alpha^2\epsilon^5+81536\alpha^3\epsilon^5-17280S^7
$$$$
-32894400S^6\alpha+55296S^6\epsilon-28480512S^5\alpha^2-11973200S^5\alpha\epsilon-56736S^5\epsilon^2$$$$
-15187968S^4\alpha^3
-98371104S^4\alpha^2\epsilon+3873668S^4\alpha\epsilon^2+16416S^4\epsilon^3$$$$-22984704S^3\alpha^3\epsilon
+12038560S^3\alpha^2\epsilon^2$$$$
+1023168S^3\alpha\epsilon^3+8101888S^2\alpha^3\epsilon^2-7206168S^2\alpha^2\epsilon^3
-75270S^2\alpha\epsilon^4$$$$+748544S\alpha^3\epsilon^3
-75920S\alpha^2\epsilon^4+171776\alpha^3\epsilon^4+10780\alpha^2\epsilon^5+12096S^6$$$$
-12323904S^5\alpha-221472S^5\epsilon-16200192S^4\alpha^2
+5234736S^4\alpha\epsilon$$$$+35712S^4\epsilon^2
+5898240S^3\alpha^3$$$$+33378368S^3\alpha^2\epsilon+6822456S^3\alpha\epsilon^2+8928S^3\epsilon^3+
39736320S^2\alpha^3\epsilon$$$$
+5423296S^2\alpha^2\epsilon^2-428180S^2\alpha\epsilon^3+5335040S\alpha^3\epsilon^2+1540864S\alpha^2\epsilon^3
$$$$-6396S\alpha\epsilon^4
-489984\alpha^3\epsilon^3$$$$+87592\alpha^2\epsilon^4-127872S^5+5319504S^4\alpha-150912S^4\epsilon+25857024S^3\alpha^2$$$$
+15261744S^3\alpha\epsilon+127296S^3\epsilon^2+14155776S^2\alpha^3+21374816S^2\alpha^2\epsilon$$$$-96820S^2\alpha\epsilon^2
-5760S^2\epsilon^3$$$$
-4657152S\alpha^3\epsilon+2878592S\alpha^2\epsilon^2+65284S\alpha\epsilon^3+249856\alpha^3\epsilon^2+33424\alpha^2\epsilon^3$$$$
+529\alpha\epsilon^4
-363456S^4+11275968S^3\alpha+481248S^3\epsilon+6987264S^2\alpha^2$$$$+1933504S^2\alpha\epsilon
-5760S^2\epsilon^2-5898240S\alpha^3$$$$
-1440896S\alpha^2\epsilon+269504S\alpha\epsilon^2-559104\alpha^3\epsilon-290048\alpha^2\epsilon^2
-2104\alpha\epsilon^3$$$$+538560S^3
+1619968S^2\alpha+61920S^2\epsilon-4012032S\alpha^2$$$$-313152S\alpha\epsilon
+9504S\epsilon^2+589824\alpha^3-154048\alpha^2\epsilon
-31360\alpha\epsilon^2+116928S^2$$$$-796288S\alpha+8640S\epsilon
+316416\alpha^2-8000\alpha\epsilon$$$$-39744S+51904\alpha-2016\epsilon+2304
\end{equation}
\begin{equation}K=2S^3\alpha^2\epsilon^3-2S^3\alpha^2\epsilon^2-2S^3\alpha\epsilon^3+2S^2\alpha^2\epsilon^3-3S^3\alpha\epsilon^2-2S^2\alpha^2\epsilon^2
$$$$-4S^2\alpha\epsilon^3-2S\alpha^2\epsilon^3
-76S^3\alpha\epsilon+7S^3\epsilon^2-13S^2\alpha\epsilon^2
+2S\alpha^2\epsilon^2$$$$-2S\alpha\epsilon^3
-2\alpha^2\epsilon^3-120S^3\alpha+S^3\epsilon+26S^2\alpha\epsilon-2S^2\epsilon^2+5S\alpha\epsilon^2+2\alpha^2\epsilon^2$$$$-56S^3
+24S^2\alpha-11S^2\epsilon
+78S\alpha\epsilon-3S\epsilon^2+3\alpha\epsilon^2+22S^2+120S\alpha+9S\epsilon$$$$-12\alpha\epsilon+42S-24\alpha+3\epsilon-12
\end{equation}
 we defined
\begin{equation}J=-4\beta\gamma_{\|}\alpha+2\alpha\beta^2\gamma_{\perp}-2\alpha^2\gamma_{\|}\gamma_{\perp}\label{J}\end{equation}
$$-2\alpha^2\beta\gamma_{\perp}
+3\alpha\beta\gamma_{\perp}+\beta^2\gamma_{\|}\gamma_{\perp}-30\beta\gamma_{\|}\alpha^2
-5\alpha\beta^2\gamma_{\|}
$$$$+2\alpha^2\beta\gamma_{\|}^2-\alpha\beta^2\gamma_{\|}^2-2\alpha^2\beta^2\gamma_{\|}+2\alpha\beta^3\gamma_{\|}+\alpha\beta\gamma_{\|}\gamma_{\perp}
+5\beta^2\gamma_{\|}/2-18\alpha^2\gamma_{\|}$$$$
+\alpha\beta-4\alpha^2\gamma_{\perp}+\beta^2\gamma_{\perp}-48\alpha^3\gamma_{\|}-46\alpha^2\beta+
\alpha\beta^2-\beta^3\gamma_{\|}^2$$$$-\beta^3\gamma_{\|}-32\alpha^2+5\beta^2/2-72\alpha^3$$
\begin{equation}E_1=(1/2)\alpha^2\gamma_{\|}\gamma_{\perp}-(15/4)\alpha\beta\gamma_{\perp}-\beta^2\gamma_{\|}\gamma_{\perp}
-4\alpha\beta^2\gamma_{\perp}-\gamma_{\|}^2\beta\alpha+4
\alpha^2\beta\gamma_{\perp}$$$$-2\alpha^2\beta\gamma_{\|}^2
+\alpha\beta^2\gamma_{\|}^2+9\alpha\beta^2\gamma_{\|}+
(33/8)\alpha\beta\gamma_{\|}+62\alpha^2\beta\gamma_{\|}-4\alpha\beta^3\gamma_{\|}$$$$+(1/8)\beta^3\gamma_{\|}\gamma_{\perp}
+96\alpha^3\gamma_{\|}
-(3/2)\gamma_{\|}^2\alpha^2-2\alpha\beta^2+92\alpha^2\beta$$$$-(1/8)\gamma_{\|}^2\beta^2+4\alpha^2\beta^2\gamma_{\|}+144\alpha^3+35
\alpha^2-(23/8)\beta^2$$$$
+\alpha^2\gamma_{\perp}-(11/8)\beta^2\gamma_{\perp}+\beta^3\gamma_{\|}+(11/4)\alpha\beta+\beta^3\gamma_{\|}^2+(39/2)
\alpha^2\gamma_{\|}$$$$-(33/16)\beta^2\gamma_{\|}
-(1/8)\alpha\beta^2\gamma_{\|}\gamma_{\perp}-(17/8)\alpha\beta\gamma_{\|}\gamma_{\perp}\end{equation}
\begin{equation}E_2=
98\alpha^2\beta\gamma_{\perp}+16\alpha\beta^2\gamma_{\perp}+5\beta^3\gamma_{\|}\gamma_{\perp}-44
\alpha^2\gamma_{\|}^2\beta-12\alpha\gamma_{\|}^2\beta^2
+24\alpha^3\beta\gamma_{\|}^2$$$$-6\alpha^2\beta^2\gamma_{\|}^2
-15\alpha\beta^3\gamma_{\|}^2+24\alpha^3\beta\gamma_{\|}
+6\alpha^2\beta^2\gamma_{\|}-33\alpha\beta^3\gamma_{\|}+6\alpha^2\beta\gamma_{\|}-
\alpha\beta^2\gamma_{\|}$$$$+24\alpha^3\gamma_{\|}\gamma_{\perp}-
\beta^4\gamma_{\|}\gamma_{\perp}
-48\alpha^3\gamma_{\|}^2$$$$-\beta^3\gamma_{\|}^2+120\alpha^3+3
\beta^3\gamma_{\perp}-15\alpha\beta^2+150\alpha^2\beta+4\beta^3\gamma_{\|}+3\beta^4\gamma_{\|}+96\alpha^3\gamma_{\|}
$$$$
-3\beta^4\gamma_{\|}^2+48\alpha^3\gamma_{\perp}-3\alpha\beta^3\gamma_{\|}\gamma_{\perp}+4\alpha^2\beta^2\gamma_{\|}\gamma_{\perp}+30
\alpha\beta^2\gamma_{\|}\gamma_{\perp}$$$$+46\alpha^2\beta\gamma_{\|}\gamma_{\perp}\end{equation}
\begin{equation}
E_3=12\alpha^3\beta\gamma_{\|}^2-6\alpha\beta^3\gamma_{\|}^2-6\alpha^2\beta^2\gamma_{\|}^2+24\alpha\beta^2\gamma_{\perp}-26\alpha^2\beta\gamma_{\perp}
-12\alpha\beta^2\gamma_{\|}\gamma_{\perp}$$$$
+18\alpha^2\beta\gamma_{\|}\gamma_{\perp}+12\alpha^3\beta\gamma_{\|}-12\alpha\beta^3\gamma_{\|}
-27\alpha\beta^2\gamma_{\|}-140\alpha^2\beta\gamma_{\|}+36\alpha^3\gamma_{\|}\gamma_{\perp}$$$$+
2\beta^4\gamma_{\|}\gamma_{\perp}
 +8\alpha^2\gamma_{\|}^2\beta
-12\alpha\gamma_{\|}^2\beta^2+48\alpha^3\gamma_{\|}^2-2\beta^3\gamma_{\|}^2
+5\beta^3\gamma_{\|}-4\beta^3\gamma_{\perp}-3\beta^3$$$$
-384\alpha^3 +93\alpha\beta^2 -324\alpha^3\gamma_{\|}
-342\alpha^2\beta+4\alpha^2\beta^2\gamma_{\|}\gamma_{\perp}$$$$-6\alpha\beta^3\gamma_{\|}\gamma_{\perp}+72\alpha^3\gamma_{\perp}
\end{equation}
\begin{equation}
E_4=12\alpha^2\gamma_{\perp}-3\beta^2\gamma_{\perp}
+96\alpha^3\gamma_{\|}^2+468\alpha^3\gamma_{\|} +118\alpha^2\beta
-102\alpha\beta^2+4\beta^3\gamma_{\perp}$$$$
-(21/2)\beta^3\gamma_{\|}+5\beta^3\gamma_{\|}^2+54\alpha^2\gamma_{\|}
-(15/2)\beta^2\gamma_{\|}+8\alpha^2\beta^2\gamma_{\|}\gamma_{\perp}
-6\alpha\beta^3\gamma_{\|}\gamma_{\perp}$$$$
-3\alpha\beta\gamma_{\|}\gamma_{\perp}
+96\alpha^2\gamma_{\|}\beta\gamma_{\perp}+
18\alpha\gamma_{\|}\beta^2\gamma_{\perp}-(11/2)\beta^3+576\alpha^3-(15/2)\beta^2
$$$$+96\alpha^2-2\beta^4\gamma_{\|}\gamma_{\perp}+17\alpha^2\beta^2\gamma_{\|}-(51/2)\alpha\beta^3\gamma_{\|}+96\alpha^3\gamma_{\|}\gamma_{\perp}
+126\alpha^2\beta\gamma_{\perp}$$$$+8\alpha\beta^2\gamma_{\perp}+260\alpha^2\beta\gamma_{\|}-32\alpha\beta^2\gamma_{\|}+27\alpha\beta^2\gamma_{\|}^2
+82\alpha^2\beta\gamma_{\|}^2+12\alpha\beta\gamma_{\|}$$$$-3\beta^2\gamma_{\|}\gamma_{\perp}
+6\alpha^2\gamma_{\|}\gamma_{\perp}-9\alpha\beta\gamma_{\perp}-3\alpha\beta
+192\alpha^3\gamma_{\perp}+(17/2)\beta^4\gamma_{\|}
\end{equation}\begin{equation}
E_5=\beta^3\gamma_{\perp}+504\gamma_{\|}\alpha^3-72\gamma_{\|}^2\alpha^3
-\gamma_{\|}^2\beta^3-3\beta^4\gamma_{\|}^2
-6\beta^4\gamma_{\|}+708\alpha^2\beta$$$$ -30\alpha\beta^2
+13\beta^3\gamma_{\|} -6\alpha^2\beta^2\gamma_{|}\gamma_{\perp}
+5\alpha\beta^3\gamma_{\|}\gamma_{\perp}$$$$-72\alpha^2\beta\gamma_{\|}\gamma_{\perp}
-6\alpha\beta^2\gamma_{\|}\gamma_{\perp}-144\alpha^3\gamma_{\perp}+576\alpha^3+9\beta^3$$$$
-72\alpha^3\gamma_{\|}\gamma_{\perp}+\beta^4\gamma_{\|}\gamma_{\perp}-60\gamma_{\|}^2\alpha^2\beta-14\gamma_{\|}^2\alpha\beta^2
+6\alpha^2\beta^2\gamma_{\|}^2-3\alpha\beta^3\gamma_{\|}^2$$$$
+6\alpha^2\beta^2\gamma_{\|}
+276\alpha^2\beta\gamma_{\|}+62\alpha\beta^2\gamma_{\|}
-36\alpha^2\beta\gamma_{\perp}$$$$-52\alpha\beta^2\gamma_{\perp}+3\beta^3\gamma_{\|}\gamma_{\perp}
\end{equation}
\begin{equation}
E_6=4\alpha\beta^2\gamma_{\perp}-168\alpha^3\gamma_{\|}^2
-696\alpha^3\gamma_{\|} -\beta^3\gamma_{\|}^2
-(13/2)\beta^3\gamma_{\|} -716\alpha^2\beta$$$$ -52\alpha\beta^2
+\beta^3\gamma_{\perp}-14\alpha^2\beta^2\gamma_{\|}\gamma_{\perp}+13\alpha\beta^3\gamma_{\|}\gamma_{\perp}$$$$
-138\alpha^2\gamma_{\|}\beta\gamma_{\perp}
-9\alpha\gamma_{\|}\beta^2\gamma_{\perp}-336\alpha^3\gamma_{\perp}
+6\beta^4\gamma_{\|}-720\alpha^3-3\beta^3$$$$
-168\alpha^3\gamma_{\|}\gamma_{\perp}+
\beta^4\gamma_{\|}\gamma_{\perp}-16\alpha^2\beta^2\gamma_{\|}+10\alpha\beta^3\gamma_{\|}
-124\alpha^2\gamma_{\|}^2\beta-22\alpha\gamma_{\|}^2\beta^2
$$$$-670\alpha^2\gamma_{\|}\beta
-102\alpha\gamma_{\|}\beta^2-164\alpha^2\beta\gamma_{\perp}
\label{E}\end{equation}
\begin{equation}
M_1=-(1/4)\alpha\beta\gamma_{\perp}+(17/4)\alpha\beta\gamma_{\|}
+12\alpha^2+(7/2)\alpha\beta+8\alpha^2\gamma_{\|}-(1/8)\gamma_{\|}^2\beta$$$$-(1/4)\alpha\gamma_{\|}\gamma_{\perp}
-(1/8)\beta\gamma_{\|}\gamma_{\perp}+(1/2)\alpha
-(1/2)\alpha\gamma_{\perp}-
(1/8)\beta\gamma_{\perp}-(1/4)\gamma_{\|}^2\alpha$$$$+(1/4)\gamma_{\|}\alpha+(1/8)\gamma_{\perp}\beta+(1/4)\beta
\end{equation}
\begin{equation}M_2=8\alpha^2\beta\gamma_{\perp}+2\alpha\beta^2\gamma_{\perp}
+80\alpha^2\beta\gamma_{\|}+14\alpha\beta^2\gamma_{\|}+144\alpha^3-2\alpha\beta^2
$$$$+72\alpha^2\beta+96\alpha^3\gamma_{\|}-8\alpha^2\gamma_{\|}^2+8\alpha^2\gamma_{\|}\gamma_{\perp}
+8\alpha\beta\gamma_{\perp}+\beta^2\gamma_{\|}\gamma_{\perp}+14\alpha\gamma_{\|}\beta
$$$$-6\alpha\gamma_{\|}^2\beta+6\alpha\beta\gamma_{\|}\gamma_{\perp}+84\alpha^2-\beta^2\gamma_{\|}^2
+16\alpha^2\gamma_{\perp}+\beta^2\gamma_{\perp}+8\alpha\beta$$$$-6\beta^2\gamma_{\|}+52\alpha^2\gamma_{\perp}-5\beta^2
\end{equation}
\begin{equation}
M_3=8\alpha^2\beta\gamma_{\perp}-4\alpha\beta^2\gamma_{\perp}+16\alpha^2\beta\gamma_{\|}+4\alpha\beta^2\gamma_{\|}
-4\alpha\beta^2+48\alpha^3\gamma_{\|}+92\alpha^2\beta$$$$+8\alpha^2\gamma_{\|}^2+8\alpha^2\gamma_{\|}\gamma_{\perp}
-4\alpha\beta\gamma_{}\perp-2\beta^2\gamma_{\|}
\gamma_{\perp}-18\alpha\gamma_{\|}
\beta-32\alpha^2+16\alpha^2\gamma_{\perp}$$$$-2\beta^2\gamma_{\perp}-
18\alpha
\beta-2\beta^2\gamma_{\perp}^2-36\alpha^2\gamma_{\|}+2\beta^2+72\alpha^3
\end{equation}
\begin{equation}
M_4=32\alpha^2\gamma_{\perp}+2\beta^2\gamma_{\perp}
+6\alpha^2\beta-13\alpha\beta^2+24\alpha^2\beta+54\alpha^2\gamma_{\|}-19\alpha\beta$$$$-
(17/2)\beta^2\gamma_{\|}+12\alpha\beta\gamma_{\|}\gamma_{\perp}-(21/2)\beta^2+
60\alpha^2+16\alpha^2\beta\gamma_{\perp}+4\alpha\beta^2\gamma_{\perp}
$$$$-16\alpha^2\beta\gamma_{\|}-4\alpha\beta^2\gamma_{\|}+10\alpha\beta\gamma_{\|}+16\alpha^2\gamma_{\|}\gamma_{\perp}
+16\alpha\beta\gamma_{\perp}
+2\beta^2\gamma_{\|}\gamma_{\perp}$$$$+16\alpha^2\gamma_{\|}^2+2\beta^2\gamma_{\|}^2+12\alpha\beta\gamma_{\|}^2
\end{equation}
\begin{equation}
M_5=-60\alpha^2\beta+16\alpha\beta^2+36\alpha^2\beta\gamma_{\|}+14\alpha\beta^2\gamma_{\|}-12\alpha^2\beta\gamma_{\perp}
-2\alpha\beta^2\gamma_{\perp}$$$$+96\alpha^2-8\alpha\beta\gamma_{\|}\gamma_{\perp}-24\alpha^2\gamma_{\perp}
-\beta^2\gamma_{\perp}+84\alpha^2\gamma_{\|}+3\beta^2\gamma_{\|}+62\alpha\beta-\beta^2\gamma_{\|}^2
$$$$+48\alpha\gamma_{\|}\beta-8\alpha\gamma_{\|}^2\beta
-12\alpha^2\gamma_{\|}\gamma_{\perp}
-10\alpha\beta\gamma_{\perp}-\beta^2\gamma_{\|}\gamma_{\perp}$$$$+4\beta^2-12\gamma_{\|}^2\alpha^2
\end{equation}
\begin{equation}M_6=-8\alpha\beta^2+28\alpha^2\gamma_{\|}\beta
+2\alpha\gamma_{\|}\beta^2-28\alpha^2\beta\gamma_{\perp}
-2\alpha\beta^2\gamma_{\perp}
-120\alpha^2$$$$-16\alpha\beta\gamma_{\|}\gamma_{\perp}-56\alpha^2\gamma_{\perp}
-\beta^2\gamma_{\perp}-9\beta^2\gamma_{\|}-72\alpha\beta
-\beta^2\gamma_{\|}^2 -76\alpha\gamma_{\|}
\beta-16\alpha\gamma_{\|}^2\beta$$$$-28\alpha^2\gamma_{\|}\gamma_{\perp}-18\alpha\beta\gamma_{\perp}
-\beta^2\gamma_{\|} \gamma_{\perp}-8\beta^2-116\alpha^2\gamma_{\|}
-28\alpha^2\gamma_{\|}^2\end{equation}
\begin{equation}
N_1=-(7/4)\alpha\beta\gamma_{\perp}+(1/8)\beta^2\gamma_{\|}\gamma_{\perp}
+2\alpha^2\beta\gamma_{\|}-2\alpha\beta^2\gamma_{\|}-(1/4)\alpha\beta\gamma_{\|}
+13\alpha^2$$$$+(7/2)\alpha^2\gamma_{\perp}-(9/2)\alpha\beta
-(3/2)\alpha^2\gamma_{\|}-(1/8)\alpha\beta\gamma_{\|}\gamma_{\perp}-(1/8)\alpha\gamma_{\|}\gamma_{\perp}$$$$
+(1/8)\beta\gamma_{\|}\gamma_{\perp}+(7/4)\alpha+(1/2)\alpha\gamma_{\perp}-(1/4)\beta\gamma_{\perp}
-(7/8)\gamma_{\|}\alpha$$$$+(5/16)\gamma_{\|}\beta-(5/8)\beta
\end{equation}
\begin{equation}
N_2=-38\alpha^2\beta\gamma_{\perp}-8\alpha\beta^2\gamma_{\perp}
-\beta^3\gamma_{\|}\gamma_{\perp}+24\alpha^3\beta\gamma_{\|}
-18\alpha^2\beta^2\gamma_{\|}-6\alpha\beta^3\gamma_{\|}$$$$-8\alpha^2\beta\gamma_{\|}
-20\alpha\beta^2\gamma_{\|}+192\alpha^3-24\alpha^3\gamma_{\perp}-\alpha\beta^2
+66\alpha^2\beta+6\beta^3\gamma_{\|}
-48\alpha^3\gamma_{\|}$$$$+4\alpha^2\beta\gamma_{\|}\gamma_{\perp}-3\alpha\beta^2\gamma_{\|}\gamma_{\perp}
-4\alpha^2\gamma_{\|} \gamma_{\perp}
-19\alpha\beta\gamma_{\perp}+\beta^2\gamma_{\|}\gamma_{\perp}+3\alpha\gamma_{\|}\beta+3\alpha
\beta\gamma_{\|}\gamma_{\perp}$$$$+102\alpha^2+10\alpha^2\gamma_{\perp}
-\beta^2\gamma_{\perp}-26\alpha\beta+(5/2)\beta^2\gamma_{\|}-28\alpha^2\gamma_{\|}-(5/2)\beta^2
\end{equation}
\begin{equation}
N_3=+4\alpha^2\beta\gamma_{\perp}+4\alpha\beta^2\gamma_{\perp}
+12\alpha^3\beta\gamma_{\|}-10\alpha^2\beta\gamma_{\|}-10\alpha\beta^2\gamma_{\|}
-348\alpha^3$$$$+4\alpha\beta^2-36\alpha^3\gamma_{\perp}+48\alpha^3\gamma_{\|}-74\alpha^2\beta+4\alpha^2\beta\gamma_{\|}\gamma_{\perp}
+2\beta^3\gamma_{\|}\gamma_{\perp}+4\alpha^2\gamma_{\|}\gamma_{\perp}
$$$$+34\alpha\beta\gamma_{\perp}+2\beta^2\gamma_{\|}\gamma_{\perp}-24\alpha\gamma_{\|}\beta
-6\alpha\beta\gamma_{\|}\gamma_{\perp}-208\alpha^2-36\alpha^2\gamma_{\perp}
-2\beta^2\gamma_{\perp}+84\alpha
\beta$$$$+5\beta^2\gamma_{\|}+28\alpha^2\gamma_{\|}-5\beta^2-12\alpha^2\beta^2\gamma_{\|}-6\alpha\beta^2\gamma_{\|}\gamma_{\perp}
\end{equation}
\begin{equation}
N_4=-8\alpha^2\gamma_{\perp}+96\alpha^3\gamma_{\|}
+72\alpha^2\beta
+13\alpha\beta^2+(17/2)\beta^3\gamma_{\|}+56\alpha^2\gamma_{\|}-17\alpha\beta$$$$-
5\beta^2\gamma_{\|}-6\alpha\beta\gamma_{\|}\gamma_{\perp}+8\alpha^2\gamma_{\|}\beta\gamma_{\perp}$$$$
-6\alpha\gamma_{\|}\beta^2\gamma_{\perp}-96\alpha^3\gamma_{\perp}-36\alpha^3+5\beta^2+34\alpha^2-40\alpha^2\beta\gamma_{\perp}
$$$$-4\alpha\beta^2\gamma_{\perp}+51\alpha^2\beta\gamma_{\|}-(31/2)\alpha\beta^2\gamma_{\|}
-6\alpha\beta\gamma_{\|}+8\alpha^2\gamma_{\|}\gamma_{\perp}$$$$-14\alpha\beta\gamma_{\perp}
-2\beta^2\gamma_{\|}\gamma_{\perp}-2\beta^3\gamma_{\|}\gamma_{\perp}
\end{equation}
\begin{equation}
N_5=-72\gamma_{\|}\alpha^3+102\alpha^2\beta-19\alpha\beta^2-
3\beta^3\gamma_{\|}-6\alpha^2\beta\gamma_{\|}\gamma_{\perp}+5\alpha\beta^2\gamma_{\|}\gamma_{\perp}$$$$+72\alpha^3\gamma_{\perp}
+576\alpha^3+6\alpha^2\beta^2\gamma_{\|}-24\alpha^2\beta\gamma_{\|}
+\alpha\beta^2\gamma_{\|}+30\alpha^2\beta\gamma_{\perp}-4\alpha\beta^2\gamma_{\perp}
$$$$+\beta^3
\gamma_{\|}\gamma_{\perp}+336\alpha^2+5\alpha\beta\gamma_{\|}\gamma_{\perp}+60\alpha^2\gamma_{\perp}$$$$
-\beta^2\gamma_{\perp}-42\alpha^2\gamma_{\|}
+(5/2)\beta^2\gamma_{\|}-119\alpha\beta+8\alpha\gamma_{\|}\beta-6\alpha^2\gamma_{\|}\gamma_{\perp}$$$$-45\alpha\beta\gamma_{\perp}
+\beta^2\gamma_{\|}\gamma_{\perp}
-(5/2)\beta^2-6\alpha\beta^3\gamma_{\|}
\end{equation}
\begin{equation}
N_6=-168\alpha^3\gamma_{\|}+6\beta^3\gamma_{\|}+24\alpha^2\beta+8\alpha\beta^2-14\alpha^2\gamma_{\|}
\beta\gamma_{\perp}
+13\alpha\gamma_{\|}\beta^2\gamma_{\perp}$$$$+168\alpha^3\gamma_{\perp}
-144\alpha^3-56\alpha^2\gamma_{\|}\beta+8\alpha\gamma_{\|}\beta^2
+40\alpha^2\beta\gamma_{\perp}+2\alpha\beta^2\gamma_{\perp}-164\alpha^2
$$$$+13\alpha\beta\gamma_{\|}\gamma_{\perp}+\beta^3\gamma_{\|}\gamma_{\perp}
+2\beta^2\gamma_{\perp}+(5/2)\beta^2\gamma_{\|}+68\alpha\beta+28\alpha\gamma_{\|}\beta-14\alpha^2\gamma_{\|}\gamma_{\perp}
$$$$+28\alpha\beta\gamma_{\perp}+\beta^2\gamma_{\|}\gamma_{\perp}+5\beta^2-98\alpha^2\gamma_{\|}
\label{N}\end{equation}
\newpage
\begin{center}
Table 1: Numerical values of the critical points from
$g(\alpha,\beta)=0$ \vskip .5cm \noindent
\begin{tabular}{|c|c|c|c|c|c|c|c|}
  \hline
  $\epsilon$ & $\alpha$ & $\beta$ & $\frac{X_c}{H}$ & $\frac{V_c}{H}$ & $\frac{\lambda_{+}}{H}$ & $\frac{\lambda_-}{H}$ &  $Nature$ \\
  \hline
-10& 2.83& -28.33& -0.16& -1.05& $3.44\times10^{-7}$&
  $1.02\times10^7$&$unstable$\\
  -9.5& 1.83& -17.39& -0.15& -0.94&
  $2.41\times10^{-6}$& $1.15\times10^6$& $unstable$\\
  -9.0&  1.47& -13.21& -0.13& -0.81& $6.73\times10^{-6}$&
  302388& $unstable$\\
  -8.5& 1.31& -11.12& -0.11& -0.66&
  $1.09\times10^{-5}$& 125336& $unstable$\\
  -8.0&  1.25& -9.97& -0.09& -0.51& $1,07\times10^{-5}$&
  74690.2& $unstable$\\
  -7.5& 1.25& -9.34& -0.07& -0.34&
  $5.96\times10^{-6}$& 59820.2& $unstable$\\
  -7.0&  1.29& -9.04& -0.04& -0.12& $1.20\times10^{-6}$&
  40429.9& $unstable$\\
  -6.5& 1.38& -8.96& -0.13& -1.11&
  $3.38\times10^{-6}$& $1.15\times10^6$& $unstable$\\
  -6.0& 1.5& -9& -&
-& -& -&-\\
-5.5&  1.64& -9.02& 0.04& 0.22& 210576&
$7.46\times10^{-7}$& $unstable$\\
-5.0& 1.76& -8.79& 0.09& 0.44&
  326848& $1.85\times10^{-6}$& $unstable$\\
  -4.5& 1.78& -7.98& 0.14& 0.68&
   311103& $4.66\times10^{-6}$& $unstable$\\
   -4.0& 1.61& -6.45& 0.21& 0.95&
  160192& $1.8\times10^{-5}$& $unstable$\\
  -3.5& 1.29& -4.50& 0.28& 1.20&
   30243.1& $1.5\times10^{-4}$& $unstable$\\
   -3.0& 0.91& -2.72& 0.32&
  1.17& -0.002& -2104.02& $stable$\\
  -2.5& 0.58& -1.44&
  0.31& 0.85& -0.002& -1087.69& $stable$\\
  -2.0&  0.33& -0.67& 0.29&
  0.47& -0.01& -104.12& $stable$\\
  -1.5& 0.16& -0.25&
  0.31& 0.25& -0.03& -8.04& $stable$\\
  -1.0&  0.05& -0.05& 0.33&
  0.06& -0.0006& -20.34& $stable$\\
  -0.5& -0.037& 0.019&
  0.38& -0.04& 0.0001& 44.14& $unstable$\\
  0.0& -0.10 &0&
  0.47& -0.09& 0.0004&
  65.34& $unstable$\\
  0.5& -0.16& -0.08&
  0.61& -0.09& -42.27& -0.0006& $stable$\\
  1.0& -0.22& -0.22&
  0.69& -0.38& -16.09& -0.03& $stable$\\
  1.5& -0.31&-0.46&
  0.79& -0.68& -63.82& -0.02& $stable$\\
  2.0& -0.46&-0.92&
  0.75& -1.25& -684.79& -0.007& $stable$\\
  2.5& -0.91&
-2.27& 0.49& -1.89& -15144.2& -0.0007& $stable$\\
3.0& 57& 171&
  0.11& -2.25& $-1.84\times10^{11}$& 0& $quasistable$\\
  3.5& 0.80&
  2.81& -0.19& -2.28& 0.007& 2392.19& $unstable$\\
  4.0& 0.38&
  1.53& -0.35& -2.15& 0.02& 954.45& $unstable$\\
  4.5& 0.24&  1.08& -0.43& -1.98& 0.04& 327.43& $unstable$\\
  5.0& 0.17&
  0.85& -0.13& -2.49& 0.10& 187.57& $unstable$\\
  5.5& 0.13&
  0.70& -0.47& -1.71& 0.14& 67.84& $unstable$\\
  6.0& 0.1&  0.6& -0.50& -1.50& 0.21& 34.20& $unstable$\\
  6.5& 0.08&
  0.52& -0.52& -1.30& 0.29& 18.54& $unstable$\\
\hline
\end{tabular}
\end{center}
\begin{center}
Continue of the table 1 \vskip .5cm \noindent
\begin{tabular}{|c|c|c|c|c|c|c|c|}
  \hline
  $\epsilon$ & $\alpha$ & $\beta$ & $\frac{X_c}{H}$ & $\frac{V_c}{H}$ & $\frac{\lambda_{+}}{H}$ & $\frac{\lambda_-}{H}$ &  $Nature$ \\
  \hline
  7.0& 0.07&
  0.46& -0.53& -1.14& 0.38& 10.78& $unstable$\\
  7.5& 0.06&
  0.41& -0.53& -0.99& 0.46& 6.79& $unstable$\\
  8.0& 0.047&
  0.37& -0.53& -0.87& 0.51& 4.67& $unstable$\\
  8.5&0.04&
   0.34& -0.52& -0.78& 0.53& 3.57& $unstable$\\
   9.0& 0.03&
  0.31& -0.51& -0.69& 0.49& 3.05& $unstable$\\
  9.5& 0.03&
   0.28& -0.50& -0.62& 0.43& 2.85& $unstable$\\
   10&
  0.03& 0.26& -0.49& -0.56& 0.35& 2.84& $unstable$\\
  \hline
\end{tabular}
\end{center}
\begin{center}
Table 2: Numerical values of the critical points from
$h(\alpha,\beta)=0$ \vskip .5cm \noindent
\begin{tabular}{|c|c|c|c|c|c|c|c|}
  \hline
  $\epsilon$ & $\alpha$ & $\beta$ & $\frac{X_c}{H}$ & $\frac{V_c}{H}$ & $\frac{\lambda_{+}}{H}$ & $\frac{\lambda_-}{H}$ &  $Nature$ \\
  \hline
-9.5&  0.0002& -0.002& -0.07& -0.0007& $-1.64\times10^{19}$&
  $1.64\times10^{19}$& $quasistable$\\
   -9&  0.0003& -0.003& -0.06& -0.0007& $-3.43\times10^{17}$&
  $3.43\times10^{17}$& $quasistable$\\
  -8.5&  0.0006& -0.005& -0.05& -0.0007& $-1.17\times10^{16}$&
  $1.17\times10^{16}$& $quasistable$\\
  -8&  0.012& -0.095& -0.05& -0.002& $-1.15\times10^{13}$&  $1.15\times10^{13}$& $quasistable$\\
  2& 0.023& 0.06&
  1.83& -0.17& -13.83& 29.76& $quasistable$\\
\hline
\end{tabular}
\end{center}
\begin{center}
Table. 3: Directional barotropic indexes for stable critical
points \vskip .5cm \noindent
\begin{tabular}{|c|c|c|c|}
  \hline
  $\epsilon$ & $\gamma_{\|}$ & $\gamma_{\perp}$& $\bar{\gamma}$\\
 \hline
-3.0& -1.94411& -0.871808&-1.22924\\
-2.5& -1.91385& -0.870343&-1.21818\\
-2.0& -1.83111& -0.867365&-1.18861\\
-1.5& -1.8807& -0.868963&-1.20621\\
-1.0&-1.97877& -0.873716&-1.24207\\
0.5& -4.16604& -1.22154&-2.20304\\
1.0& -5.35907& -1.47972&-2.77284\\
1.5& -8.73365& -2.26929&-4.42408\\
2.0& -6.88267& -1.82964&-3.51398\\
2.5& -2.89546& -0.98289&-1.62041\\
3.0& -1.24608& -0.913408&-1.0243\\
\hline
\end{tabular}
\end{center}
\newpage
\begin{center} Table 4: Numerical values of the critical points
from $Q_1=0=Q_2$ \vskip .5cm \noindent
\begin{tabular}{|c|c|c|c|c|}
  \hline
  $\epsilon$ & $\alpha$ & $\beta$ & $S=\frac{X_c}{H}$ & $W=\frac{U_c}{H}$  \\
  \hline
-10& -0.5008283783& 5.008283783& 5.672177113& -2.540701569
\\
  -9.5&0.008588129709&-0.08158723224& 3.023252473& 19.19903027
  \\
  -9.0& -0.3419657915& 3.077692124& 5.866672536& -2.870775434\\
  -8.5& 0.03111637642&-0.2644891996& 6.991106084& -3.825445223\\
  -8.0& -0.2021739560& 1.617391648& -0.2921491238& -73.02353124\\
  -7.5&-0.2013895375& 1.510421531& -0.3042075592& -69.56093193\\
  -7.0& 0.01762796647&-0.1233957653& 0.9880386922& -2.852346682\\
  -6.5& 0.001318156955& -0.008568020208& 4.334744006& 3.092910056\\
  -6.0& 0.007708235615& -0.04624941369& 7.682244662& 17.46535314\\
-5.5&  0.02101479362& -0.1155813649&1.015056666& 0.9609554350\\
-5.0& 0.02273691836& -0.1136845918& 1.011411199&.9009535194\\
  -4.5& 0.3358036901& -1.511116605& -0.8025296715& 1.084975158\\
   -4.0& -0.6825737268& 2.730294907& 1.000000000& 1.500000000\\
  -3.5& -0.07331914408& 0.2566170043& -0.6999346528& 1.422764741\\
   -3.0& 0.03296306120& -0.09888918360& 1.009596676&
  -3.000322516\\
  -2.5& 0.002500151696&-0.006250379240&
  0.8768125637& -0.1587420570\\
  -2.0&  -0.1976806957& 0.3953613914& -0.6505670194&
  -43.03853291\\
  -1.5& -1.841640730& 2.762461095&
 1.264060040& 0.08366129551\\
  -1.0&  0.03578282272& -0.03578282272& 1.108321598&
 -3.148271005\\
  -0.5&-0.09366501789& 0.04683250894&
  0.9640215072& 4.949684114\\
  0.0& -0.1147744417 &0&
  0.9430243343&5.452307324\\
  0.5& -1.613998791& -0.8069993955&
  0.1402507731& 5.065940205\\
  1.0& -0.07762607582&-0.07762607582&
  0.8724488468& -2.226578342\\
  1.5& -0.04309607660&-0.06464411490&
  -0.2659450362& -0.6480856992\\
  2.0&-0.06386237809&-0.1277247562&
  1.077334291& 4.815759124\\
  2.5& -0.01048460772&
-0.02621151930& 1.673294323& 4.766717191\\
3.0& -0.01085650863& -0.03256952589&
  1.763225100& 6.111778140\\
  3.5& -0.01114082063&
 -0.03899287220& 1.832465308& 7.405531418\\
  4.0& -0.01125014540&
  -0.04500058160& 1.889026726& 8.596966328\\
  4.5& -0.6420938245&  -2.889422210& -0.1830711047& -1.269557693\\
  5.0& -0.03639923426&
  -0.1819961713& 1.045354179& 3.237078633\\
  5.5& -0.03293268964&
  -0.1811297930&1.044732163& 3.091150882\\
  6.0& -0.02998645789&  -0.1799187473& 1.044262782& 2.968940037\\
  6.5& -0.02747136773&
 -0.1785638902& 1.043890582& 2.865702647\\
\hline
\end{tabular}
\end{center}
\begin{center}
Continue of the table 4 \vskip .5cm \noindent
\begin{tabular}{|c|c|c|c|c|}
  \hline
  $\epsilon$ & $\alpha$ & $\beta$ & $S=\frac{X_c}{H}$ & $W=\frac{U_c}{H}$  \\
  \hline
  7.0&-0.3749835875&
  -2.624885112&-0.3090067417& -1.022704328\\
  7.5& -0.02344217531&
  -0.1758163148&1.043321302& 2.702014357\\
  8.0& -0.01760041930&
  -0.1408033544& 0.9575196268& -2.555981985\\
  8.5&-0.5317122979&
  -4.519554532& -0.2784902851& -1.120341334\\
   9.0& -0.009107176783&
  -0.08196459105&2.175421464&16.12062525\\
  9.5& -0.01800644937&
   -0.1710612690& 1.042536599& 2.483679346\\
   10&
  -0.3484878224& -3.484878224& -0.3724881052& -0.9969579023\\
  \hline
\end{tabular}
\end{center}
\begin{center} Table. 5: Numerical eigenvalues of the critical points given in the table 4
\vskip .5cm \noindent
\begin{tabular}{|c|c|c|c|c|}
  \hline
  $\epsilon$ & $\frac{\lambda_{1}}{H}$ & $\frac{\lambda_2}{H}$ &$\frac{\lambda_3}{H}$&  $Nature$ \\
  \hline
-10&-56.09553611& -2.966073405& 14.32858140& $quasistable$
\\
  -9.5&-16.42079332&-16.42079332& -16.42079332& $stable$
  \\
  -9.0& -39.41075939& -4.469490793& 23.55890209& $quasistable$\\
  -8.5& -10.47026903&-10.47026903& -10.47026903&$stable$\\
  -8.0&-188.5915478& 230.3312676& 1382.767250& $quasistable$\\
  -7.5&-269.8366691&236.5906765& 1622.736873& $quasistable$\\
  -7.0& 3.127075042&3.127075042& 3.127075042& $unstable$\\
  -6.5& -23.36759615& -23.36759615& -23.36759615& $stable$\\
  -6.0& -117.5763827& -117.5763827& -117.5763827&$stable$\\
-5.5&  4.369038156& 4.369038156&4.369038156& $unstable$\\
-5.0& 4.718381182& 4.718381182& 4.718381182&$unstable$\\
  -4.5& -13.92106764& -13.92106764& -13.92106764& $stable$\\
   -4.0& -& -& -&-\\
  -3.5& -6.137455414& -6.137455414& -6.137455414&$stable$\\
   -3.0&33.83392885&33.83392885& 33.83392885&
  $unstable$\\
  -2.5& -269.2680738&-26.66528946&
  -2.560732253&$stable$\\
  -2.0&  -2011.747812& 281.2079672& 997.8665124&
 $quasistable$\\
  -1.5& 26.07908996& 26.07908996&
 26.07908996& $stable$\\
  -1.0&  2.537984967& 2.537984967& 2.537984967&
 $unstable$\\
  -0.5&-106.3899284& -9.725278664&
 16.98321000& $quasistable$\\
  0.0& -437.4483747 &-10.68657937&
  9.866459397&$quasistable$\\
  0.5& -46.57836240& -46.57836240&
  -46.57836240& $stable$\\
  1.0& -146.8128188&-0.3600111454&
 1.135133633& $quasistable$\\
  1.5&-21.50135808&-3.408923614&
  8.344092637&$qusistable$\\
\hline
\end{tabular}
\end{center}
\begin{center}
Continue of the table 5 \vskip .5cm \noindent
\begin{tabular}{|c|c|c|c|c|}
  \hline
  $\epsilon$ & $\alpha$ & $\beta$ & $S=\frac{X_c}{H}$ & $W=\frac{U_c}{H}$  \\
  \hline
  2.0&-9.190843385&1.280167729&
  162.7602089& $quasistable$\\
  2.5& -899.2040823&
-21.47910549&23.40368260& $quasistable$\\
3.0& 272.0139052& 272.0139052&
  272.0139052& $unstable$\\
  3.5& 156.2493894&
 156.2493894& 156.2493894& $unstable$\\
  4.0& 124.6155935&
  124.6155935& 124.6155935& $unstable$\\
  4.5& -71.05639837&  3.834442359& 15.33932781& $qusistable$\\
  5.0&-6.431832124&
  3.046146802& 19.39156395& $quasistable$\\
  5.5&-5.981912994&
  3.375362941&15.81024393& $quasistable$\\
  6.0& -5.577163990& 3.797156875& 12.88056976& $qusistable$\\
  6.5& -5.210831649&
4.413814117& 10.27377589& $quasistable$\\
  7.0&-7.654910829&
  -7.654910829&-7.654910829& $stable$\\
  7.5&-4.570699075&
  -4.570699075&-4.570699075& $stable$\\
  8.0&-43.27248566&
  0.09713401841& 1.118632369& $quasistable$\\
  8.5&-3.292817705&
  8.807091991& 36.08526440& $quasistable$\\
   9.0&83.46916396&
  83.46916396&83.46916396&$unstable$\\
  9.5& -3.550382482&
  -3.550382482& -3.550382482& $stable$\\
   10&
  3.693672829& 3.693672829& 3.693672829& $unstable$\\
  \hline
\end{tabular}
\end{center}
 \begin{center} Table 6: Numerical values of the directional barotropic indexes for stable critical points for case $A_{\|}=0,A_{\perp}\neq0$
 \vskip .5cm \noindent
\begin{tabular}{|c|c|c|c|}
  \hline
  $\epsilon$ & $\gamma_{\|}$ & $\gamma_{\perp}$ & $\bar{\gamma}$ \\
  \hline
  -9.5 & 1.988507379 & 0.8742941653 & 1.245698570 \\
 -8.5& -0.5334982425&-1.539189875& -1.203959330 \\
  -6.5 & 1.599746186 & 0.8687609176 & 1.112422674 \\
 -6.0 &1.299300624 & 0.9020587737 & 1.034472724\\
 -4.5 & -0.1095517779&-6.873465855 & -4.618827829 \\
  -3.5 & -0.1765158132 & -4.293040071 & -2.920865318  \\
   -2.5 & -15.23542190 & -3.858082860 & -7.650529207\\
 -1.5& 8.574035053 & 2.230982164 & 4.345333127 \\
  0.5& -1.326259725 & -0.8970650260 & -1.040129926  \\
  7.0 & -0.5278760119 &-1.552757186 & -1.211130128 \\
 7.5 & 47.16666426 & 11.80756712 & 23.59393283\\
  9.5 & 48.01833369 &12.02020245 & 24.01957953\\
  \hline
\end{tabular}
\end{center}
\end{document}